# Development of realistic simulations for the polarization of the cosmic microwave background

**Marta Monelli**

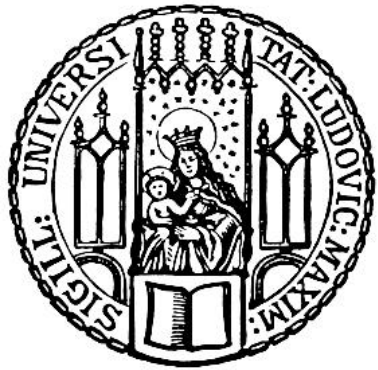

München 2024

# Development of realistic simulations for the polarization of the cosmic microwave background

**Marta Monelli**

Dissertation
an der Fakultät für Physik
der Ludwig–Maximilians–Universität
München

vorgelegt von
Marta Monelli
aus Poggibonsi (SI), Italia

München, den 29.02.2024

# Contents

# List of Figures

# List of Tables

# Zusammenfassung

Die Polarisation des kosmischen Mikrowellenhintergrunds (CMB) kann zur Erforschung der kosmischen Inflation (durch Messung der primordialen $B$-Moden) und zur Untersuchung der paritätsverletzenden Physik (durch Nachweis der kosmischen Doppelbrechung) beitragen. Diese vielversprechenden Möglichkeiten treiben die Entwicklung einer Reihe neuer boden-, ballon- und weltraumgestützter CMB-Experimente voran. Für den Erfolg dieser ehrgeizigen Missionen müssen jedoch systematische Effekte genau kontrolliert und reduziert werden.

Zu diesem Zweck werden einige CMB-Experimente der nächsten Generation (einschließlich LiteBIRD) rotierende $\lambda/2$-Plättchen (engl. half-wave plates oder HWPs) als Polarisationsmodulatoren verwendet. Im Idealfall sollte diese Wahl die $1/f$-Rauschkomponente in der beobachteten Polarisation vollständig unterdrücken und den Intensitäts-Polarisations-Leckstrom reduzieren, wodurch zwei wichtige systematische Effekte abgeschwächt werden. Jede reale HWP hat jedoch Nichtidealitäten, die, wenn sie in der Analyse nicht richtig behandelt werden, zu zusätzlichen systematischen Effekten führen können.

Nach einer kurzen Einführung in die wissenschaftliche Fragestellung diskutiert diese Doktorarbeit die allgemeinen Charakteristika jedes CMB-Experiments, stellt die HWP vor und präsentiert eine neue Simulationspipeline für zeitlich geordnete Daten (time-ordered data, TOD), die auf ein LiteBIRD-ähnliches Experiment zugeschnitten ist und TOD und Himmelskarten für realistische Detektoren und HWPs liefern kann.

Wir zeigen, dass die Simulationsmethode verwendet werden kann, um zu untersuchen, wie nicht-ideale HWPs den gemessenen kosmischen Doppelbrechungswinkel beeinflussen, was bei einer realistischen Wahl der HWP zu einer Abweichung von einigen Grad führt. Wir leiten auch analytische Formeln her, die die beobachteten Temperatur- und Polarisationskarten modellieren, und validieren sie anhand der Simulationsergebnisse.

Schließlich stellen wir ein einfaches semi-analytisches Modell vor, um die HWP-Nichtidealitäten durch die Analyseschritte zu propagieren, die für jedes CMB-Experiment notwendig sind (Beobachtung von Multifrequenzkarten, Vordergrundkorrektur und Schätzung von Leistungsspektren), und berechnen die HWP-bedingte Veränderung des geschätzten Tensor-zu-Skalar-Verhältnisses $r$, wobei wir feststellen, dass die HWP zu einer Unterschätzung von $r$ führt. Wir zeigen auch, wie die Kalibrierung der Signalverstärkung der CMB-Temperatur verwendet werden kann, um die negative Auswirkungen der HWP teilweise zu kompensieren, und geben eine Reihe von Empfehlungen für das HWP-Design, die dazu beitragen können, die Vorteile der Verstärkungskalibrierung zu maximieren.

# Abstract

Polarization of the cosmic microwave background (CMB) can help probe cosmic inflation (via the measurement of primordial $B$ modes) and test parity-violating physics (via the detection of cosmic birefringence). These promising opportunities are driving the development of a number of new ground-based, balloon-borne and space-based CMB experiments. However, for these ambitious missions to be successful, systematic effects must be precisely controlled and accurately mitigated.

To this end, some next-generation CMB experiments (including LiteBIRD) will use rotating half-wave plates (HWPs) as polarization modulators. Ideally, this choice should completely remove the $1/f$ noise component in the observed polarization and reduce the intensity-to-polarization leakage, thus mitigating two important systematic effects. However, any real HWP is characterized by non-idealities which, if not properly treated in the analysis, can lead to additional systematics.

In this thesis, after briefly introducing the science case, we discuss the macro steps that make up any CMB experiment, introduce the HWP, and present a new time-ordered data (TOD) simulation pipeline tailored to a LiteBIRD-like experiment that can return TOD and binned maps for realistic beams and HWPs.

We show that the simulation framework can be used to study how the HWP non-idealities affect the measured cosmic birefringence angle, resulting in a bias of a few degrees for a realistic choice of HWP. We also derive analytical formulae that model the observed temperature and polarization maps and test them against the output of the simulation.

Finally, we present a simple, semi-analytical end-to-end model to propagate the HWP non-idealities through the macro-steps that make up any CMB experiment (observation of multi-frequency maps, foreground cleaning, and power spectra estimation) and compute the HWP-induced bias on the estimated tensor-to-scalar ratio, $r$, finding that the HWP leads to an underestimation of $r$. We also show how gain calibration of the CMB temperature can be used to partially mitigate the non-idealities' impact and present a set of recommendations for the HWP design that can help maximize the benefits of gain calibration.

# Chapter 1

# Introduction

**Summary:** In this chapter, we review some of the key results obtained from observations of the cosmic microwave background (CMB) since its discovery, discuss what kind of new physics can be extracted from CMB polarization, and emphasize the importance of controlling systematic effects in order to achieve this goal. Finally, we outline the content of the rest of this thesis.

Over the past 60 years, observations of the cosmic microwave background (CMB) have been an invaluable source of cosmological information and have played a key role in shaping our understanding of the universe, helping to solidify three central ideas in modern cosmology:

**Big Bang model:** The Big Bang model describes an expanding universe that is homogeneous and isotropic on large scales. Because of the expansion, the density and temperature of the universe increase as we go back in time, eventually leading to a singularity at some early enough time, called the Big Bang.

**$\Lambda$CDM model:** According to the $\Lambda$CDM model, the universe is filled with three main components: *i) dark energy*, a dark component with negative pressure typically modeled by a cosmological constant, $\Lambda$, *ii) cold dark matter* (CDM), which has so far only been detected by its gravitational interaction, and *iii) ordinary matter*. The $\Lambda$CDM model is a particular Big Bang model and is often referred to as the standard cosmological model.

**Inflationary paradigm:** Inflation is a postulated phase of accelerated expansion believed to have occurred at very early times, immediately after the Big Bang. Originally proposed to address some shortcomings of the Big Bang model (the horizon and flatness problems), inflation provides a natural mechanism for generating initial conditions for cosmological perturbations from primordial vacuum quantum fluctuations.

A rigorous mathematical description of the standard cosmological model and the inflationary paradigm goes beyond the scope of this thesis, but some excellent references are [1–3]. In the next paragraph, we will instead focus on the observational features of the CMB supporting this theoretical framework.

**Evidence supporting the Big Bang model**   In 1964, while attempting to remove all spurious signals from a super-sensitive communications antenna, Penzias and Wilson were left with an isotropic antenna temperature excess of $3.5 \pm 1.0\,\mathrm{K}$ at $4080\,\mathrm{MHz}$. Their results were published in 1965 [4], along with a companion letter by Dicke, Peebles, Roll, and Wilkinson, who interpreted the excess as a signature of the Big Bang model [5]. They argued that a homogeneous, isotropic and expanding universe should be filled by relic black-body radiation coming from the high temperature stage.

To test their interpretation, a number of experiments attempted to measure the black-body spectrum over a wide range of frequencies. Among them, the Far Infrared Absolute Spectrophotometer (FIRAS) instrument, operated from 1989 to 1993 aboard the COsmic Background Explorer (COBE) satellite, provided the best data, confirming that the CMB spectrum is well explained by a black-body with temperature $T_0 = 2.72548 \pm 0.00057\,\mathrm{K}$ [6].

The observational evidence that the CMB is a (nearly) isotropic black-body is one of the 'pillars' of the Big Bang model, since it supports the idea that our universe is homogeneous and isotropic on large scales, and that it is expanding.

**Cosmological parameters from anisotropies**   Penzias and Wilson's discovery was followed by a series of crucial theoretical predictions about what cosmological signatures might be imprinted in the CMB as temperature anisotropies (see for example [7–9]). The measurement of the CMB anisotropies became an important observational goal, driving the development of a number of experiments. The first full-sky observation was made in 1992, and better and better data followed over the next two decades.

**1990s:**  In 1992, the Differential Microwave Radiometers (DMR) instrument aboard COBE detected low-resolution ($\theta_{\mathrm{res}} \sim 7°$) relative temperature fluctuations $\delta T(\hat{n})/T_0 \sim 10^{-5}$ [10], that were compatible with the predictions from inflationary models [11–13]. COBE's success led to the proposal of a follow-up mission in 1995: the Wilkinson Microwave Anisotropy Probe (WMAP).

**2000s:**  WMAP operated from 2001 to 2010 with significant improvements in both angular resolution and noise level over COBE (see Table 1.1), resulting in better temperature measurements and the observation of the polarized CMB. Overall, the WMAP data fit a six-parameter[1] $\Lambda$CDM model remarkably well [14–16].

**2010s:**  The *Planck* Surveyor, launched in 2009, followed WMAP and provided us with the most precise full-sky maps of CMB temperature and polarization to date. *Planck*'s technical improvements over WMAP (see again Table 1.1), made possible to constrain the $\Lambda$CDM parameters with greater precision [17–19].

---

[1]The six parameters of the basic $\Lambda$CDM model are the physical baryon density, $\Omega_b h^2$; the physical CDM density, $\Omega_c h^2$; the dark energy density $\Omega_\Lambda$; the amplitude of primordial scalar curvature perturbations, $\Delta_{\mathcal{R}}^2$ at $k = 0.002\,\mathrm{Mpc}^{-1}$; the power-law spectral index of primordial density (scalar) perturbations, $n_s$; and the reionization optical depth, $\tau$ [14].

Several ground-based and balloon-borne experiments have provided valuable complementary temperature and polarization data [20–28] that support the observations from space by COBE, WMAP and *Planck*. To date, we have very precise measurements of the CMB temperature at all angular scales and good polarization data, especially on small angular scales. This can be seen in Figure 1.1, where past observations are plotted together with the predic-tions for the polarization measurement by the future Lite (Light) satellite for the studies of $B$-mode polarization and Inflation from cosmic background Radiation Detection (Lite-BIRD) [29]. Together with several other future CMB experiments, LiteBIRD aims to extract the valuable information still encoded in the polarization of the CMB.

|  | $n_{\mathrm{det}}$ | $n_{\mathrm{chan}}$ | $\theta_{\mathrm{res}}$ | NET [mK$\sqrt{\mathrm{s}}$] |
|---|---|---|---|---|
| COBE | 6 | 3 | 7° | ~ 30 @ 90 GHz |
| WMAP | 20 | 5 | 0.2° | ~ 0.7 @ 94 GHz |
| *Planck* | 74 | 9 | 0.1° | ~ 0.04 @ 100 GHz |

Table 1.1: Instrument specifics for COBE, WMAP and *Planck*: number of detectors, $n_{\mathrm{det}}$, number of frequency channels, $n_{\mathrm{chan}}$, and angular resolution, $\theta_{\mathrm{res}}$, together with noise equivalent temperature (NET) of the channels centered at ~ 100 GHz. The NET is defined in Appendix A, where we also show how to compute the specific values for the three experiments.

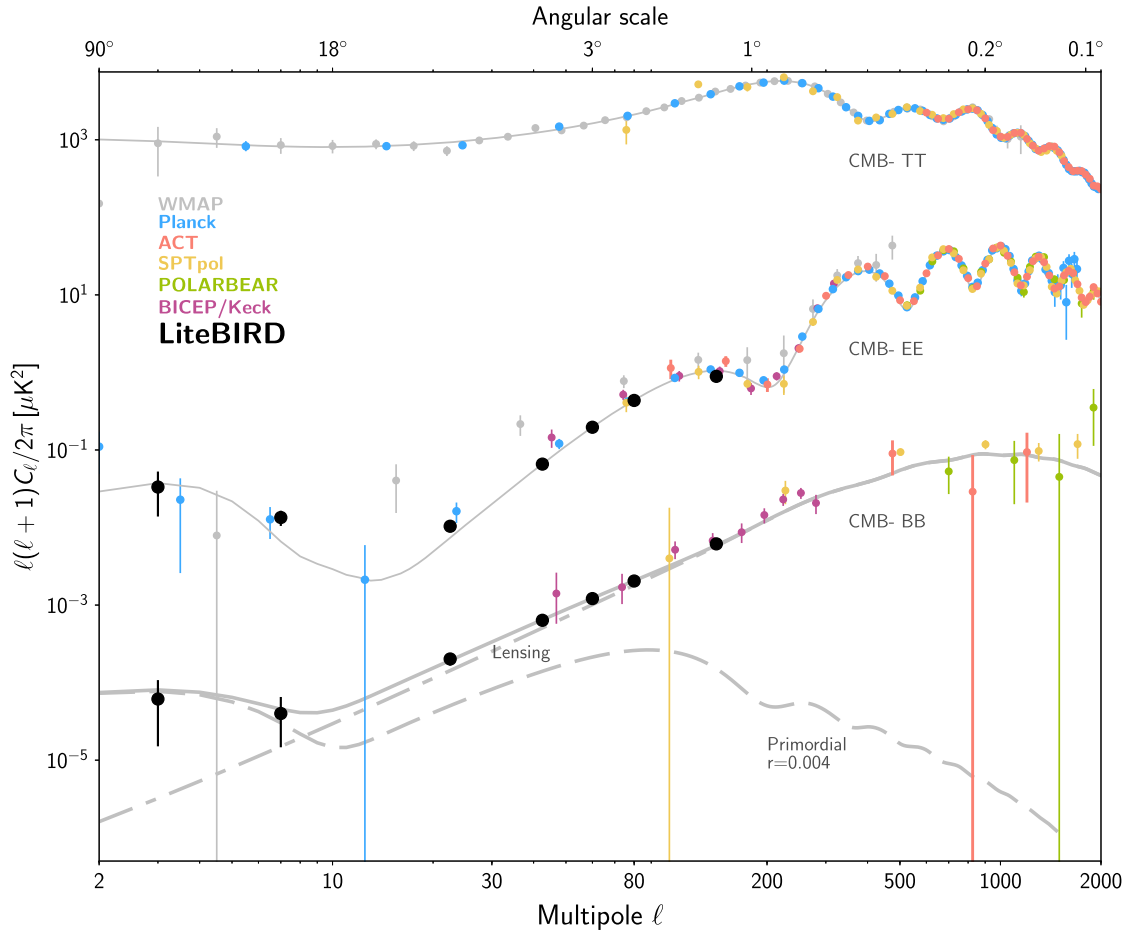

Figure 1.1: CMB temperature, *E* modes and *B* modes angular power spectra. The gray lines show theoretical spectra (assuming the best-fit ΛCDM parameters and *r* = 0.004), the colored points represent past measurements [14, 15, 18, 22–28], and the black points show LiteBIRD's expected polarization sensitivity. Image and caption adapted from [29].

## 1.1   New physics from CMB polarization

Future observations of the polarized CMB aim to shed light on some open questions in modern cosmology and particle physics. A detection of the primordial $B$-mode signal could help us constrain inflationary models, and a measurement of the cosmic birefringence angle could help us test parity-violating physics. Here we briefly introduce these two possibilities (see Sections 2.2 and 2.3 for a more detailed discussion).

**Constraining inflationary models from primordial $B$ modes**   Inflation is expected to source initial conditions for scalar [30–34], vector, and tensor [35, 36] cosmological perturbations. However, vector modes are expected to decay rapidly after horizon re-entry and we therefore neglect them. As for scalar and tensor perturbations, many inflationary models predict their power spectra to obey power laws: $P_s(k) = A_s k^{n_s-1}$ and $P_t(k) = A_t k^{n_t}$, respectively. The relative amplitude of scalar and tensor primordial perturbations is quantified in terms of the tensor-to-scalar ratio, $r = A_s/A_t$, whose value is model dependent.

Since tensor perturbations [35, 36] would leave a distinct $B$-mode signature on the CMB polarization [37–40], $r$ could be inferred from the angular power spectrum of the primordial $B$ modes. To date, CMB observations have only placed upper bounds on $r$, the tightest being $r < 0.032$ (95% CL) [41] (see also [28, 42, 43]), but future surveys, involving both ground-based (Simons Observatory (SO) [44], South Pole Observatory [45] and CMB Stage-4 [46]) and spaceborne (LiteBIRD [29] and the Probe of Inflation and Cosmic Origins (PICO) [47]) missions, aim for unprecedentedly low overall uncertainties, which, depending on the true value of $r$, would lead to a detection or a tightening of the upper bounds. Both these outcomes would allow us to place strong constraints on inflationary models [48, 49].

**Probing parity-violating physics from cosmic birefringence**   The unknown nature of dark matter and dark energy is one of the most elusive mysteries in modern physics, and some models consider the possibility that a parity-violating pseudoscalar field, $\chi$, could be responsible for both [50, 51]. If this is the case, and if $\chi$ has a parity-violating coupling to the electromagnetic field, then the CMB photons would certainly be affected by $\chi$. In particular, if $\chi$ is time-dependent, the linear polarization plane of CMB photons would rotate as they travel toward us [52–54].

Because of its similarity with photon propagation through a birefringent material, this phenomenon is referred to as cosmic birefringence. The cosmic birefringence angle, $\beta$, denotes the overall rotation angle from last scattering to today. Although the effect of $\beta$ on the observed CMB angular power spectra is degenerate with an instrumental miscalibration of the polarization angle [55–58], the methodology proposed in [59–61], which relies on the polarized Galactic foreground emission to determine miscalibration angles, allowed to infer $\beta = 0.35 \pm 0.14°$ at 68% C.L. [62] from nearly full-sky *Planck* polarization data [63]. Subsequent works [64–66] reported more precise measurements of $\beta$.

The statistical significance of $\beta$ is expected to improve with the next generation of CMB experiments, given the high precision at which they aim to calibrate the absolute position angle of linear polarization. This will make it unnecessary to rely on the Galactic foreground to calibrate angles and measure $\beta$ [49], thus avoiding the potential complications highlighted in [67]. If the detection of a non-zero $\beta$ were to be confirmed, it would directly probe parity violation and would help constrain dark matter and dark energy models.

### 1.1.1   The need to study systematic effects

The unprecedented sensitivity goals of future surveys can only be achieved if systematic effects are well understood and kept under control. For example, an unmitigated $1/f$ noise component could prevent us from detecting the primordial $B$ modes on large scales. Another problematic effect is the miscalibration of the polarization angle which, being degenerate with cosmic birefringence, would prevent us from measuring $\beta$ directly.

A promising strategy to reduce some of these systematic effects is to employ a rotating half-wave plate (HWP) as a polarization modulator. As shown by previous analyses [68–75], a rotating HWP can both mitigate the $1/f$ noise component [68] and reduce a potential temperature-to-polarization ($I \rightarrow P$) leakage due to the pair differencing of orthogonal detectors' readings [76, 77]. Because of these advantages, HWPs are used in the design of some next-generation experiments, including SO [44] and LiteBIRD [29]. However, non-idealities in realistic HWPs induce additional systematic effects which should be well understood for future experiments to meet their sensitivity requirements.

In this thesis, we present two different approaches to study the systematic effects induced by the HWP non-idealities. In one case, the effect of the HWP non-idealities is accurately simulated at the level of the time-ordered data (TOD), and in the other, it is instead approximately modeled in the observed maps. These two methods are both valuable and nicely complementary: realistic simulations can account for systematic effects in their (at least partial) complexity, while approximate models are extremely helpful to gain some intuition about the problem at hand and represents the first step to develop efficient mitigation strategies.

## 1.2   Content of this thesis

We provide a brief overview of CMB polarization in Chapter 2, where we introduce $E$ and $B$ modes, show how they can arise from scalar and tensor perturbations, and give a few more details on the new physics that can be probed with CMB polarization (inflation and cosmic birefringence). In Chapter 3 we review some of the main steps that make up any CMB experiments (data acquisition, map-making, foreground cleaning and parameter estimation) and introduce the mathematical framework that we use in the rest of the thesis.

Chapter 4 focuses on HWPs: we discuss the motivations that led to their use as polarization modulators and introduce the non-idealities. In Chapter 5 we present a simulation pipeline to include HWP non-idealities at the level of the detected data for a LiteBIRD-like experiment. In chapters 6 and 7, we study how the HWP non-idealities would affect the measured cosmic birefringence angle and tensor-to-scalar ratio, respectively, employing both TOD simulations and semi-analytical models. Conclusions and future perspectives are presented in Chapter 8.

The content of this thesis is based on the following publications:

[78] **M. Monelli**, E. Komatsu, A. E. Adler, M. Billi, P. Campeti, N. Dachlythra, A. J. Duivenvoorden, J. E. Gudmundsson and M. Reinecke: *"Impact of half-wave plate systematics on the measurement of cosmic birefringence from CMB polarization"*. Preprint: 2211.05685. Published in JCAP, Volume 2023, March 2023.

[79] **M. Monelli**, E. Komatsu, T. Ghigna, T. Matsumura, G. Pisano and R. Takaku: *"Impact of half-wave plate systematics on the measurement of CMB B-mode polarization"*. Preprint: 2311.07999. Sumbitted to JCAP.

# Chapter 2

# CMB polarization

**Summary:** In this chapter, we provide a brief introduction to CMB anisotropies (focusing on polarization). We show how they can be decomposed into spherical harmonics, introduce $E$ and $B$ modes in harmonic space and how to visualize them in flat-sky approximation. We discuss how CMB polarization can be produced via Thomson scattering of quadrupole anisotropies and how primordial $B$ modes are produced by gravitational waves. Finally, we briefly review the new physics that can be extracted from CMB polarization (inflation and parity violation).

*Section 2.1 is loosely inspired by some lecture notes on physical cosmology I have contributed drafting (the other authors are Andrea Ferrara and Luca Marchetti; front matter and summary are available at* `http://cosmology.sns.it/physical_cosmology_book.html`*; to be published by Edizioni della Normale). Sections 2.2 and 2.3 are instead adapted from [49].*

## 2.1 Introduction

The CMB was emitted about $380\,000$ years after the Big Bang, when the temperature of the universe became low enough to allow electrons and protons to recombine into hydrogen atoms. As the number of free electrons decreased, the scattering rate of $e^- + \gamma \to e^- + \gamma$ eventually fell below the expansion rate, causing the photons to stop interacting with the other constituents of the cosmic fluid, i.e. to decouple. Since then, they have traveled (almost) freely through space, their energy redshifting with the cosmic expansion.

This relic radiation fills all space in the observable universe and reaches us from all directions. In particular, for a given direction of observation $\hat{n}$, what we receive is linearly polarized[1] radiation, which can be characterized in terms of three quantities: $T_{\mathrm{CMB}}(\hat{n})$, $Q_{\mathrm{CMB}}(\hat{n})$, and $U_{\mathrm{CMB}}(\hat{n})$. These are related to the Stokes parameters introduced in Appendix B, with two main differences.

---

[1]The CMB is not expected to have significant circular polarization, so we neglect it.

- In Appendix B, we adopted the usual optics convention of defining the Stokes parameters in a right-handed coordinate system with the $z$ axis along the direction of photon propagation, $\hat{k}$. In CMB literature, Stokes parameters are also defined in a right-handed coordinate system, but the $z$ axis is typically taken along the direction of observation $\hat{n} = -\hat{k}$, while the $x$ and $y$ axes are taken along the increment directions of, respectively, $\theta$ and $\varphi$ (see Figure 2.1). As a consequence, the Stokes $U$ parameter flips sign as one switches from the optics to the CMB convention.

- According to their definition in eq. (B.8), Stokes parameters have units (N/C)$^2$, while $T_{\text{CMB}}(\hat{n})$, $Q_{\text{CMB}}(\hat{n})$, and $U_{\text{CMB}}(\hat{n})$ have temperature units, often K or $\mu$K.

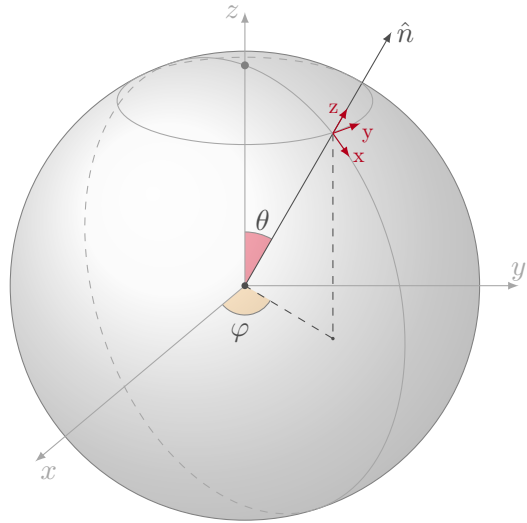

Figure 2.1: Coordinate systema in CMB convention. For a given $\hat{n}$, identified by $\theta$ and $\varphi$, the $z$ axis lies along $\hat{n}$. The $x$ and $y$ axes are taken along the increment directions of, respectively, $\theta$ and $\varphi$.

By integrating $T_{\text{CMB}}(\hat{n})$ over the whole sky, one obtains the average CMB temperature $T_0$:

$$T_0 \equiv \int \frac{\mathrm{d}^2 \hat{n}}{4\pi} T_{\text{CMB}}(\hat{n}) \simeq 2.725 \text{ K} , \qquad (2.1)$$

while $Q_{\text{CMB}}(\hat{n})$ and $U_{\text{CMB}}(\hat{n})$ average to zero. Anisotropies in temperature and polarization contain a great wealth of cosmological information that can be compactly packed into the angular power spectra, $C_\ell^{XY}$. In the remainder of this section, we briefly introduce the $C_\ell^{XY}$, and discuss how scalar perturbations can generate CMB polarization. For the sake of compactness, we will drop the CMB subscripts from now on.

## 2.1.1 Temperature anisotropies

Being a scalar field on the sphere, $T(\hat{n})$ can be decomposed as sum of spherical harmonics, $Y_{\ell m}(\hat{n})$. These are the eigenfunctions of the Laplace operator on the sphere and form a complete set of orthonormal functions, meaning that

$$\int \mathrm{d}^2 \hat{n}\, Y_{\ell m}(\hat{n}) Y_{\ell' m'}^*(\hat{n}) = \delta_{\ell\ell'} \delta_{mm'} , \qquad (2.2)$$

where the $*$ denotes the complex conjugate. Each spherical harmonic is characterized by two integer numbers: the degree $\ell$ and the order $m$. While $\ell$ can take any non-negative integer value, $m$ takes integer values between $-\ell$ and $\ell$. Different values of $\ell$ represent different 'angular frequencies' of the spherical harmonics: $Y_{\ell m}$ draws $\ell$ positive-negative patterns on the sphere. Instead, different values of $m$ represent different 'orientations' of the $\ell$ positive-negative patterns. The spherical harmonics decomposition of $T(\hat{n})$ reads

$$T(\hat{n}) = \sum_{\ell=0}^{\infty} \sum_{m=-\ell}^{\ell} a_{\ell m}^T Y_{\ell m}(\hat{n}) \,, \tag{2.3}$$

where the $a_{\ell m}^T$ are the coefficients of the decomposition, which can be computed by multiplying both sides of (2.3) by $Y_{\ell' m'}$ and integrating over $\mathrm{d}^2 \hat{n}$:

$$a_{\ell m}^T = \int \mathrm{d}^2 \hat{n}\, T(\hat{n}) Y_{\ell m}^*(\hat{n}) \,. \tag{2.4}$$

**Angular power spectrum** If we had an ensemble of universes, we could treat the $a_{\ell m}^T$ coefficients as complex stochastic variables and take ensemble averages to summarize their properties. Under the assumptions that CMB anisotropies are Gaussian, the $a_{\ell m}^T$ would also be (complex) Gaussian variables and all their statistical information would be encoded in $\langle a_{\ell m}^T a_{\ell' m'}^{T*} \rangle$ covariances. In particular, assuming that temperature deviations are statistically isotropic, such an ensemble average can be written as

$$\langle a_{\ell m}^T a_{\ell' m'}^{T*} \rangle = C_\ell^{TT} \delta_{\ell \ell'} \delta_{m m'} \,, \tag{2.5}$$

where $C_\ell^{TT}$ is the temperature angular power spectrum. Since we can only observe our own universe, we have no way of taking the 'ensemble average', but statistical isotropy guarantees that all $a_{\ell m}^T$ coefficients for a given $\ell$ can be regarded as independent quantities drawn from a statistical distribution. This means that the $C_\ell^{TT}$ can be estimated by replacing the ensemble average by an average over $m$:

$$C_\ell^{TT} = \frac{1}{2\ell+1} \sum_{m=-\ell}^{\ell} a_{\ell m}^T a_{\ell m}^{T*} \,. \tag{2.6}$$

Note that as $\ell$ decreases, the number of samples $N = 2\ell + 1$ that we can use to estimate the angular power spectrum becomes smaller and smaller. This introduces an uncertainty known as the cosmic variance.

## 2.1.2 $E$- and $B$-mode polarization

Describing CMB polarization in terms of the Stokes $Q$ and $U$ parameters is problematic, as they are not invariant under rotations of the $x$-$y$ coordinate system. Under a counterclockwise rotation of the basis vectors of an angle $\psi$, $Q(\hat{n})$ and $U(\hat{n})$ transform according to eq. (B.13), or equivalently

$$Q(\hat{n}) \pm iU(\hat{n}) \to [Q(\hat{n}) \pm iU(\hat{n})]\, e^{\mp 2i\psi} \,. \tag{2.7}$$

The combinations $Q \pm iU$ are clearly still coordinate-dependent, but they transform as the $_{\pm 2}Y_{\ell m}$ spin-weighted spherical harmonics[2], which allows us to decompose

$$Q(\hat{n}) \pm iU(\hat{n}) = \sum_{\ell=0}^{\infty} \sum_{m=-\ell}^{\ell} {}_{\pm 2}a_{\ell m} \, {}_{\pm 2}Y_{\ell m}(\hat{n}) \, , \tag{2.8}$$

with coefficients

$$_{\pm 2}a_{\ell m} = \int \mathrm{d}^2\hat{n} \left[ Q(\hat{n}) \pm iU(\hat{n}) \right] {}_{\pm 2}Y_{\ell m}^*(\hat{n}) \, . \tag{2.9}$$

It turns out that the linear combinations

$$a_{\ell m}^E \equiv -\frac{_{+2}a_{\ell m} + {}_{-2}a_{\ell m}}{2} \, , \tag{2.10a}$$

$$a_{\ell m}^B \equiv i\frac{_{+2}a_{\ell m} - {}_{-2}a_{\ell m}}{2} \, , \tag{2.10b}$$

are the spherical harmonic coefficients of two fields, $E(\hat{n})$ and $B(\hat{n})$, which are invariant under rotations, and therefore well suited to describe polarization.

$E$ **and** $B$ **under parity**   In spherical coordinates $(r, \theta, \varphi)$, the parity transformation $\hat{n} \to -\hat{n}$ maps $r \to r$, $\theta \to \pi - \theta$ and $\varphi \to \pi + \varphi$, which causes $U$ to flip sign while leaving $Q$ unchanged. Instead, the spin-weighted spherical harmonics transform according to $_sY_{\ell m} \to {}_{-s}Y_{\ell m}(-1)^{\ell+s}$. By plugging these transformations into eq. (2.9), we find that, under parity, the spin-2 spherical harmonics coefficients change according to

$$_{\pm 2}a_{\ell m} \to {}_{\mp 2}a_{\ell m}(-1)^\ell \, . \tag{2.11}$$

In turn, by plugging this transformation rule into eqs. (2.10), we find that

$$a_{\ell m}^E \to a_{\ell m}^E (-1)^\ell \, , \tag{2.12a}$$

$$a_{\ell m}^B \to a_{\ell m}^B (-1)^{\ell+1} \, , \tag{2.12b}$$

implying that $E$ and $B$ have opposite parity.

**Visualizing** $E$ **and** $B$   Although they are straightforward to define in harmonic space, understanding how $E(\hat{n})$ and $B(\hat{n})$ are related to the values of the Stokes parameters on the sphere can be tricky. They are a bit easier to visualize in flat-sky approximation.

Consider a small portion of the sky around the North pole. In flat-sky approximation, instead of using $(\theta, \varphi)$ pairs, points are identified by a vector $\boldsymbol{\theta}$ on the tangent plane to the sphere at the North pole (see Figure 2.2). The vector has length $\theta$ and forms an angle $\varphi$ with the $x$ axis of the tangent plane. Deriving the expressions for $E(\boldsymbol{\theta})$ and $B(\boldsymbol{\theta})$ goes beyond the scope of this thesis, but the interested reader can take a look at [80]. They show that

---

[2]In general, under a counter-clockwise rotation of the basis vectors of an angle $\psi$, a spin-weighted spherical harmonic $_sY_{\ell m}$ transform as $_sY_{\ell m} \to {}_sY_{\ell m}e^{-is\psi}$.

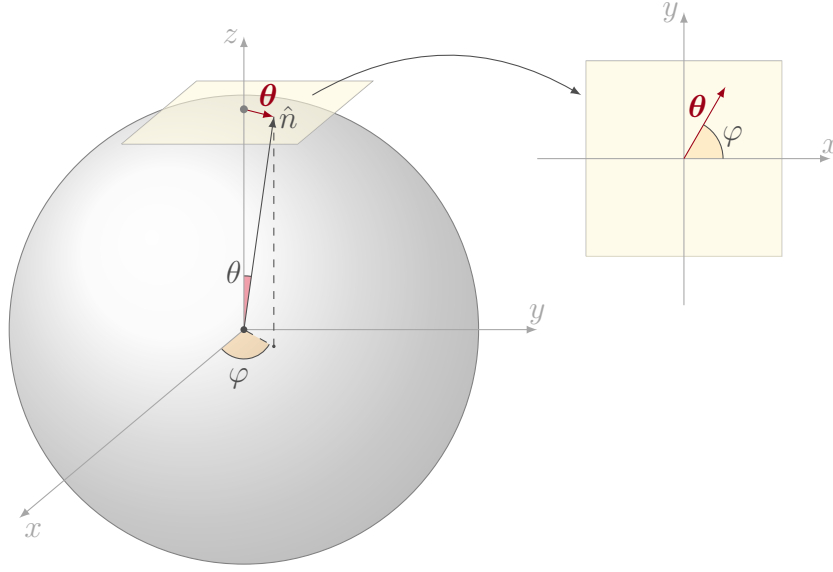

Figure 2.2: A small enough neighborhood of the North pole is approximately flat. The direction $\hat{n}$ identifies therefore a vector $\boldsymbol{\theta}$ on the plane tangent to the North pole. The modulus of the vector is given by $\theta$, while its direction is determined by $\varphi$.

$$E(\boldsymbol{\theta}) = \int \mathrm{d}^2\epsilon\, \omega(\epsilon) \left[ -\mathcal{Q}(\boldsymbol{\theta}+\boldsymbol{\epsilon})\cos 2\varphi_\epsilon - \mathcal{U}(\boldsymbol{\theta}+\boldsymbol{\epsilon})\sin 2\varphi_\epsilon \right], \quad (2.13\mathrm{a})$$

$$B(\boldsymbol{\theta}) = \int \mathrm{d}^2\epsilon\, \omega(\epsilon) \left[ -\mathcal{U}(\boldsymbol{\theta}+\boldsymbol{\epsilon})\cos 2\varphi_\epsilon + \mathcal{Q}(\boldsymbol{\theta}+\boldsymbol{\epsilon})\sin 2\varphi_\epsilon \right], \quad (2.13\mathrm{b})$$

where $\boldsymbol{\epsilon}$ is a displacement vector, $\omega(\epsilon) \equiv -1/(\pi\epsilon^2)$, $\mathcal{Q} \equiv Q/(4I)$ and $\mathcal{U} \equiv U/(4I)$ are dimensionless Stokes parameters, and $\varphi_\epsilon$ denotes the angle between $\boldsymbol{\epsilon}$ and the $x$ axis (see Figure 2.3). The quantities in square brackets

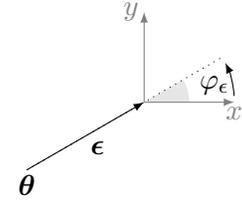

Figure 2.3: Definition of $\varphi_\epsilon$.

$$\mathcal{Q}_r(\boldsymbol{\theta}+\boldsymbol{\epsilon}) \equiv -\mathcal{Q}(\boldsymbol{\theta}+\boldsymbol{\epsilon})\cos 2\varphi_\epsilon - \mathcal{U}(\boldsymbol{\theta}+\boldsymbol{\epsilon})\sin 2\varphi_\epsilon\,, \quad (2.14\mathrm{a})$$

$$\mathcal{U}_r(\boldsymbol{\theta}+\boldsymbol{\epsilon}) \equiv -\mathcal{U}(\boldsymbol{\theta}+\boldsymbol{\epsilon})\cos 2\varphi_\epsilon + \mathcal{Q}(\boldsymbol{\theta}+\boldsymbol{\epsilon})\sin 2\varphi_\epsilon\,, \quad (2.14\mathrm{b})$$

are dimensionless Stokes parameters evaluated in a rotated coordinate system $(x, y)_r$, where $x_r$ and $y_r$ are tangential and radial to $\boldsymbol{\epsilon}$, respectively. The integrals in (2.13) can then be interpreted as follows: at $\boldsymbol{\theta}$, the quantity $E(\boldsymbol{\theta})$ gets (weighted) contributions by $\mathcal{Q}_r$ at *any* point in space, while $\mathcal{U}_r$ contributes to $B(\boldsymbol{\theta})$.

Consider, for instance, the left polarization pattern of Figure 2.4. How does this pattern contribute to $E$ and $B$ in $\boldsymbol{\theta}$? According to the above interpretation, each polarization rod in the figure is parallel to $\hat{\mathbf{y}}_r$, hence corresponding to a negative value of $\mathcal{Q}_r$. Summing all the contributions together, $E(\boldsymbol{\theta})$ ends up to be negative, while $B(\boldsymbol{\theta}) = 0$. On the contrary, the

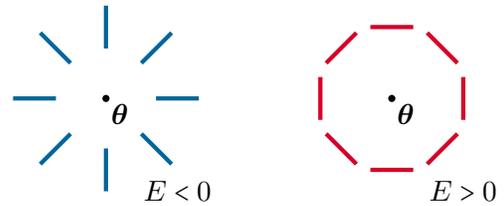

Figure 2.4: Examples of polarization patterns around a point resulting in pure $E(\boldsymbol{\theta})$.

right pattern of Figure 2.4 has $E(\boldsymbol{\theta}) > 0$ and
$B(\boldsymbol{\theta}) = 0$, since all the polarization rods are
characterized by a positive $Q_r$ and vanishing $U_r$.

With similar considerations one can show
that the left (right) pattern of Figure 2.5 is char-
acterized by vanishing $E(\boldsymbol{\theta})$ and negative (pos-
itive) $B(\boldsymbol{\theta})$.

The opposite parity of $E$ and $B$ is clear from
Figures 2.4 and 2.5.

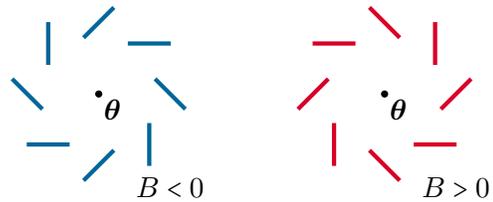

Figure 2.5: Examples of polarization pat-
terns around a point resulting in pure $B(\boldsymbol{\theta})$.

**Angular power spectra**   As for the power spectra, we can generalize eq. (2.6) to

$$C_\ell^{XY} = \frac{1}{2\ell + 1} \sum_{m=-\ell}^{\ell} a_{\ell m}^X a_{\ell m}^{Y*},$$

(2.15)

where $X$ and $Y$ can be $T$, $E$, or $B$. The information encoded in the temperature and
polarization of the CMB is therefore summarized in six angular power spectra: three auto-
correlations ($TT, EE, BB$), and three cross-correlations ($TE, EB, TB$). Since the $a_{\ell m}^E$ and
$a_{\ell m}^B$ transform under parity according to eqs. (2.12), the $TT$, $EE$, $BB$ and $TE$ correlations
are parity-even while $TB$ and $EB$ are parity-odd, and can therefore be used to probe new
parity-violating physics [81].

### 2.1.3   Origin of CMB polarization

Although radiation in the early universe is generally unpolarized, Thomson scatterings
between low-energy photons and electrons, $e^- + \gamma \to e^- + \gamma$, provide a natural way to induce
some degree of polarization at the time of photon decoupling. In fact, free electrons can
act as polarizers, converting unpolarized radiation into linearly polarized radiation. As a
concrete example, consider a coordinate system with a free electron in the origin. If some
unpolarized radiation traveling along the $x$ axis gets deflected by the electron along an
orthogonal direction, say along the $z$ axis, the outgoing radiation will be polarized, as only
the $y$-component of the incoming electric field survives the scattering.

In general, the radiation that last scatters off of a free electron comes from all directions
and whether the scattered radiation is polarized or not depends on the local intensity
pattern around the electron itself. To understand why this is the case, consider a free
electron in the origin which is hit by radiation coming from four different directions at the
same time, $\pm\hat{x}$ and $\pm\hat{y}$, and assume the scattered radiation to travel along $\hat{z}$.

**Monopole pattern**   In this case, the intensity of the incoming unpolarized radiation is
    isotropic around the free electron. As shown in the left panel of Figure 2.6, the
    outgoing radiation is constituted by the components of the incoming rays that are
    transverse to the $z$-axis, i.e. the $y$-component for the light coming from $\pm\hat{x}$ and the
    $x$-component for the light coming from $\pm\hat{y}$. Since their intensity is identical, the
    outgoing radiation is unpolarized.

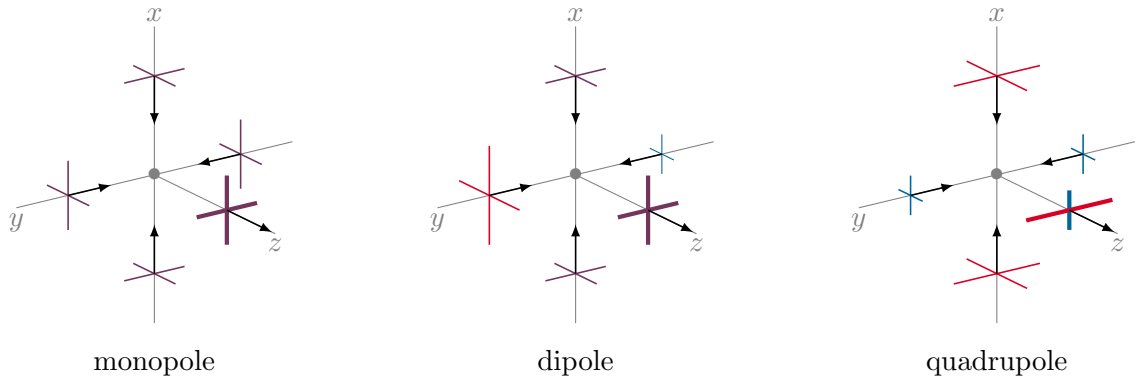

Figure 2.6: Comparison of the polarization state of Thomson-scattered radiation for three different local intensity patterns. Monopole and dipole (left and central panels, respectively), result in unpolarized radiation. For a quadrupole pattern, the incoming radiation is characterized by higher (lower) intensity from $\pm\hat{x}$ ($\pm\hat{y}$), and the scattered radiation shows some linear polarization.

**Dipole pattern** In this first example of anisotropic pattern the intensity is higher (lower) from $+\hat{y}$ ($-\hat{y}$), and average from $\pm\hat{x}$. As sketched in the central panel of Figure 2.6, the $y$-component of the incoming radiation from $\pm\hat{x}$ is transmitted, which has average intensity. The outgoing intensity along the $x$-direction is also average, since it comes from a colder spot ($-\hat{y}$-direction) and a hotter spot ($+\hat{y}$-direction). As a result, we see that this pattern does not induce any polarization either, despite being anisotropic.

**Quadrupole pattern** Again, radiation is anisotropic in this case: less intense from $\pm\hat{x}$ and more intense from $\pm\hat{y}$. The $y$-component of the scattered light will therefore be less intense than its $x$-component, as shown in the right panel of Figure 2.6. The outgoing radiation is linearly polarized along the $y$-axis.

Although these situations are not realistic (one should consider radiation incoming from all directions and being deflected along an arbitrary outgoing direction), they give us an intuitive understanding of why polarization can arise only if the local intensity distribution has a non-vanishing quadrupole moment.

**CMB polarization from scalar modes** Scalar perturbations to the metric and energy-momentum tensor result in perturbations, $\delta T$, to the temperature of the photon fluid at decoupling, which can be written as

$$T(\mathbf{x}) = \bar{T} + \delta T(\mathbf{x}), \tag{2.16}$$

where $\bar{T}$ denotes the photon background temperature. Our goal here is to gain some intuition about what polarization patterns can be induced by scalar perturbations. To keep things simple and use the definitions of $E$ and $B$ modes provided in eqs. (2.13), we will work in flat-sky approximation.

Given an arbitrary direction of observation, $\hat{n}$, we can always define a flat-sky coordinate system $(\hat{x}, \hat{y})$ such that the point $(0,0)$ corresponds to the point on the sphere identified by $\hat{n}$. Consider a temperature monochromatic plane wave with wavenumber $\mathbf{k}$ along $\hat{x}$:

$$\delta T_{\mathbf{k}}(x) = A_{\mathbf{k}} \cos(kx + \phi_{\mathbf{k}}),\qquad(2.17)$$

where $A$ is the wave's amplitude and $\phi_{\mathbf{k}}$ represents its phase. Given a point near the origin identified by the vector $\boldsymbol{\epsilon}$, the temperature at $\boldsymbol{\epsilon}$ will only depend on $x_\epsilon \equiv \epsilon \cos(\varphi_\epsilon)$ (see Figure 2.7), i.e.

$$\delta T_{\mathbf{k}}(\boldsymbol{\epsilon}) = A_{\mathbf{k}} \cos(k\epsilon \cos\varphi_\epsilon + \phi_{\mathbf{k}}).\quad(2.18)$$

Since $\delta T(\boldsymbol{\epsilon})$ only depends on $x$, the intensity coming from the $\pm\hat{y}$ directions will always be identical. Whether there is a nonzero local quadrupole then only depends on the intensity coming from the $\pm\hat{x}$ directions. In points where $\delta T_{\mathbf{k}}(\boldsymbol{\epsilon}) = 0$, the local quadrupole vanishes and no polarization is produced. In points corresponding to a maximum (minimum) of the cosine, polar-ization oriented along the $x$ ($y$) axis can be produced. Overall, the polarization induced by the $\delta T_{\mathbf{k}}$ mode can be described as

$$\mathcal{Q}(\boldsymbol{\epsilon}) = f(\epsilon \cos\varphi_\epsilon),\qquad \mathcal{U}(\boldsymbol{\epsilon}) = 0,\qquad(2.19)$$

where $f$ is some function of $x = \epsilon \cos\varphi_\epsilon$. To see how this polarization pattern contribute to $E$ and $B$, we use the pair of eqs. (2.13), and write $E(\hat{n})$ and $B(\hat{n})$ in terms of $\mathcal{Q}(\boldsymbol{\epsilon})$ as in eq. (2.19). Since $\mathcal{Q}$ is an even function of $\varphi_\epsilon$, regardless of the specific functional form of $f$ in eq. (2.19), when we plug it in eq. (2.13), only the terms multiplied by even functions survive, i.e. $B(\boldsymbol{\theta}) = 0$. Scalar perturbations cannot produce $B$ modes at decoupling.

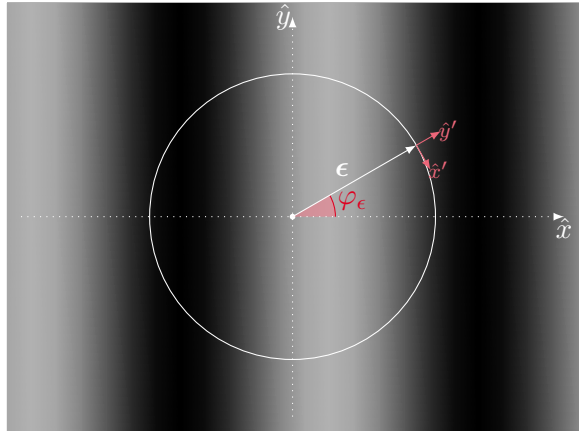

Figure 2.7: Framework to study the polarization pattern induced by an energy density monochromatic plane wave with $\mathbf{k}$ along $x$ in flat sky approximation. Darker (lighter) regions correspond to higher (lower) temperatures.

This result can be easily generalized to arbitrary $\mathbf{k}$, as long as they are orthogonal to the direction of observation $\hat{n}$. If $\mathbf{k}$ forms an angle $\alpha$ with the $\hat{x}$ axis, one can use a new coordinate system $(\hat{x}', \hat{y}')$ with $\hat{x}'$ parallel to $\mathbf{k}$ and repeat the same steps as before, obtaining again a vanishing $B(\hat{n})$. More general $\mathbf{k}$, i.e. non orthogonal to $\hat{n}$, can only be studied beyond the flat-sky approximation, but they still confirm our intuition [2].

The only mechanism that can indirectly produce $B$ modes from scalar perturbations in the $\Lambda$CDM model is gravitational lensing [82]. As the CMB photons travel toward us, their trajectory can be deflected by the gravitational potential of large scale structures. This results in a warping of the temperature and polarization maps, which can cause $E$ modes becoming $B$ modes, and vice versa. This is shown for instance in Figure 1.1, where the lensing $BB$ correlation, $C_\ell^{BB,\mathrm{lens}}$, is shown (dash-dotted gray line).

## 2.2 New physics: constraining inflationary models

Inflation is expected to source initial conditions for both density fluctuations (scalar perturbations [30–34]) and primordial gravitational waves (tensor perturbations [35, 36]). Gravitational waves are a propagating disturbance in the metric tensor:

$$g_{\mu\nu}(\mathbf{x}) = \bar{g}_{\mu\nu} + h_{\mu\nu}(\mathbf{x}) \,, \tag{2.20}$$

where $\bar{g}_{\mu\nu}$ is the background metric tensor, and $h_{\mu\nu}(\mathbf{x})$ denotes the perturbation to the metric tensor caused by the gravitational wave.

Given a gravitational wave with wavenumber $\mathbf{k}$, its effect is to periodically stretch and compress space in the two directions orthogonal to $\hat{k}$. Gravitational waves have two linear polarizations, usually called plus (+) and cross (×), with reference to the pattern of stretching and compression they cause (see Figure 2.8).

Primordial gravitational waves leave their signatures imprinted on the CMB temperature anisotropies [83–87], as well as on polarization [87–92]. The spin-2 nature of gravitational waves leads to both $E$ and $B$ modes.

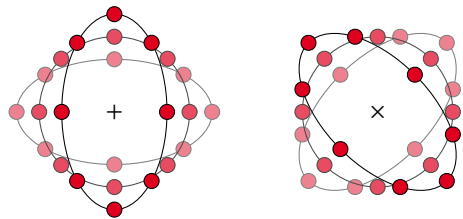

Figure 2.8: Stretching and compression patterns caused by + (left) and × (right) gravitational waves. Red dots indicate configurations of noninteracting inertial test masses.

**CMB polarization from tensor modes**  Here we consider a framework similar to what we used to discuss scalar modes: a primordial gravitational wave that propagates with wavenumber $\mathbf{k}$ parallel to the $x$ axis of a flat-sky coordinate system. The gravitational wave can be decomposed into $h_+$ and $h_\times$, which result in a pure $\mathcal{Q}$ and pure $\mathcal{U}$ polarization pattern, respectively, as sketched in Figures 2.9 and 2.10. As for the scalar case, the resulting Stokes parameters will only depend on $x = \epsilon \cos \varphi_\epsilon$ and counterparts of eq. (2.19) read

$$\mathcal{Q}_+(\boldsymbol{\epsilon}) = f_+(\epsilon \cos \varphi_\epsilon) \,, \qquad \mathcal{U}_+(\boldsymbol{\epsilon}) = 0 \,, \tag{2.21a}$$

$$\mathcal{Q}_\times(\boldsymbol{\epsilon}) = 0 \,, \qquad \mathcal{U}_\times(\boldsymbol{\epsilon}) = f_\times(\epsilon \cos \varphi_\epsilon) \,. \tag{2.21b}$$

By plugging the expressions for $\mathcal{Q}_+$ and $\mathcal{U}_+$ into eq. (2.13) and using that $\mathcal{Q}_+$ is an even function of $\varphi_\epsilon$, we see that $B_+$ identically vanishes, as for the scalar modes case. However, by repeating the same steps for the $\mathcal{Q}_\times$ and $\mathcal{U}_\times$, one can see that $B_\times$ does not identically vanish! While scalar modes can only produce $E$ modes in CMB polarization, tensor modes can result in $B$-mode polarization[3].

---

[3]As for the scalar case, generalizing these results to arbitrary wavenumbers goes beyond the scope of this thesis. The interested reader can take a look at [2].

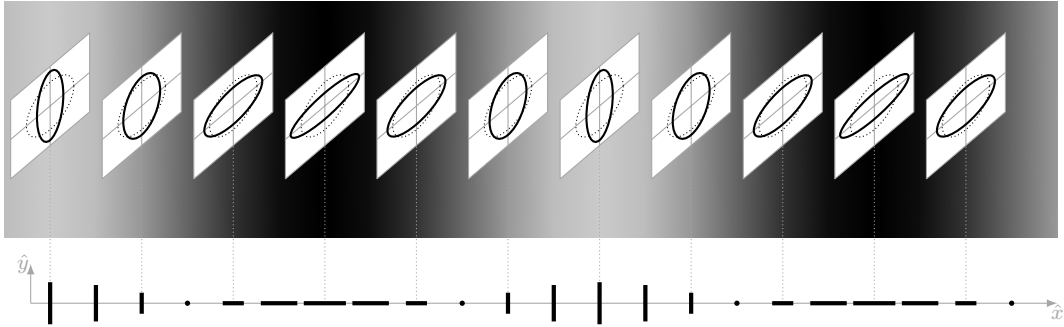

Figure 2.9: Polarization pattern induced by a $h_+$ tensor mode with $\mathbf{k}$ along the $x$ axis.

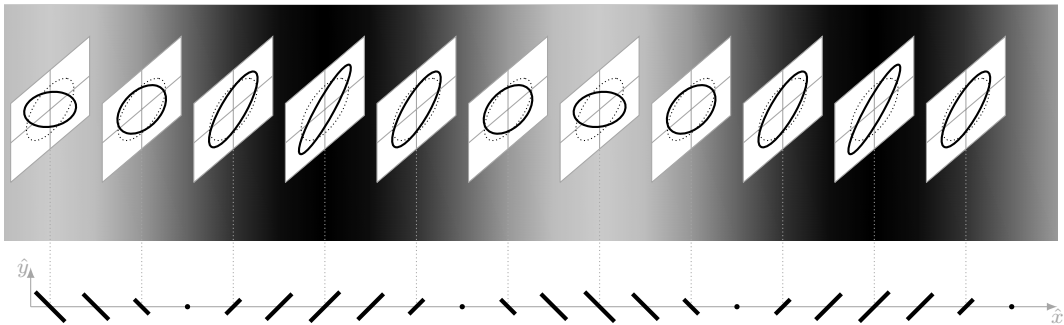

Figure 2.10: Polarization pattern induced by a $h_\times$ tensor mode with $\mathbf{k}$ along the $x$ axis.

**Extracting $r$ from primordial $B$ modes**   As mentioned in Section 1.1, many inflationary models predict the scalar and tensor power spectra to obey power laws: $P_s(k) = A_s k^{n_s-1}$ and $P_t(k) = A_t k^{n_t}$, respectively. The amplitude of gravitational waves relative to that of density fluctuations is model-dependent and is usually quantified in terms of the tensor-to-scalar ratio, $r \equiv A_t/A_s$. The $BB$ angular power spectrum can then be modeled as

$$C_\ell^{BB} = C_\ell^{BB,\text{lens}} + r C_\ell^{BB,\text{GW}}. \tag{2.22}$$

where $C_\ell^{BB,\text{GW}}$ is the primordial $B$-mode power spectrum with $r = 1$ [39, 40], and $C_\ell^{BB,\text{lens}}$ is the lensed $B$-mode power spectrum [82]. By inferring $r$ from CMB obsevations, one can constrain inflationary theories and hope to learn something about inflationary physics, which is one of the major open questions in modern cosmology [48, 49].

## 2.3   New physics: probing parity-violating physics

As mentioned in Section 1.1, a time-dependent parity-violating pseudoscalar, $\chi$, would cause the plane of CMB photons' linear polarization to rotate as they travel toward us, a phenomenon known as cosmic birefringence. In this section, we start by studying light propagation in an expanding vacuum, then add a homogeneous parity-violating pseudoscalar and see how it affects light propagation. Finally, we discuss the signatures in the observed CMB angular power spectra that are imprinted by cosmic birefringence.

**EM field propagating in expanding vacuum**   Under the assumptions of flatness, homogeneity and isotropy, one can choose a set of coordinates where the metric $g_{\mu\nu}$ takes the simple Friedmann-Lemaître-Robertson-Walker (FLRW) form [93–95]:

$$ds^2 = g_{\mu\nu}dx^\mu dx^\nu = a^2(\eta)\left(-d\eta^2 + d\mathbf{x}^2\right),\tag{2.23}$$

where $\eta$ is the conformal time, $\mathbf{x}$ denotes comoving coordinates, and $a(\eta)$ is the scale factor. In this coordinate system, the action for the free electromagnetic field, $A_\mu$, reads

$$S = \int d^4x'\sqrt{-g}\,\mathcal{L}_A = -\frac{1}{4}\int d^4x'\sqrt{-g}\,F_{\mu\nu}F^{\mu\nu},\tag{2.24}$$

where $g = -a^8(\eta)$ is the determinant of the FLRW metric, and $F_{\mu\nu} \equiv \partial_\mu A_\nu - \partial_\nu A_\mu$ is the antisymmetric electromagnetic tensor. Given the action, $S$, one can derive the equations of motion for $A_\mu$ by imposing

$$\frac{\delta S}{\delta A_\rho} = -\frac{1}{2}\int d^4x'\sqrt{-g}\,F^{\mu\nu}\frac{\delta F_{\mu\nu}}{\delta A_\rho} = 0\,,\tag{2.25}$$

where $\delta/\delta A_\mu$ denotes the functional derivative with respect to the electromagnetic field and satisfies $\delta A_\mu(x)/\delta A_\nu(y) = \delta_\mu^\nu \delta(x-y)$. The functional derivative at the right-hand side of eq. (2.25) then reads

$$\frac{\delta F_{\mu\nu}(x')}{\delta A_\rho(x)} = \partial_\mu \frac{\delta A_\nu(x')}{\delta A_\rho(x)} - \partial_\nu \frac{\delta A_\mu(x')}{\delta A_\rho(x)} = \left(\partial_\mu \delta_\nu^\rho - \partial_\nu \delta_\mu^\rho\right)\delta(x'-x)\,,\tag{2.26}$$

which, when plugged back into eq. (2.25), gives

$$0 = -\frac{1}{2}\int d^4x'\sqrt{-g}\,F^{\mu\nu}\left(\partial_\mu \delta_\nu^\rho - \partial_\nu \delta_\mu^\rho\right)\delta(x'-x) = \partial_\mu\left(\sqrt{-g}F^{\mu\rho}\right).\tag{2.27}$$

Given the FLRW metric of eq. (2.23), the last equality reads

$$\partial_\mu\left(\sqrt{-g}\,F^{\mu\rho}\right) = \partial_\mu\left(\sqrt{-g}\,g^{\mu\nu}g^{\rho\sigma}F_{\nu\sigma}\right) = \eta^{\mu\nu}\eta^{\rho\sigma}\partial_\mu F_{\nu\sigma} = 0\,,\tag{2.28}$$

where $\eta_{\mu\nu}$ is the Minkowski metric tensor. With gauge conditions $A_0 = 0$ and $\nabla \cdot \mathbf{A} = 0$, eq. (2.28) translates into an equation of motion for $\mathbf{A}$:

$$\mathbf{A}'' - \nabla^2 \mathbf{A} = 0\,,\tag{2.29}$$

where $\nabla \equiv \partial/\partial\mathbf{x}$ and the prime denotes $\partial/\partial\eta$. In Fourier space, eq. (2.29) becomes

$$\mathbf{A}_\mathbf{k}'' + k^2 \mathbf{A}_\mathbf{k} = 0\,,\tag{2.30}$$

where $\mathbf{k}$ is the comoving wavenumber of the Fourier mode $\mathbf{A}_\mathbf{k}$.

**Adding a homogeneous parity-violating pseudoscalar field**   A pseudoscalar can couple to the electromagnetic field via a Chern-Simons term in the Lagrangian

$$\mathcal{L}_{\text{CS}} = -\frac{\alpha}{4f}\chi F\widetilde{F} \equiv -\frac{\alpha}{4f}\chi F_{\mu\nu}\frac{\varepsilon^{\mu\nu\rho\sigma}}{2\sqrt{-g}}F_{\rho\sigma}\,, \tag{2.31}$$

where $\alpha$ is a coupling constant, $f$ is a decay constant with dimensions of energy and $\varepsilon^{\mu\nu\rho\sigma}$ is a totally antisymmetric symbol with $\varepsilon^{0123} = 1$. We can see how the presence of $\chi$ affects light propagation by repeating the same steps as before, now starting from the action $S = \int \mathrm{d}^4 x'\sqrt{-g}\,(\mathcal{L}_A + \mathcal{L}_{\text{CS}})$. The counterpart of eq. (2.27) then reads

$$\partial_\mu\left[\sqrt{-g}\left(F^{\mu\rho} + \frac{\alpha}{f}\chi\frac{\varepsilon^{\mu\rho\nu\sigma}}{2\sqrt{-g}}F_{\nu\sigma}\right)\right] = 0\,, \tag{2.32}$$

where the first term on the left-hand side is the same as for the vacuum case, while the second one can be written as[4]

$$\frac{\alpha}{2f}\varepsilon^{\mu\rho\nu\sigma}\partial_\mu(\chi F_{\nu\sigma}) = \frac{\alpha}{2f}\varepsilon^{\mu\rho\nu\sigma}F_{\nu\sigma}\partial_\mu\chi + \frac{\alpha}{2f}\chi\varepsilon^{\mu\rho\nu\sigma}\partial_\mu F_{\nu\sigma} = \delta_i^\rho\frac{\alpha}{f}\chi'\varepsilon^{0ijk}\partial_j A_k\,. \tag{2.33}$$

Thus, the equations of motion for $A_\mu$ read $\eta^{\mu\nu}\eta^{\rho\sigma}\partial_\mu\partial_\nu A_\sigma + \delta_i^\rho\frac{\alpha}{f}\chi'\varepsilon^{0ijk}\partial_j A_k = 0$ or, equivalently, in terms of $\mathbf{A}$ (again with gauge conditions $A_0 = 0$ and $\nabla\cdot\mathbf{A} = 0$),

$$\mathbf{A}'' - \nabla^2\mathbf{A} - \frac{\alpha}{f}\chi'\nabla\times\mathbf{A} = 0\,. \tag{2.34}$$

**Projecting on helicity components**   It is interesting to project the above equation into components of $\pm$ helicity. Consider a Fourier mode $\mathbf{A_k}$ with comoving wavenumber $\mathbf{k}$ along $\hat{z}$. We can introduce the vectors $\boldsymbol{\epsilon}_+ \equiv (\hat{x} - i\hat{y})/\sqrt{2}$ and $\boldsymbol{\epsilon}_- \equiv (\hat{x} + i\hat{y})/\sqrt{2}$, and define the components of $\mathbf{A_k}$ with positive and negative helicity, respectively, by $A_\pm \equiv \mathbf{A_k}\cdot\boldsymbol{\epsilon}_\pm$. Using that

$$\boldsymbol{\epsilon}_\pm\cdot\nabla\times\mathbf{A_k} = \frac{1}{\sqrt{2}}(\varepsilon^{132}ikA_\mathbf{k}^2 \pm \varepsilon^{231}kA_\mathbf{k}^1) = \pm\frac{k}{\sqrt{2}}(A_\mathbf{k}^1 \mp iA_\mathbf{k}^2) = \pm kA_\pm\,, \tag{2.35}$$

we can project equation (2.34) into helicity components to obtain

$$A_\pm'' + \left(k^2 \mp \frac{\alpha}{f}\chi'k\right)A_\pm = 0\,. \tag{2.36}$$

Eq. (2.36) tells us the two helicity components of the electromagnetic field have different dispersion relations, i.e. the plane of linear polarization rotates by the angle

$$\beta = \frac{\alpha}{2f}(\chi_0 - \chi_{\text{dec}})\,, \tag{2.37}$$

where $\chi_0$ is today's value of the $\chi$ field, while the subscript dec specifies $\chi$ at the time of photon decoupling, when the CMB was emitted. We usually refer to $\beta$ as the cosmic birefringence angle.

---

[4]In the last equality, we have used that $\frac{\alpha}{2f}\chi\varepsilon^{\mu\rho\nu\sigma}\partial_\mu F_{\nu\sigma} = \frac{\alpha}{f}\chi\varepsilon^{\mu\rho\nu\sigma}\partial_\mu\partial_\nu A_\sigma = 0$ by symmetry and that, since $\chi$ depends only on time, $\frac{\alpha}{2f}\varepsilon^{\mu\rho\nu\sigma}\partial_\mu(\chi F_{\nu\sigma}) = \delta_i^\rho\frac{\alpha}{f}\chi'\varepsilon^{0ijk}\partial_j A_k$.

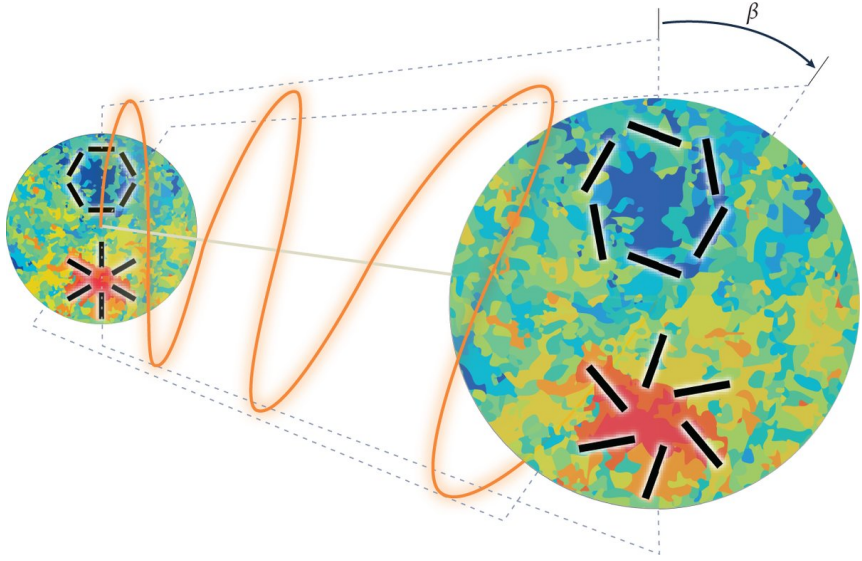

Figure 2.11: Cosmic birefringence causes the plane of linear polarization (orange line) to rotate by an angle $\beta$ between last scattering and the present day. In turn, this leads to the mixing of $E$ and $B$ modes. Image credit: Yuto Minami

**Effects on the angular power spectra** With the plane of linear polarization rotating by an angle $\beta$, the Stokes parameters transform according to (2.7) with $\psi = -\beta$:

$$Q^{\text{obs}}(\hat{n}) \pm iU^{\text{obs}}(\hat{n}) = \left[Q(\hat{n}) \pm iU(\hat{n})\right] e^{\pm 2i\beta} \,. \tag{2.38}$$

Retracing the steps to get to eqs. (2.10), one can write the observed spherical harmonic coefficients in terms of the intrinsic ones and the cosmic birefringence angle:

$$a_{\ell m}^{E,\text{obs}} = a_{\ell m}^{E} \cos(2\beta) - a_{\ell m}^{B} \sin(2\beta) \,, \tag{2.39a}$$

$$a_{\ell m}^{B,\text{obs}} = a_{\ell m}^{B} \cos(2\beta) + a_{\ell m}^{E} \sin(2\beta) \,, \tag{2.39b}$$

and, by plugging these expressions in eq. (2.15), the observed angular power spectra read

$$C_{\ell}^{EE,\text{obs}} = C_{\ell}^{EE} \cos^2(2\beta) + C_{\ell}^{BB} \sin^2(2\beta) - C_{\ell}^{EB} \sin(4\beta) \,, \tag{2.40a}$$

$$C_{\ell}^{BB,\text{obs}} = C_{\ell}^{BB} \cos^2(2\beta) + C_{\ell}^{EE} \sin^2(2\beta) + C_{\ell}^{EB} \sin(4\beta) \,, \tag{2.40b}$$

$$C_{\ell}^{EB,\text{obs}} = C_{\ell}^{EB} \cos(4\beta) + \frac{C_{\ell}^{EE} - C_{\ell}^{BB}}{2} \sin(4\beta) \,. \tag{2.40c}$$

From the above equations is it clear that a non-zero cosmic birefringence mixes $E$ and $B$ modes, as also shown schematically in Figure 2.11. Interestingly, even if $C_{\ell}^{EB} = 0$ at the surface of last scattering, cosmic birefringence can produce a non-zero observed $EB$ correlation, which can be written in terms of observed $C_{\ell}^{EE,\text{obs}}$ and $C_{\ell}^{BB,\text{obs}}$:

$$C_{\ell}^{EB,\text{obs}} = \frac{\tan(4\beta)}{2} \left(C_{\ell}^{EE,\text{obs}} - C_{\ell}^{BB,\text{obs}}\right) \,. \tag{2.41}$$

**Measuring the cosmic birefringence angle**   Although the observed $EB$ correlation is a sensitive probe of cosmic birefringence, inferring $\beta$ from $C_\ell^{EB,\mathrm{obs}}$ is not as simple as eq. (2.41) suggests. Unless we know exactly how the incoming polarization is rotated by the elements of the telescope's optical chain and how the polarization-sensitive orientations of the detectors are related to the sky coordinates, the polarization angle we measure will not be the intrinsic one, but will be shifted by a non-zero miscalibration angle, $\alpha$. The problem is that the effect of a non-zero $\alpha$ on the observed $C_\ell^{EB,\mathrm{obs}}$ is completely degenerate with cosmic birefringence, since they both shift the polarization angle. As a result, in the absence of any other information, we can only determine the sum of the two angles, $\alpha + \beta$.

The methodology proposed in [59–61] solves this issue by calibrating $\alpha$ with Galactic foreground emission. Their idea is that, as Galactic photons are not affected by cosmic birefringence, they are only rotated by $\alpha$ and can therefore used to measure $\alpha$, which can be then substracted from the overall rotation angle of the CMB signal. This strategy allowed to infer $\beta = 0.35 \pm 0.14°$ at 68% C.L. [62] from nearly full-sky *Planck* polarization data [63] and obtain more precise measurements in subsequent works [64–66].

The problem with these approaches is that $\alpha$ can be calibrated from the foregrounds only if we know the intrinsic $EB$ and $TB$ correlations of the foreground signal, which is not the case. More robust estimates of $\beta$ can then be achieved if we have a better understanding of foreground emission, or if we have a better calibration strategy which makes it possible not to rely on foreground emission altogether.

# Chapter 3

# CMB experiments: end-to-end

**Summary:** Data acquisition and data analysis are the two macro-steps in any CMB experiment. The output of data acquisition is the time-ordered data (TOD), and the goal of data analysis is to process the TOD to extract relevant cosmological information. This is usually done by

- making temperature and polarization maps from the TOD,
- performing some foreground cleaning routine,
- estimating the angular power spectra,
- inferring the cosmological parameters.

In this chapter, we discuss data acquisition and data analysis separately, providing a concise review of the building blocks of any end-to-end CMB experiment.

*This is an adaptation of some personal notes I took during the course of my PhD. Section 3.3 follows [96, 97]. The method of maximum likelihood is introduced in 3.4 following [98].*

## 3.1   Data acquisition

Data acquisition is the first macro-step in any CMB experiment and its output consists of the time-ordered data (TOD), i.e. the collection of the signals detected by all detectors during the entire duration of the mission. For an instrument with $n_{\mathrm{det}}$ detectors, each of which takes $n_{\mathrm{obs}}$ measurements during the mission, we represent the TOD as the vector

$$\mathbf{d} = \begin{pmatrix} \rule{0.6em}{2.2em} \\ \vdots \\ \rule{0.6em}{2.2em} \end{pmatrix} \begin{matrix} \updownarrow n_{\mathrm{obs}} \\ \\ \updownarrow n_{\mathrm{obs}} \end{matrix} \tag{3.1}$$

where each rectangle represents the signals collected by a single detector. One can think of the TOD as the superposition of a noiseless and a noise-only component:

$$\mathbf{d} = \mathbf{d}_{\text{noiseless}} + \mathbf{n}\,. \tag{3.2}$$

Various effects can contribute to $\mathbf{n}$: intrinsic thermal noise of detectors and amplifiers; detectors, amplifiers, and readout electronics instabilities; as well as environmental effects and atmospheric fluctuations (for sub-orbital experiments) [99]. Accurately modeling the noise is very complicated and goes beyond the scope of this thesis. However, if the noise is Gaussian with zero mean and stationary, its statistical properties are fully captured by its frequency power spectrum, which we introduce in Section 3.1.1. In Section 3.1.2 we will instead focus on the noiseless TOD.

### 3.1.1   Noise frequency power spectrum

Typically, CMB detectors take samples regularly over time, say every $1/f_{\text{samp}}$ interval starting from some initial time, $t_0$. For a single detector, the $j$-th element of the noise TOD, $n_j$, corresponds then to the noise measured at the time $t_j = t_0 + j/f_{\text{samp}}$. The index $j$ takes integer values in the $[0, n_{\text{obs}} - 1]$ interval.

We can define the discrete Fourier transform of $n_j$ as

$$\tilde{n}_k = \sum_{j=0}^{n_{\text{obs}}-1} n_j\, e^{-i2\pi \frac{k}{n_{\text{obs}}} j}\,, \tag{3.3}$$

where the index $k$ also takes integer values in the $[0, n_{\text{obs}} - 1]$ interval. Each $k$ corresponds to a physical frequency $f_k = k f_{\text{samp}}/n_{\text{obs}}$.

Because of stationarity, each $\tilde{n}_k$ can be treated as a complex independent random variable and, assuming the noise to be Gaussian with zero mean, we can write its probability density function (p.d.f.) as

$$\text{p.d.f.}\,(\tilde{n}_k) = \frac{1}{\sigma_k \sqrt{2\pi}} \exp\left[ -\frac{1}{2} \frac{\tilde{n}_k \tilde{n}_k^*}{\sigma_k^2} \right]\,, \tag{3.4}$$

where $\sigma_k^2 \equiv \langle \tilde{n}_k \tilde{n}_k^* \rangle$ is the variance of the distribution. Note that, because the $\tilde{n}_k$ variables are independent from each other, $\langle \tilde{n}_k \tilde{n}_{k'}^* \rangle = 0$ for $k \neq k'$. In other words $\langle \tilde{n}_k \tilde{n}_{k'}^* \rangle = \sigma_k^2 \delta_{kk'}$.

The frequency power spectrum, $P(f)$, is the counterpart of the variance $\sigma_k^2$ for continuous physical frequencies, i.e.

$$\langle \tilde{n}(f) \tilde{n}^*(f') \rangle = P(f) \delta(f - f')\,. \tag{3.5}$$

As the p.d.f. in eq. (3.4) only depends on the variance, the power spectrum $P(f)$ fully captures the statistical properties of the noise.

**Modeling the power spectrum**  Typically in CMB experiments, the noise power spectrum, $P(f)$, is accurately modeled as a superposition of a white and a $1/f$ term [99]:

$$P(f) = \sigma_0^2 \left[ 1 + \left( \frac{f_{\text{knee}}^2}{f^2} \right)^\alpha \right]. \tag{3.6}$$

Here $f$ denotes a temporal frequency; $\sigma_0$ quantifies the white noise level of the TOD; $\alpha$ is the slope of the $1/f$ noise spectrum (typically positive); and the knee frequency, $f_{\text{knee}}$, denotes the frequency at which the variance of the $1/f$ noise is equal to the white noise variance. If $\alpha$ is positive in eq. (3.6), the noise power spectrum diverges for $f = 0$, which is however impossible to observe[1]. When simulating a noise realization, it is then customary to add an additional parameter, $f_{\text{min}}$, such that the power spectrum flattens for $f < f_{\text{min}}$. The noise power spectrum can then be modeled as

$$P(f) = \sigma_0^2 \left[ \frac{f^2 + f_{\text{knee}}^2}{f^2 + f_{\text{min}}^2} \right]^\alpha. \tag{3.7}$$

The power spectra of eqs. (3.6) and (3.7), together with a white-only $P(f)$, are shown in Figure 3.1 for an arbitrary choice of the parameters.

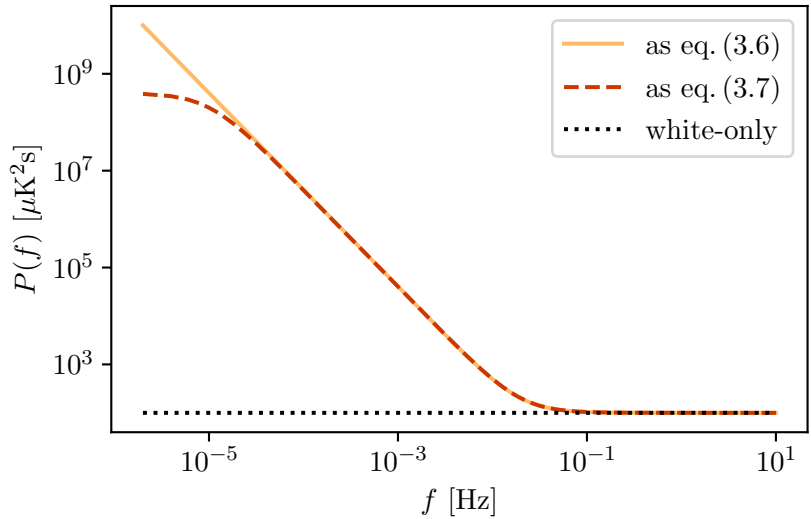

Figure 3.1: Noise power spectrum, $P(f)$, from eqs. (3.6) and (3.7) (solid yellow and dashed red lines, respectively), together with the white-only term (black dotted line). The parameters have been set to $\sigma_0 = 10\,\mu\text{K}\sqrt{s}$, $f_{\text{min}} = 10^{-5}\,\text{Hz}$, $f_{\text{knee}} = 20\,\text{mHz}$ and $\alpha = 1$.

---

[1] The lowest observable frequency is given by the inverse of the lifetime of the Universe [100]:

$$(14 \times 10^9 \text{ years})^{-1} \sim (4 \times 10^{17}\,\text{s})^{-1} \sim 10^{-17}\,\text{Hz}.$$

### 3.1.2 Modeling the noiseless TOD

In first approximation, it is safe to assume that the noiseless TOD depends only linearly on the sky signal. Depending on how the sky signal is described, i.e. whether one works in pixel or harmonic space, this results in two different modeling approaches. We discuss both in the following two paragraphs.

**Pixel space approach** By defining a pixelization scheme on the sphere, we can describe the sky signal as a $(n_{\text{pix}} \cdot n_{\text{Stokes}})$-vector, $\mathbf{m}$, where $n_{\text{Stokes}}$ is the number of Stokes parameters considered and $n_{\text{pix}}$ is the total number of pixels on the sphere. The elements of $\mathbf{m}$ are the values of the sky Stokes parameters at each pixel:

$$\mathbf{m} = \begin{pmatrix} \begin{array}{c} \end{array} \\ \vdots \\ \end{pmatrix} \begin{array}{c} n_{\text{Stokes}} \\ n_{\text{Stokes}} \end{array} \tag{3.8}$$

One can then use the linearity assumption to write

$$\mathbf{d}_{\text{noiseless}} \equiv A \cdot \mathbf{m}\,, \tag{3.9}$$

where $A$ is a $n_{\text{det}} \cdot n_{\text{obs}}$ by $n_{\text{pix}} \cdot n_{\text{Stokes}}$ matrix and is usually referred to as the response matrix. Modeling the noiseless TOD then amounts to modeling the response matrix. To do that, we need information about the scanning strategy and the instrument specifics.

- As the name suggests, the scanning strategy tells us how the telescope scans the sky as it observes. The relevant information can be summarized in three sets of angles, $\theta_t$, $\varphi_t$, and $\psi_t$, where the index $t$ takes integer values in the $[0, n_{\text{obs}} - 1]$ interval[2]. The first two identify the telescope's boresight direction, $\hat{n}_{\text{bore},t}$, i.e. the direction normal to the focal plane at its center. In turn, $\hat{n}_{\text{bore},t}$ identifies a pixel $p_{\text{bore},t}$ on the sphere. Instead, $\psi_t$ denotes the angle between the sky and the telescope coordinates on the plane orthogonal to the direction of observation (see Figure 3.2).

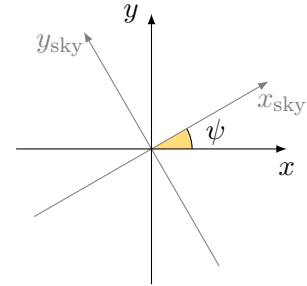

Figure 3.2: The telescope angle, $\psi$ is defined as the angle between the sky and telescope coordinates.

Note that knowing $p_{\text{bore},t}$ is not the same as knowing where each detector is pointing: detectors that are not located at boresight could be observing some other pixel.

- The 'instrument specifics', are all the characteristics that determine the detectors' pointings (i.e. the position of the detectors on the focal plane), and encode how the incoming sky signal, $\mathbf{S}_{\mathrm{sky}}$, is converted into detected signal. For example, information about the beams, the frequency bands, and the detectors' orientation.

---

[2]Note that we changed notation from Section 3.1.1, where we denoted the time index with $j$.

In the remaining of this section, we will provide some more concrete information that can help to better understand how the response matrix is structured. To keep things simple, we start by restricting ourselves to pencil beams only and single-frequency.

First, note that $A$ is a sparse matrix. In fact, at the time sample $t$, the detector $i$ will be pointing to some pixel $p_{it}$, implying that the $(i \cdot n_{\mathrm{obs}} + t)$ row of $A$ will be filled with zeros, except for the $n_{\mathrm{Stokes}}$ elements corresponding to the pixel $p_{it}$. Schematically, the response matrix then looks something like

$$A = \begin{pmatrix} & \blacksquare & \cdots & \\ & & \cdots & \blacksquare \\ \vdots & \vdots & \ddots & \vdots \\ \blacksquare & & \cdots & \\ & & \cdots & \blacksquare \end{pmatrix} \quad \tag{3.10}$$

where the black rectangles represent the sets of $n_{\mathrm{Stokes}}$ non-zero elements, while the empty rectangles represent the vanishing elements. The position of the non-zero elements depends on the scanning strategy and the detector's displacements, which determines $p_{it}$ given the time sample and detector indices.

Since each of the black rectangles has $n_{\mathrm{Stokes}}$ elements, we can think of it as a (transposed) Stokes vector, $\mathbb{S}^T$. The values of $\mathbb{S}^T$ encode how the sky signal, $\mathbf{S}_{\mathrm{sky}}$ is translated into detected signal, and they can be modeled, provided that we have a good understanding of the instrument's characteristics. This means that we should know the physical properties of the elements that make up the telescope's optical chain, and be able to transform between their coordinate systems.

As a first concrete example, consider an ideal detector oriented along the $x$ axis in the telescope's coordinate system. The action of an ideal detector on a Stokes vector is to take the arithmetical average of the $I$ and $Q$ parameters, which can be modeled as $d = \mathbf{a}^T \mathbf{S}_{\mathrm{sky}}$, where $d$ is the detected signal and $\mathbf{a}^T = \frac{1}{2}\begin{pmatrix} 1 & 1 & 0 & 0 \end{pmatrix}$. Writing $d = \mathbf{a}^T \mathbf{S}_{\mathrm{sky}}$, however, only makes sense if both $\mathbf{a}^T$ and $\mathbf{S}_{\mathrm{sky}}$ are defined in the same coordinate system, which is not necessarily the case. To fix this, one has to rotate the incoming Stokes vector from sky to telescope's coordinates, and this can be done by knowing the angle between them, i.e. the telescope angle, $\psi_t$, as shown in Figure 3.2:

$$\mathbb{S}_{it}^T = \frac{1}{2}\begin{pmatrix} 1 & 1 & 0 & 0 \end{pmatrix} \mathcal{R}_{\psi_t}. \tag{3.11}$$

If the detector is not oriented along the $x$ axis in telescope coordinates, but forms an angle $\xi_i$ with it, this rotation should also be included in the data model:

$$\mathbb{S}_{it}^T = \frac{1}{2}\begin{pmatrix} 1 & 1 & 0 & 0 \end{pmatrix} \mathcal{R}_{\xi_i + \psi_t}. \tag{3.12}$$

**Harmonic space approach** Instead of describing the sky signal as a set of pixelized maps, we can encode all the relevant information into spherical harmonic coefficients: $a_{\ell m}^I$, $_{+2}a_{\ell m}^P$, and $_{-2}a_{\ell m}^{P*}$, where $P \equiv Q + iU$. These coefficients are the same as the ones defined in eqs. (2.4) and (2.9), although we are using a slightly different notation to match [101]. One could also include information about circular polarization by introducing the $a_{\ell m}^V$ coefficients, defined as in eq. (2.4) since $V$ is a scalar. In a given reference frame (say the instrument frame), the beams can also be described in terms of spherical harmonic coefficients, which we will denote $b_{\ell s}^{\mathbb{I}}$, $_{+2}b_{\ell m}^{\mathbb{P}}$, $_{-2}b_{\ell m}^{\mathbb{P}*}$, and $b_{\ell m}^{\mathbb{V}}$.

By assuming that the only difference between the optical response at samples $t$ and $t'$ is the direction and orientation of the telescope with respect to the sky, one can write the beam-convolved TOD as (equation 10 of [101]):

$$d_t = \sum_{s\ell m} \left[ b_{\ell s}^{\mathbb{I}} a_{\ell m}^I + \frac{1}{2} \left( _{-2}b_{\ell s}^{\mathbb{P}*} {}_{+2}a_{\ell m}^P + {}_{+2}b_{\ell s}^{\mathbb{P}} {}_{-2}a_{\ell m}^{P*} \right) + b_{\ell s}^{\mathbb{V}} a_{\ell m}^V \right] \sqrt{\frac{4\pi}{2\ell+1}} e^{-is\psi_t} {}_sY_{\ell m}(\theta_t, \varphi_t) , \quad (3.13)$$

where $\theta_t$ and $\varphi_t$ determine the direction of the telescope, and $\psi_t$ its orientation.

## 3.2 Map-making

The map-making step translates the raw TOD[3], $\mathbf{d}$, into a set of pixelized maps, $\widehat{\mathbf{m}}$, with the goal of recovering the sky maps, $\mathbf{m}$. The estimated map, $\widehat{\mathbf{m}}$, can then be further analyzed to extract cosmological information. Many different map-making methods have been used in the past decades [102], the simplest being linear methods:

$$\widehat{\mathbf{m}} = M \cdot \mathbf{d} , \quad (3.14)$$

where $M$ is some matrix that specifies the method. The scalar product on the right-hand side spans all detectors and all observations.

### 3.2.1 Bin-averaging

The bin-averaging (or binning) is the simplest map-maker one can use and the only one we will discuss in this thesis[4]. It is a linear method specified by the matrix $M = (\widehat{A}^T \widehat{A})^{-1} \widehat{A}^T$, so that

$$\widehat{\mathbf{m}} = (\widehat{A}^T \widehat{A})^{-1} \widehat{A}^T \cdot \mathbf{d} , \quad (3.15)$$

where $\widehat{A}$ is the response matrix assumed by the map-maker. When $\widehat{A} = A$, one is able to recover the sky maps, up to a noise term:

$$\begin{aligned} \widehat{\mathbf{m}} = M \cdot \mathbf{d} &= (A^T A)^{-1} A^T \cdot (A \cdot \mathbf{m} + \mathbf{n}) \\ &= \mathbf{m} + (A^T A)^{-1} A^T \cdot \mathbf{n} . \end{aligned} \quad (3.16)$$

---

[3]Often, the TOD is pre-processed (for example, it may be calibrated, or filtered) before it goes through the map-maker. These intermediate steps are extremely important for the overall performance of the experiment, but we will neglect them here for the sake of simplicity.

[4]The interested reader can take a look at [102], where other methods are also discussed.

Regardless of the specific model used for $\widehat{A}$, it will have a similar structure to the one shown in eq. (3.10):

$$\widehat{A} = \begin{pmatrix} & & \cdots & \\ & & \cdots & \\ \vdots & \vdots & \ddots & \vdots \\ & & \cdots & \end{pmatrix} \quad \longrightarrow \quad \widehat{A}^T = \begin{pmatrix} & & \cdots & & \\ & & \cdots & & \\ \vdots & \vdots & \ddots & \vdots & \vdots \\ & & \cdots & & \end{pmatrix}, \tag{3.17}$$

which means that the $\widehat{A}^T \widehat{A}$ scalar product and its inverse, $(\widehat{A}^T \widehat{A})^{-1}$, will be block-diagonal:

$$\widehat{A}^T \widehat{A} = \begin{pmatrix} & & \cdots & \\ & & \cdots & \\ \vdots & \vdots & \ddots & \vdots \\ & & \cdots & \end{pmatrix} \begin{pmatrix} & & \cdots & \\ & & \cdots & \\ \vdots & \vdots & \ddots & \vdots \\ & & \cdots & \end{pmatrix} = \begin{pmatrix} & & \cdots & \\ & & \cdots & \\ \vdots & \vdots & \ddots & \vdots \\ & & \cdots & \end{pmatrix}. \tag{3.18}$$

In particular, the non-zero block corresponding to the pixel $p$ is given by

$$\sum_{it \in \{it\}_p} \widehat{\mathbb{S}}_{it} \widehat{\mathbb{S}}_{it}^T = \sum_{it \in \{it\}_p} \begin{pmatrix} \widehat{\mathbb{I}}^2 & \widehat{\mathbb{I}}\widehat{\mathbb{Q}} & \widehat{\mathbb{I}}\widehat{\mathbb{U}} & \widehat{\mathbb{I}}\widehat{\mathbb{V}} \\ \widehat{\mathbb{Q}}\widehat{\mathbb{I}} & \widehat{\mathbb{Q}}^2 & \widehat{\mathbb{Q}}\widehat{\mathbb{U}} & \widehat{\mathbb{Q}}\widehat{\mathbb{V}} \\ \widehat{\mathbb{U}}\widehat{\mathbb{I}} & \widehat{\mathbb{U}}\widehat{\mathbb{Q}} & \widehat{\mathbb{U}}^2 & \widehat{\mathbb{U}}\widehat{\mathbb{V}} \\ \widehat{\mathbb{V}}\widehat{\mathbb{I}} & \widehat{\mathbb{V}}\widehat{\mathbb{Q}} & \widehat{\mathbb{V}}\widehat{\mathbb{U}} & \widehat{\mathbb{V}}^2 \end{pmatrix}_{it}, \tag{3.19}$$

where $\{it\}_p$ denotes the set of detectors $i$ that are observing the pixel $p$ at the time $t$. Explicitly, the reconstructed Stokes parameters at the pixel $p$ then read

$$\begin{pmatrix} \widehat{I} \\ \widehat{Q} \\ \widehat{U} \\ \widehat{V} \end{pmatrix}_p = \sum_{it \in \{it\}_p} \left[ \sum_{i't' \in \{it\}_p} \begin{pmatrix} \widehat{\mathbb{I}}^2 & \widehat{\mathbb{I}}\widehat{\mathbb{Q}} & \widehat{\mathbb{I}}\widehat{\mathbb{U}} & \widehat{\mathbb{I}}\widehat{\mathbb{V}} \\ \widehat{\mathbb{Q}}\widehat{\mathbb{I}} & \widehat{\mathbb{Q}}^2 & \widehat{\mathbb{Q}}\widehat{\mathbb{U}} & \widehat{\mathbb{Q}}\widehat{\mathbb{V}} \\ \widehat{\mathbb{U}}\widehat{\mathbb{I}} & \widehat{\mathbb{U}}\widehat{\mathbb{Q}} & \widehat{\mathbb{U}}^2 & \widehat{\mathbb{U}}\widehat{\mathbb{V}} \\ \widehat{\mathbb{V}}\widehat{\mathbb{I}} & \widehat{\mathbb{V}}\widehat{\mathbb{Q}} & \widehat{\mathbb{V}}\widehat{\mathbb{U}} & \widehat{\mathbb{V}}^2 \end{pmatrix}_{i't'} \right]^{-1} \begin{pmatrix} \widehat{\mathbb{I}} \\ \widehat{\mathbb{Q}} \\ \widehat{\mathbb{U}} \\ \widehat{\mathbb{V}} \end{pmatrix}_{it} d_{it}. \tag{3.20}$$

Eq. (3.20) is equivalent to eq. (3.15), albeit less compact. It can however help to understand what performing a bin-averaging actually means.

## 3.3   Foreground cleaning

The presence of Galactic foregrounds is particularly problematic for observations of primordial $B$ modes, since they are expected to be at least $\sim 100$ times fainter than the foreground emission [96, 97]. The brightest polarized contaminants are synchrotron and thermal dust emission[5], which dominate at lowest and highest frequencies, respectively (see Figure 3.3, where the spectral energy distributions (SEDs) of CMB, dust and synchrotron are shown; see also Figure 3.4, where we show the total intensity and polarization maps for several *Planck* frequency channels). From Figure 3.3, it is clear that the brightness of each component depends on frequency in a specific way. In particular, the specific intensity of CMB anisotropies follows a differential black-body, while dust and synchrotron can be modeled as a modified black-body and a power law, respectively [104]

$$\delta I_{\mathrm{CMB},\nu} = \frac{2\nu^2}{c^2}\frac{x_0^2 e^{x_0}}{(e^{x_0}-1)^2}k_B\,\delta T, \qquad (3.21a)$$

$$\delta I_{\mathrm{dust},\nu} = A_{\mathrm{dust}}\left(\frac{\nu}{\nu_\star}\right)^{\beta_{\mathrm{dust}}}B_\nu(T_{\mathrm{dust}}), \quad (3.21b)$$

$$\delta I_{\mathrm{sync},\nu} = A_{\mathrm{sync}}\left(\frac{\nu}{\nu_\ast}\right)^{\beta_{\mathrm{sync}}}. \qquad (3.21c)$$

where $x_0 \equiv h\nu/(k_B T_0)$ with $T_0$ the average temperature of the CMB, and $B_\nu(T) = 2h\nu^3/[c^2(e^x - 1)]$ is a black-body spectrum with $x \equiv h\nu/(k_B T)$ [6]. The fact that the SEDs of CMB and foreground emission de-pend on frequency in different ways can be used to disentangle these two signals and many methods have been proposed so far to achieve this goal. These different methods can however be classified into three main classes which we present next (following [97]).

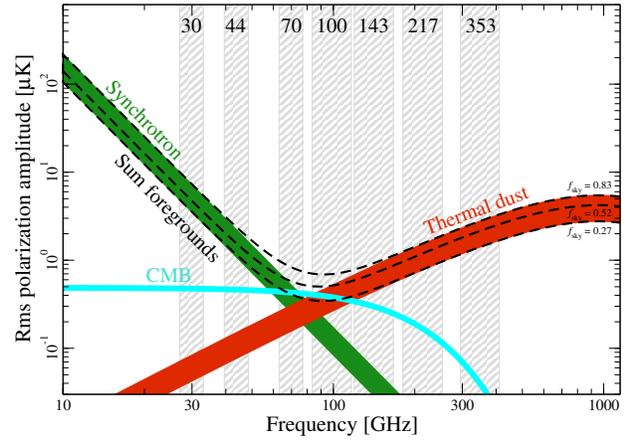

Figure 3.3: Polarization brightness temperature rms as a function of frequency and astrophysical component. Each component is smoothed to an angular resolution of 40 arcmin full-width-at-half-maximum (FWHM), and the lower and upper edges of each line are defined by masks retaining 73 and 93 % of the sky, respectively. Image and caption adapted from [103].

**Template fitting** Template fitting is one of the simplest foreground cleaning methods, based on the assumption that the spatial distribution $X_i(\hat{n})$ of the $i$ foreground component is known. The overall observed signal is then modeled as

$$T(\hat{n},\nu) = \sum_i \alpha_i(\nu)X_i(\hat{n}) + n(\hat{n},\nu)\,, \qquad (3.22)$$

where the $\alpha_i(\nu)$ are the template coefficients, encoding the frequency dependence of the template emission $X_i$. Template fitting was used successfully to remove the foreground emission in WMAP data. The main issue with this method is that it assumes the SEDs of the foreground components to be uniform throughout the sky, which is not the case.

---

[5]Synchrotron radiation is emitted by relativistic cosmic ray electrons that are accelerated by magnetic fields and start spiraling. As for thermal dust emission, it consists of radiation re-emitted by interstellar dust grains which are heated by the interstellar radiation field [96, 97].

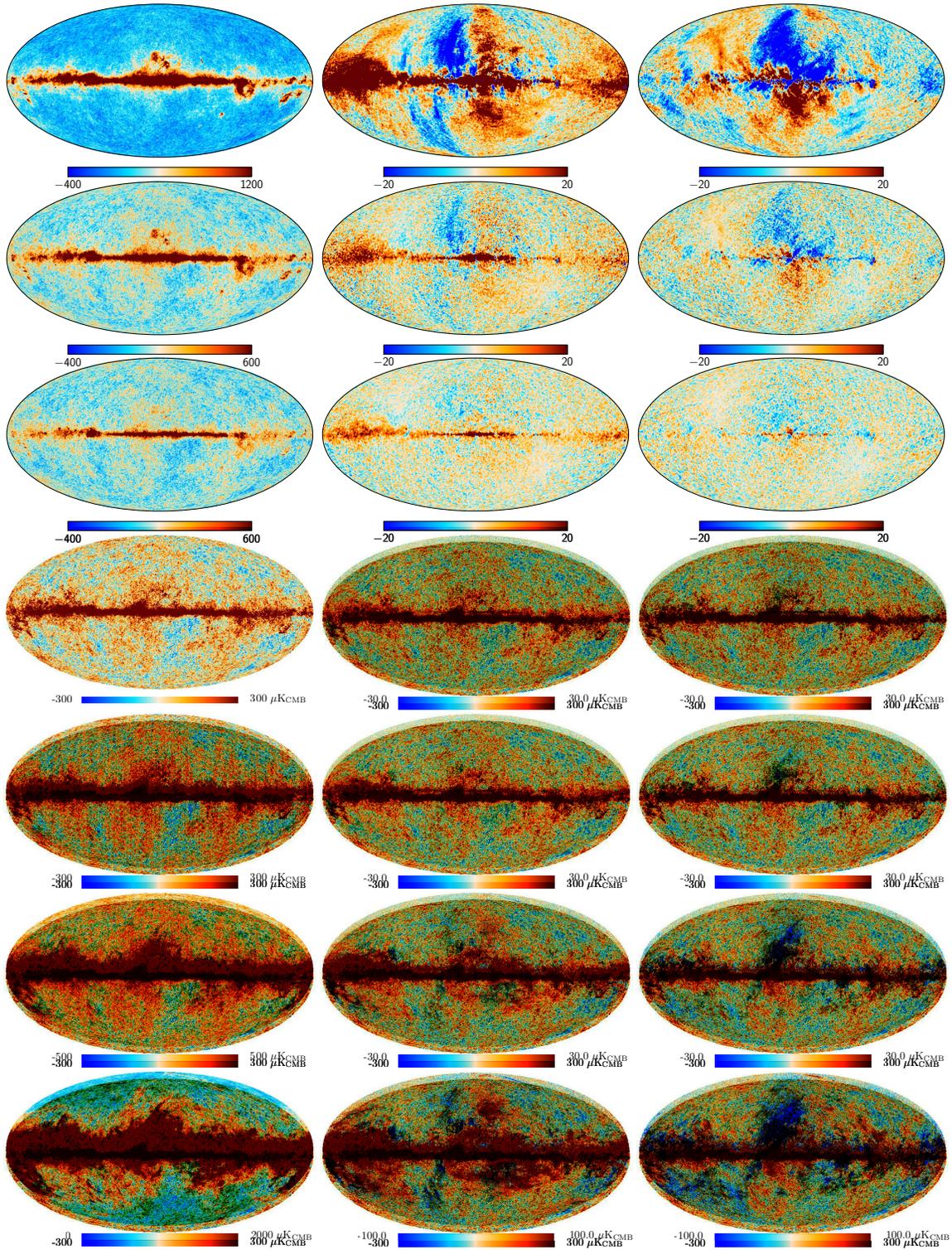

Figure 3.4: *Planck*-LFI maps at 30, 44 and 70 GHz [105] (first three rows), and HFI maps at 100, 143, 217, and 353 GHz [63] (last four rows), for Stokes $I$, $Q$, and $U$ (in columns).

**Parametric methods** The general assumption behind all parametric methods is that the functional form of the frequency scalings is known, and our ignorance can be quantified by means of relatively few spectral parameters, which can be determined by a fitting procedure. Parametric methods are widely used in CMB analyses and perform very well as long as the sky signal is modeled correctly. In case the assumed frequency scalings are wrong, however, one ends up with biased estimates for the spectral parameters, which can translate into foreground contamination.

**Blind methods** This class of methods is useful when we have poor knowledge about the SEDs of the foreground emission, since it does not make any assumption about their spatial distribution nor frequency dependence. The simplest blind methods are the internal linear combination (ILC) and its implementation in harmonic space, the harmonic ILC (HILC), which we will briefly present in Section 7.2.2.

Regardless of what method one chooses, the output of the foreground cleaning step will be an estimate of the CMB signal, in the form of a set of Stokes $I$, $Q$, and $U$ maps, or their spherical harmonics coefficients. The last step of any CMB analysis pipeline then is to analyze the resulting maps (or $a_{\ell m}^X$, or $C_\ell^{XY}$) to constrain cosmological parameters.

## 3.4   Parameter inference

Likelihood approaches are the most widespread methods used for statistical inference in CMB analyses [106]. In this section we provide an intuitive description of the maximum likelihood method following [98], and then move to a more concrete example of likelihood function to estimate the tensor-to-scalar ratio, $r$, from the observed $C_\ell^{BB}$.

**The method of maximum likelihood**   Consider a random variable $x$ distributed according to the p.d.f. $f(x; \theta)$ and assume that the functional form of $f(x; \theta)$ is known, while the value of the parameter $\theta$ is unknown. The method of maximum likelihood is a technique for estimating the unknown parameter from a finite sample of data.

Suppose that the random variable $x$ has been measured $n$ times, yielding the values $x_1, x_2, \ldots, x_n$. Given the hypothesis $f(x; \theta)$, including the value of $\theta$, the probability $P$ that $x_i \in [x_i, x_i + dx_i]$ for all $i$ is given by

$$P = \prod_{i=1}^{n} f(x_i, \theta) dx_i.$$  (3.23)

If the hypothesized p.d.f. and the parameter value are correct, one expects a high probability for the measured data. On the contrary, a parameter value far from the true value should give a low probability for the measurements obtained. The same reasoning applies to the likelihood function

$$L(\theta) \equiv \prod_{i=1}^{n} f(x_i, \theta) \,. \tag{3.24}$$

Then the maximum likelihood (ML) estimators for the parameter $\theta$, often denoted by a hat, $\hat{\theta}$, are defined as those that maximize the likelihood function. As long as the likelihood function is a differentiable function of $\theta$ and the maximum is not at the boundary of the parameter range, the estimator satisfies

$$\left. \frac{\partial L}{\partial \theta} \right|_{\hat{\theta}} = 0 \,. \tag{3.25}$$

If more than one local maximum exists, the highest one is taken. Note that the motivation of the ML principle presented above does not necessarily guarantee any optimal properties for the resulting estimators.

### 3.4.1 Concrete example: likelihood for the tensor-to-scalar ratio

Given a theoretical model for the $BB$ angular power spectrum as a function of the tensor-to-scalar ratio, $C_\ell^{BB}(r)$, the probability to observe the $a_{\ell m}^{B,\mathrm{obs}}$ coefficient reads

$$P\left(a_{\ell m}^{B,\mathrm{obs}} | C_\ell^{BB}(r)\right) = \frac{1}{\sqrt{2\pi C_\ell^{BB}(r)}} \exp\left[-\frac{1}{2} \frac{a_{\ell m}^{B,\mathrm{obs}} a_{\ell m}^{B,\mathrm{obs}*}}{C_\ell^{BB}(r)}\right] \,, \tag{3.26}$$

where we have used that $a_{\ell m}^{B,\mathrm{obs}}$ is a complex Gaussian variable with zero mean and $C_\ell^{BB}(r)$ variance. The probability of observing a whole set of $\{a_{\ell m}^{B,\mathrm{obs}}\}$ coefficients with the same $\ell$ then satisfies

$$\begin{aligned} \log P\left(\{a_{\ell m}^{B,\mathrm{obs}}\} | C_\ell^{BB}(r)\right) &= \log\left[\prod_{m=-\ell}^{\ell} P\left(a_{\ell m}^{B,\mathrm{obs}} | C_\ell^{BB}(r)\right)\right] \\ &= -\frac{2\ell+1}{2} \log C_\ell^{BB}(r) + \sum_{m=-\ell}^{\ell}\left[-\frac{1}{2} \frac{a_{\ell m}^{B,\mathrm{obs}} a_{\ell m}^{B,\mathrm{obs}*}}{C_\ell^{BB}(r)}\right] + \mathrm{const.} \\ &= -\frac{1}{2}\left[(2\ell+1)\log C_\ell^{BB}(r) + \frac{\sum_{m=-\ell}^{\ell}\left(a_{\ell m}^{B,\mathrm{obs}} a_{\ell m}^{B,\mathrm{obs}*}\right)}{C_\ell^{BB}(r)}\right] + \mathrm{const.} \,. \end{aligned} \tag{3.27}$$

Note that the right-hand side depends on the observed spherical harmonic coefficients only via the combination $a_{\ell m}^{B,\mathrm{obs}} a_{\ell m}^{B,\mathrm{obs}*} = r_{\ell m}^2$, where $r_{\ell m}$ denotes the modulus of the complex number $a_{\ell m}^{B,\mathrm{obs}} = r_{\ell m} e^{i\theta_{\ell m}}$. This makes straightforward to translate eq. (3.27) into a probability of observing $C_\ell^{BB,\mathrm{obs}} = (2\ell+1)^{-1} \sum_m r_{\ell m}^2$ given the model[6]:

$$\log P\left(C_\ell^{BB,\mathrm{obs}}|C_\ell^{BB}(r)\right) = -\frac{2\ell+1}{2}\left[\log C_\ell^{BB}(r) + \frac{C_\ell^{BB,\mathrm{obs}}}{C_\ell^{BB}(r)} - \frac{2\ell-1}{2\ell+1}\log C_{\ell,\mathrm{obs}}^{BB}\right] + \mathrm{const.}. \quad (3.28)$$

The likelihood function, $L(r)$, can then be obtained by evaluating $P$ for a set of values of the tensor-to-scalar ratio.

---

[6]To change variables from the set $\{r_{\ell m}\}$ to the angular power spectra $C_\ell^{BB,\mathrm{obs}}$, one has to compute

$$P\left(C_\ell^{BB,\mathrm{obs}}|C_\ell^{BB}(r)\right) = \int \mathrm{d}r_{\ell,-\ell} \dots \mathrm{d}r_{\ell,\ell} P\left(\{r_{\ell m}\}|C_\ell^{BB}(r)\right)\delta\left(C_\ell^{BB,\mathrm{obs}} - \frac{1}{2\ell+1}\sum_{m=-\ell}^{\ell} r_{\ell m}^2\right)$$

$$= \int R_\ell^{2\ell}\,\mathrm{d}R_\ell\,\mathrm{d}\Omega_{2\ell+1} P\left(R_\ell|C_\ell^{BB}(r)\right)\delta\left(C_\ell^{BB,\mathrm{obs}} - \frac{R_\ell^2}{2\ell+1}\right),$$

where $R_\ell^2 \equiv \sum_m r_{\ell m}^2$ and $\mathrm{d}\Omega_{2\ell+1}$ represents the $(2\ell+1)$-dimensional angular line element. Now, since the integrand does not depend on $\Omega_{2\ell+1}$, it will only result in a multiplicative factor. The $\delta$ function can be expressed in terms of $R_\ell$ itself by using that $\delta(g(x)) = \delta(x - x_0)/g'(x_0)$, where $x_0$ is a root of $g$, i.e.

$$\delta\left(C_\ell^{BB,\mathrm{obs}} - \frac{R_{\ell^2}}{2\ell+1}\right) = -\frac{2\ell+1}{2R_\ell}\delta\left(R_\ell - \sqrt{(2\ell+1)C_\ell^{BB,\mathrm{obs}}}\right).$$

Plugging this in the expression above returns

$$P\left(C_\ell^{BB,\mathrm{obs}}|C_\ell^{BB}(r)\right) \propto \int R_\ell^{2\ell-1}\,\mathrm{d}R_\ell\,P\left(R_\ell|C_\ell^{BB}(r)\right)\delta\left[R_\ell - \sqrt{(2\ell+1)C_\ell^{BB,\mathrm{obs}}}\right].$$

Eq. (3.28) follows after replacing $P\left(R_\ell|C_\ell^{BB}(r)\right)$ with the expression on the right-hand side of eq. (3.27).

# Chapter 4

# HWPs as polarization modulators

**Summary:** Some of the next-generation CMB experiments will use rotating half-wave plates (HWPs) as polarization modulators. In this chapter, we show how this choice can help to achieve better polarization measurements by mitigating the systematic effects due to the presence of $1/f$ noise and induced by pair-differencing the readings of orthogonal detectors. We also introduce the HWP non-idealities and briefly discuss some of the reasons why they should be carefully studied.

*This is an adaptation of some personal notes I took during the course of my PhD. In particular, Section 4.1.1 is based on a project that I started during my time at MPA that did not result in a publication.*

## 4.1   The ideal half-wave plate

A waveplate, or retarder, is a polarization-altering device made of a birefringent material whose index of refraction depends on the polarized state of the incoming radiation. Light polarized along the fast axis has a lower index of refraction and travels faster through the waveplate, while light polarized along the slow axis (orthogonal to the fast axis) has a higher index of refraction. Graphically, we represent the fast (slow) axis with a green (red) line on the face of the waveplate, as shown in Figure 4.1. The difference in the index of refraction, $\Delta n$, results in a phase shift between the polarization components, $\Gamma$, that also depends on the thickness of the crystal, $L$, and the light's wavelength, $\lambda_0$, via the relation

$$\Gamma = \frac{2\pi\,\Delta n\,L}{\lambda_0}. \tag{4.1}$$

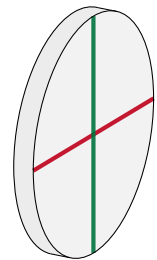

waveplate

Figure 4.1: Schematic representation of a waveplate with the fast and slow axes represented by a green and a red line, respectively.

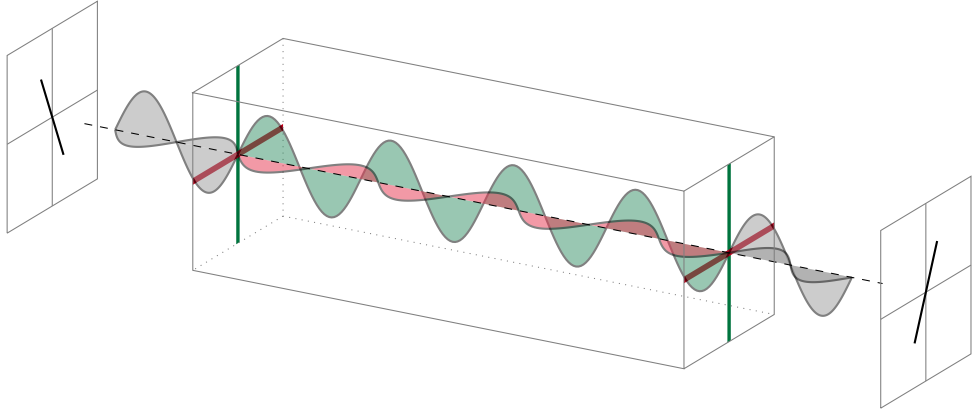

Figure 4.2: Close-up of how an HWP works. The green and red lines on the HWP faces represent the fast and slow axis, respectively. Light polarized along the fast axis has a refraction index $n_f$, while light polarized along the slow axis has $n_s > n_f$, resulting in a $\pi$ phase shift between the two orthogonal components. The incoming and outgoing polarization vectors are then one the reflection of the other with respect to the fast axis.

For a half-wave plate (HWP), the relationship between $L$, $\Delta n$, and $\lambda_0$ is chosen so that the phase shift between polarization components is $\Gamma = \pi$. As shown in Figure 4.2, this phase shift causes the incoming and outgoing polarization vectors to be one the reflection of each other with respect to the fast axis. At the level of Stokes parameters, this amounts to keeping $I$ and $Q$ fixed, while flipping the sign of $U$ and $V$. The effect of an HWP can therefore be described by the Mueller matrix (see Appendix B for its definition)

$$\mathcal{M} = \begin{pmatrix} 1 & 0 & 0 & 0 \\ 0 & 1 & 0 & 0 \\ 0 & 0 & -1 & 0 \\ 0 & 0 & 0 & -1 \end{pmatrix}. \tag{4.2}$$

Rotating HWPs are used in CMB experiments as polarization modulators, since they help mitigate systematic effects. In Sections 4.1.1 and 4.1.2 we will discuss their two main advantages: the suppression of the $1/f$ noise component and the mitigation of pair-differencing systematic effects.

### 4.1.1  Suppression of $1/f$ noise

If unmitigated, a $1/f$ noise component constitutes a critical problem for the detection of primordial $B$ modes. This is because a frequency power spectrum as the one in eqs. (3.6) or (3.7) translates in a $1/\ell$ angular power spectrum:

$$N_\ell \propto 1 + \left( \frac{\ell_{\mathrm{knee}}^2}{\ell^2} \right)^{\alpha_\ell}, \tag{4.3}$$

where $\ell_{\mathrm{knee}}$ and $\alpha_\ell$ depend on the noise properties, the scanning strategy specifics and the map-maker used to analyze the data. Regardless of the specific values taken by the

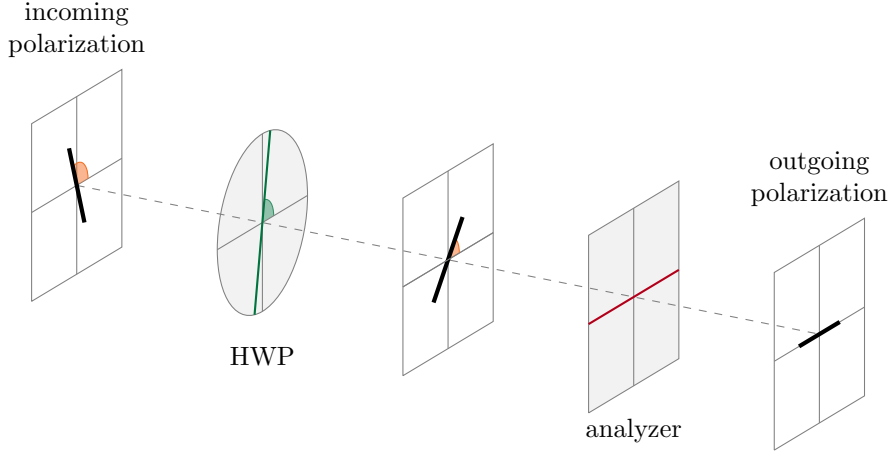

incoming
polarization

outgoing
polarization

HWP

analyzer

Figure 4.3: The incoming light first meets the HWP, that (in the ideal case) rotates the polarization direction. Then, an analyzer projects the polarized light along a given direction and the light is detected.

parameters, $N_\ell$ is larger at lower multipoles, and it could therefore keep us from detecting the primordial $B$ modes, which are best constrained at low multipoles. In the following of this section, we will see how the HWP can help in this direction.

We start by writing the observed noise angular power spectra as $\widehat{N}_\ell^{XY} = \langle \widehat{n}_{\ell m}^X \widehat{n}_{\ell m'}^{Y*} \rangle$, where $\widehat{n}_{\ell m}^X$ denotes the spherical harmonics coefficients of the $(\widehat{\mathbf{I}}, \widehat{\mathbf{Q}}, \widehat{\mathbf{U}})$ maps reconstructed from the noise TOD, $\mathbf{n}$. If we assume the simple binning map-maker introduced in Section 3.2, the reconstructed Stokes parameters at the pixel $p$ can be modeled explicitly as

$$(\widehat{I}, \widehat{Q}, \widehat{U})_p = \sum_{it} \left[ (A^T A)^{-1} A^T \right]_{pit} n_{it} , \tag{4.4}$$

where $A$ is the response matrix, and the indices $i$ and $t$ span the detectors and time samples, respectively. Note that each element of $\left[ (A^T A)^{-1} A^T \right]_{pit}$ is a 3-vector, which we denote $(\mathbb{I}, \mathbb{Q}, \mathbb{U})_{pit}$, following the same notation as in Section 3.1.2. Eq. (4.4) then reads

$$(\widehat{\mathbf{I}}, \widehat{\mathbf{Q}}, \widehat{\mathbf{U}}) = \sum_{it} (\mathbb{I}, \mathbb{Q}, \mathbb{U})_{it} n_{it} , \tag{4.5}$$

where the boldface quantities represent pixelized maps. Again, $n_{it}$ represents the noise observed by the detector $i$ at the time sample identified by the index $t$, and can therefore be written as[1]

$$n_{it} = \sum_f \tilde{n}_{if} e^{i2\pi \frac{f}{n_{\text{obs}}} t} , \tag{4.6}$$

where $n_{\text{obs}}$ denotes the total number of observations, $f$ takes values in the interval $[0, n_{\text{obs}} - 1]$, and $\tilde{n}_{if}$ is the discrete Fourier transform of $n_{it}$. Eq. (4.5) then becomes

$$(\widehat{\mathbf{I}}, \widehat{\mathbf{Q}}, \widehat{\mathbf{U}}) = \sum_i \sum_f \tilde{n}_{if} \left[ \frac{1}{n_{\text{obs}}} \sum_t (\mathbb{I}, \mathbb{Q}, \mathbb{U})_{it} e^{-i2\pi \frac{f}{n_{\text{obs}}} t} \right] . \tag{4.7}$$

---

[1] The inverse discrete Fourier transform is defined compatibly with eq. (3.3).

By taking the spherical harmonics coefficients of both sides, we find

$$\widehat{n}_{\ell m}^{X} = \sum_i \sum_f \tilde{n}_{if} [a_{\ell m}^{\mathbb{X}}]_{if} \,. \tag{4.8}$$

where the $[a_{\ell m}^{\mathbb{X}}]_{if}$ denote the spherical harmonics coefficients of the term in square brackets in eq. (4.7). To compute the angular power spectra from the spherical harmonics coefficients, one has to first take the product

$$\widehat{n}_{\ell m}^{X} \widehat{n}_{\ell m'}^{Y*} = \sum_{ii'} \sum_{ff'} \tilde{n}_{if} \tilde{n}_{i'f'}^{*} [a_{\ell m}^{\mathbb{X}}]_{if} [a_{\ell m}^{\mathbb{Y}*}]_{i'f'} \,, \tag{4.9}$$

which takes a more compact form if we average over the noise variables. In particular, assuming the noise to be stationary and uncorrelated between detectors, the product $\tilde{n}_{if}\tilde{n}_{i'f'}$ averages to

$$\tilde{n}_{if} \tilde{n}_{i'f'}^{*} \to P_{if} \delta_{ff'} \, \delta_{ii'} \,, \tag{4.10}$$

where $P_{if}$ is the noise power spectrum for the detector $i$. Plugging this expression into eq. (4.9), we get

$$\widehat{n}_{\ell m}^{X} \widehat{n}_{\ell m'}^{Y*} \to \sum_i \sum_f P_{if} [a_{\ell m}^{\mathbb{X}}]_{if} [a_{\ell m}^{\mathbb{Y}*}]_{if} \,. \tag{4.11}$$

The noise angular power spectra, $\widehat{N}_{\ell}^{XY}$ then read

$$\widehat{N}_{\ell}^{XY} = \sum_i \sum_f P_{if} [C_{\ell}^{\mathbb{XY}}]_{if} \,, \tag{4.12}$$

where $[C_{\ell}^{\mathbb{XY}}]_{if}$ are the angular power spectra of the term in square brackets in eq. (4.7).

In the following two paragraphs, we will use eq. (4.12) to show how the presence of the HWP affects the reconstructed angular power spectra in a simple case: mock noise properties, low resolution input maps, mock scanning strategy, and only a few detectors[2].

In particular, given a $n_{\text{side}} = 32$ resolution `healpy`[3] pixelization on the sphere, we set $f_{\text{samp}} = 20\,\text{Hz}$ and assume to scan the sky from the 0th to the $n_{\text{pix}}$ pixel (one observation per pixel), with four detectors located at boresight with 0, 90, 45, and 135 degrees orientations. We assume the telescope coordinates to be aligned with the sky ones at all time, i.e. $\psi_t = 0$, and the noise parameters to be $\sigma_0 = 10\,\mu\text{K}\sqrt{s}$, $f_{\text{min}} = 10^{-3}\,\text{Hz}$, $f_{\text{knee}} = 4\,\text{mHz}$ and $\alpha = 1$.

**Without HWP**  When no HWP is present, the data model is specified by eq. (3.12), and since we are assuming $\psi_t = 0$, it takes the even simpler form

$$\mathbb{S}_{it}^{T} = \frac{1}{2} \begin{pmatrix} 1 & 1 & 0 & 0 \end{pmatrix} \mathcal{R}_{\xi_i} \,. \tag{4.13}$$

---

[2]The implementation for this simplistic case can be found at `https://github.com/martamonelli/Pf2N1`. The idea behind this project was to test the formalism in this simple case and then move to more realistic scanning strategies and focal planes. The code, however, turned out to be too computationally expensive to be generalized efficiently, and ended up not to result in a publication.

[3]`https://github.com/healpy/healpy`.

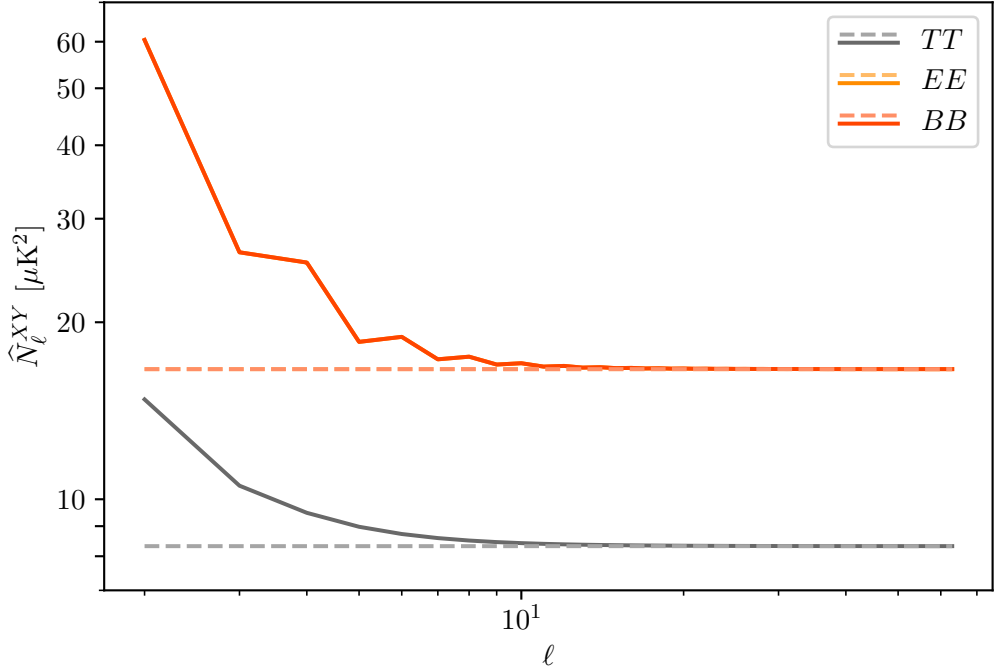

Figure 4.4: Reconstructed noise angular power spectra, $\widehat{N}_\ell^{TT}$, $\widehat{N}_\ell^{EE}$ and $\widehat{N}_\ell^{BB}$ (gray, orange and red, respectively), in the case without HWP. The dashed lines are obtained assuming white noise in input, while the solid lines correspond to a noise power spectrum as in eq. (3.7). The reconstructed spectra have a non-vanishing $1/\ell$ component.

Given $\mathbb{S}_{it}^T$, we compute the noise angular power spectra $\hat{N}_\ell^{XY}$ according to eq. (4.12). The $TT$, $EE$ and $BB$ spectra are shown in Figure 4.4, with and without the $1/f$ component. The $EE$ and $BB$ spectra overlap, and they are twice as large as their $TT$ counterpart, as expected. When the $1/f$ term is considered, the angular power spectra show a $1/\ell$ behaviour.

**With HWP** To include a rotating HWP, the model of eq. (4.13) has to be changed into

$$\mathbb{S}_{it}^T = \frac{1}{2} \begin{pmatrix} 1 & 1 & 0 & 0 \end{pmatrix} \mathcal{R}_{\xi_i - \phi_t} \mathcal{M}_{\text{HWP}} \mathcal{R}_{\phi_t} \,. \tag{4.14}$$

where $\phi_t$ represents the HWP angle, which is assumed to be rotating four times per second. By repeating the same analysis we discussed in the no-HWP case, we end up with the noise angular power spectra shown in Figure 4.5. The $EE$ and $BB$ spectra still overlap, and they are still twice as strong as the $TT$ ones, but the $1/\ell$ behaviour completely disappeared from the polarized signal.

With this simple analysis we were able to see how effectively the HWP mitigates the $1/f$ noise component in the polarized signal, in a concrete case. The intuitive reason for this behaviour is that the fast rotation of the HWP modulates the signal to high frequencies, where the $1/f$ noise is not as strong, effectively mitigating it.

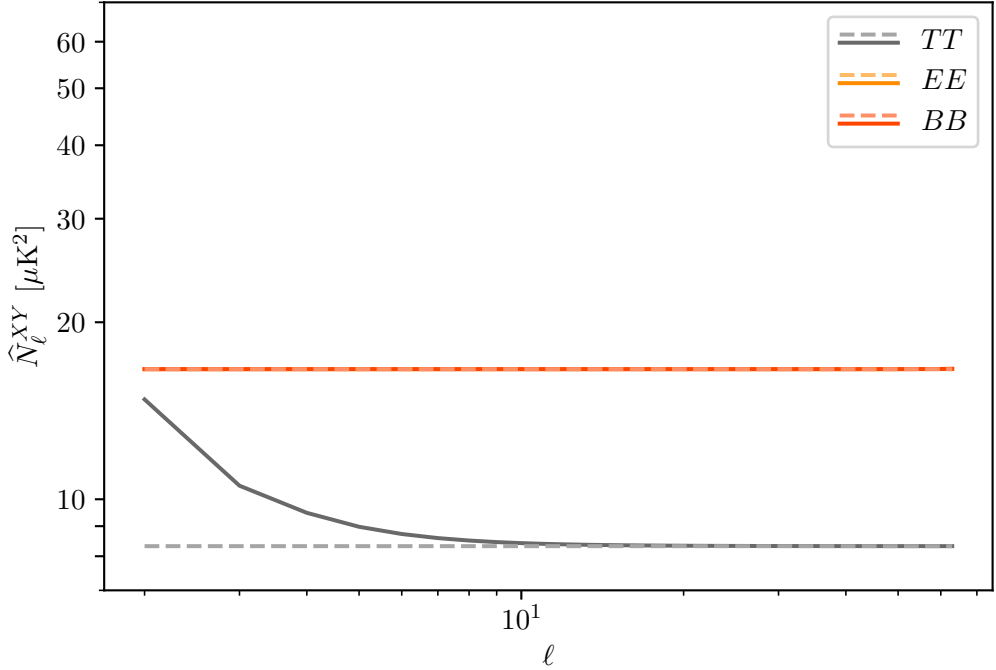

Figure 4.5: Reconstructed noise angular power spectra, $\widehat{N}_\ell^{TT}$, $\widehat{N}_\ell^{EE}$ and $\widehat{N}_\ell^{BB}$ (gray, orange and red, respectively), in the case with (ideal) HWP. The dashed lines are obtained assuming white noise in input, while the solid lines correspond to a noise power spectrum as in eq. (3.7). The HWP modulation completely suppresses the $1/\ell$ component in the polarized angular power spectra.

## 4.1.2   Pair-differencing systematics

Without HWP, the simplest method of measuring polarization (used for example by both WMAP and *Planck*) is to take the difference of the readings of pairs of orthogonal detectors, which can however induce some $I \to P$ leakage. These kind of effects are often referred to as pair-differencing systematics, and a rotating HWP can help mitigate them. Here we will show how pair-difference systematics can arise and how a rotating HWP can help mitigate them.

**Measuring polarization w/o HWP**   Consider two ideal polarimeters, oriented along the $x$ and $y$ axes, respectively. Their readings, $d_0$ and $d_{90}$, can be modeled as

$$d_0 = \mathbf{a}^T \cdot \mathbf{S} = \begin{pmatrix} \frac{1}{2} & \frac{1}{2} & 0 \end{pmatrix} \begin{pmatrix} I \\ Q \\ U \end{pmatrix} = \frac{I+Q}{2}\,, \tag{4.15a}$$

$$d_{90} = \mathbf{a}^T \cdot \mathcal{R}_{90} \cdot \mathbf{S} = \begin{pmatrix} \frac{1}{2} & \frac{1}{2} & 0 \end{pmatrix} \begin{pmatrix} 1 & 0 & 0 \\ 0 & -1 & 0 \\ 0 & 0 & -1 \end{pmatrix} \begin{pmatrix} I \\ Q \\ U \end{pmatrix} = \frac{I-Q}{2}\,. \tag{4.15b}$$

By taking the difference of the two readings, we recover $Q$. Similarly, we can use two detectors oriented along the $45°$ and $135°$ directions to measure $U$. This works perfectly as long as the detectors are identical. If the responses of the two detectors are not perfectly identical, however, one could end up with spurious polarization. For example, if the detector oriented along the $x$ axis has a slightly higher gain than the other, the difference $d_0 - d_{90}$ will have some leftover total intensity, resulting in $I \to P$ leakage, which is extremely problematic given the smallness of the primordial $B$-mode signal.

**Measuring polarization w/ HWP**   If the first element in the telescope optical chain is an HWP, the signal measured by a detector oriented along the $x$ axis reads

$$d_{0,t} = \mathbf{a}^T \cdot \mathcal{R}_{-\phi_t} \mathcal{M}_{\text{HWP}} \mathcal{R}_{\phi_t} \cdot \mathbf{S}$$

$$= \begin{pmatrix} \frac{1}{2} & \frac{1}{2} & 0 \end{pmatrix} \begin{pmatrix} 1 & 0 & 0 \\ 0 & \cos(2\phi_t) & -\sin(2\phi_t) \\ 0 & \sin(2\phi_t) & \cos(2\phi_t) \end{pmatrix} \begin{pmatrix} 1 & 0 & 0 \\ 0 & 1 & 0 \\ 0 & 0 & -1 \end{pmatrix} \begin{pmatrix} 1 & 0 & 0 \\ 0 & \cos(2\phi_t) & \sin(2\phi_t) \\ 0 & -\sin(2\phi_t) & \cos(2\phi_t) \end{pmatrix} \begin{pmatrix} I \\ Q \\ U \end{pmatrix}$$

$$= \frac{1}{2} \begin{pmatrix} 1 & \cos(2\phi_t) & -\sin(2\phi_t) \end{pmatrix} \begin{pmatrix} 1 & 0 & 0 \\ 0 & 1 & 0 \\ 0 & 0 & -1 \end{pmatrix} \begin{pmatrix} I \\ Q\cos(2\phi_t) + U\sin(2\phi_t) \\ U\cos(2\phi_t) - Q\sin(2\phi_t) \end{pmatrix}$$

$$= \frac{1}{2} \begin{pmatrix} 1 & \cos(2\phi_t) & -\sin(2\phi_t) \end{pmatrix} \begin{pmatrix} I \\ Q\cos(2\phi_t) + U\sin(2\phi_t) \\ Q\sin(2\phi_t) - U\cos(2\phi_t) \end{pmatrix}$$

$$= \frac{1}{2} \left[ I + Q\cos^2(2\phi_t) + U\sin(2\phi_t)\cos(2\phi_t) - Q\sin^2(2\phi_t) + U\cos(2\phi_t)\sin(2\phi_t) + \right]$$

$$= \frac{1}{2} \left[ I + Q\cos(4\phi_t) + U\sin(4\phi_t) \right]\,. \tag{4.16}$$

If the HWP is continuously rotating with angular frequency $\omega$, the polarized signal is modulated to $4\omega$, while the total intensity has no modulation. This can help discriminate between intrinsic and spurious polarization. For example, consider two orthogonal detectors with gains $g_0$ and $g_{90}$, respectively, the difference of their readings is

$$d_{0,t} - d_{90,t} = \frac{g_0}{2} \left[ I + Q\cos(4\omega t) + U\sin(4\omega t) \right] - \frac{g_{90}}{2} \left[ I - Q\cos(4\omega t) - U\sin(4\omega t) \right]$$

$$= \frac{g_0 - g_{90}}{2} I + \frac{g_0 + g_{90}}{2} Q\cos(4\omega t) + \frac{g_0 - g_{90}}{2} U\sin(4\omega t)\,, \tag{4.17}$$

and the leaked $I$ can be easily removed by measuring only the $4\omega$ component.

## 4.2   HWP non-idealities

From the arguments we just provided, it would seem that HWPs are extremely positive objects that have no drawbacks. However, the considerations we just made only hold for ideal HWPs, i.e. which are described by the simple Mueller matrix of eq. (4.2). No real HWP is ideal, because of a number of effects. For instance [107]:

- Waveplates have the incorrect retardance. Thus, there will be some deviation from a quarter-wave or a half-wave of retardance because of fabrication errors or a change in wavelength.

- Waveplates usually have some diattenuation because of differences in absorption coefficients (dichroism) and due to different transmission and reflection coefficients at the interfaces. For example, birefringent waveplates have diattenuation due to the difference of the Fresnel coefficients at normal incidence for the two eigenpolarizations since $n_1 \neq n_2$. This can be reduced by antireflection coatings.

- The polarization properties vary with angle of incidence. Birefringent waveplates commonly show a quadratic variation of retardance with angle of incidence; the retardance increases along one axis and decreases along the orthogonal axis.

- The polarization properties vary with wavelength.

In general, therefore, the Mueller matrix associated to a HWP has no zero components, which depend both on the frequency, $\nu$, and angle of incidence, $\theta$:

$$\mathcal{M}_{\mathrm{HWP}}(\nu,\theta) = \begin{pmatrix} m_{\mathrm{II}}(\nu,\theta) & m_{\mathrm{IQ}}(\nu,\theta) & m_{\mathrm{IU}}(\nu,\theta) & m_{\mathrm{IV}}(\nu,\theta) \\ m_{\mathrm{QI}}(\nu,\theta) & m_{\mathrm{QQ}}(\nu,\theta) & m_{\mathrm{QU}}(\nu,\theta) & m_{\mathrm{QV}}(\nu,\theta) \\ m_{\mathrm{UI}}(\nu,\theta) & m_{\mathrm{UQ}}(\nu,\theta) & m_{\mathrm{UU}}(\nu,\theta) & m_{\mathrm{UV}}(\nu,\theta) \\ m_{\mathrm{VI}}(\nu,\theta) & m_{\mathrm{VQ}}(\nu,\theta) & m_{\mathrm{VU}}(\nu,\theta) & m_{\mathrm{VV}}(\nu,\theta) \end{pmatrix}, \qquad (4.18)$$

and we refer to the deviations $\mathcal{M}_{\mathrm{HWP}}(\nu,\phi) - \mathrm{diag}(1,1,-1,-1)$ as HWP non-idealities.

While the use of rotating (perfectly ideal) HWPs as polarization modulators can certainly reduce systematic effects (due to $1/f$ noise and pair-differencing of orthogonal detectors), it is worth asking whether the non-idealities can induce 'secondary' systematics and, if so, how to keep them under control. To answer these questions, one must propagate the effect of the non-idealities through the various steps that make up a CMB experiment (which we discussed in chapter 3), and try to understand how they ultimately affect the scientific information we are trying to extract from the data.

This is the focus of this thesis, and we have approached the problem with two nicely complementary classes of tools: realistic simulations and approximate models. The simulation framework used for the former will be discussed in the next chapter.

# Chapter 5

# A `beamconv`-based simulation framework for LiteBIRD

**Summary:** Realistic TOD simulations are key to the study of systematic effects, because they can account for them in their (at least partial) complexity. In this chapter, we present a simulation framework that returns beam-convolved TOD and binned maps for a LiteBIRD-like space mission. The code is heavily based on a modified version of the `beamconv` library, which was the first publicly available implementation of a beam convolution code capable of handling realistic beams and non-ideal HWPs. Here we present the changes we have made to `beamconv` to tailor its output to a LiteBIRD-like experiment.

*This is an adaptation of some personal notes I took during the course of my PhD. Section 5.1 summarizes some of the information presented in [29], Section 5.2 is a short summary of section 4.4 of [108], and section 5.3.2 is adapted from `qpoint`'s documentation [109].*

## 5.1 The LiteBIRD experiment

LiteBIRD, the Lite (Light) satellite for the study of B-mode polarization and Inflation from cosmic background Radiation Detection, is a space mission for primordial cosmology and fundamental physics. In May 2019, LiteBIRD was selected by the Japan Aerospace Exploration Agency (JAXA) as a strategic large-class mission, and its launch is currently expected by the end of the Japanese fiscal year 2032. LiteBIRD is planned to orbit the second Sun-Earth Lagrangian point (L2), where it will remain in a Lissajous orbit for three years, while mapping the CMB polarization across the entire sky. The basic structure of the spacecraft is shown in Figure 5.1.

In the current baseline design, LiteBIRD's focal plane will host ∼ 4000 bolometers, distributed in 15 partially overlapping frequency bands over a wide frequency range: from 34

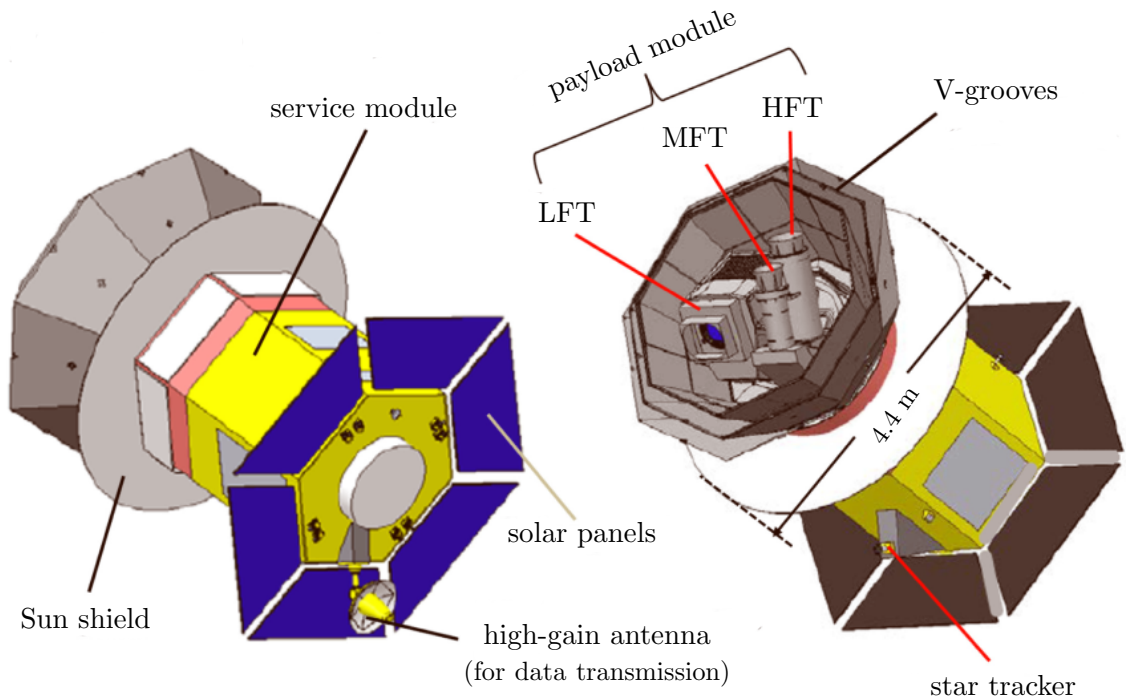

Figure 5.1: Conceptual design of the LiteBIRD spacecraft. The payload module houses the low-frequency telescope (LFT), the mid-frequency telescope (MFT), and the high-frequency telescope (HFT). Image and caption adapted from [29].

GHz to 448 GHz. The large number of detectors (compare e.g. with *Planck*'s specifications, listed in Table 1.1) will significantly reduce LiteBIRD's noise levels, while the large number of frequency bands (compare again with Table 1.1) and their frequency distribution will help to achieve a good characterization of the polarized dust and synchrotron emission.

In addition, LiteBIRD will use rotating HWPs as polarization modulators, which will help to mitigate the $1/f$ noise component and reduce pair-differencing systematics (see Chapter 4, Sections 4.1.1 and 4.1.2 in particular). Due to the limitations of current technology, it is impossible to design an HWP that operates in a controlled manner over the entire 34 - 448 GHz frequency range. Instead, LiteBIRD will use three different HWPs, each associated with a different telescope:

- The low-frequency telescope (LFT), operating between 34 and 161 GHz, will use a Pancharatnam-type [110, 111] multi-layer sapphire stack as an achromatic HWP with laser machined sub-wavelength structures anti-reflection coating [112];

- The mid-frequency telescope (MFT), operating between 89 and 224 GHz, will use a metamaterial refractive HWPs developed with metal-mesh technology [113, 114];

- The high-frequency telescope (HFT), operating between 166 and 448 GHz, will also use a metal-mesh HWP.

Thanks to these (and many other) choices, LiteBIRD is expected to achieve the unprecedented total sensitivity of $2.2\,\mu$K-arcmin, with a typical angular resolution of 0.5° at 100 GHz [29]. This will result in extremely precise polarization measurements at low multipoles, $\ell$ between 2 and 200, optimized to measure primordial $B$-modes and constrain inflationary models. LiteBIRD data is also promising for constraining cosmic birefringence and therefore probing new parity-violating physics.

In the remaining of this chapter, we will provide a brief introduction to `beamconv`[1] [101, 115] and show how we optimized it to run simulations for a LiteBIRD-like experiment.

## 5.2 Introduction to `beamconv`

In Section 3.1.2 we introduced two different approaches to model the noiseless TOD, one in pixel space and the other in harmonic space. For the latter, eq. (3.13) shows how to model the observed TOD, $d_t$, given the sky spherical harmonics coefficients, ($a^I_{\ell m}$, $_{+2}a^P_{\ell m}$, $_{-2}a^{P*}_{\ell m}$ and $a^V_{\ell m}$), the beam spherical harmonics coefficients ($b^{\mathbb{I}}_{\ell m}$, $_{+2}b^{\mathbb{V}}_{\ell m}$, $_{-2}b^{\mathbb{P}*}_{\ell m}$ and $b^{\mathbb{V}}_{\ell m}$), and the detector pointings (specified by $\theta_t$, $\varphi_t$ and $\psi_t$):

$$d_t = \sum_{s\ell m} \left[ b^{\mathbb{I}}_{\ell s} a^I_{\ell m} + \frac{1}{2} \left( _{-2}b^{\mathbb{P}*}_{\ell s}\, _{+2}a^P_{\ell m} + _{+2}b^{\mathbb{P}}_{\ell s}\, _{-2}a^{P*}_{\ell m} \right) + b^{\mathbb{V}}_{\ell s}\, a^V_{\ell m} \right] \sqrt{\frac{4\pi}{2\ell+1}} e^{-is\psi_t}\, _sY_{\ell m}(\theta_t,\phi_t)\,, \quad (5.1)$$

which is the data model at the core of the beam-convolution algorithm implemented in `beamconv`. The scanning strategy can be specified by initializing the `ScanStrategy` class:

```
1  S = ScanStrategy(duration=None, sample_rate=None,
2                   external_pointing=False, ...)
```

where we show only some arguments, with their default values in blue. Depending on the value of `external_pointing`, the boresight pointings can be loaded or, in some cases, calculated internally. Another important class of objects to specify are the beams, which can be loaded as a set of spherical harmonics coefficients or, in some cases, calculated internally. For example, a simple Gaussian beam located at boresight can be defined by

```
3  single_beam = Beam(btype='Gaussian', fwhm=None, lmax=None)
```

Once a beam (or beams) is defined, we can associate a focal plane with the `ScanStrategy` object by calling

```
4  S.add_to_focal_plane(single_beam, combine=True)
```

The information about the sky signal should instead be provided as an argument to the `scan_instrument_mpi` function in the form of spherical harmonics coefficients:

```
5  S.scan_instrument_mpi(slm, ...)
```

After running `scan_instrument_mpi`, we can recover the observed TOD and binned maps.

Because of its ability to handle realistic optics, `beamconv` is a promising framework for developing TOD simulations for the next-generation CMB experiments. In particular, it supports parallel computation through the Message Passing Interface (MPI) protocol[2] and relies on fast compiled code (`libsharp`, `qpoint`, and `numpy`) to perform critical operations, making it suitable for simulating long missions with a large number of detectors.

---

[1] https://github.com/AdriJD/beamconv.

**HWP modulation**   Besides being able to handle complex beams, `beamconv` can deal with HWP modulation in both the ideal and non-ideal cases. To model the HWP modulation, we can generalize eq. (5.1) to

$$d_t = \sum_{s\ell m} \left[ b_{\ell s}^{\mathbb{I}} a_{\ell m}^I + \frac{1}{2} \left( {}_{-2}B_{\ell s}^{\mathbb{P}} {}_2 a_{\ell m}^P + {}_2 B_{\ell s}^{\mathbb{P}} {}_{-2} a_{\ell m}^P \right) + b_{\ell s}^{\mathbb{V}} a_{\ell m}^V \right] \sqrt{\frac{4\pi}{2\ell+1}} e^{-is\psi_t} {}_s Y_{\ell m}(\theta_t, \phi_t), \quad (5.2)$$

where the effective polarized beams, ${}_{+2}B_{\ell s}^{\mathbb{P}}$ and ${}_{-2}B_{\ell s}^{\mathbb{P}*}$, depend on whether the HWP is ideal or not. For an ideal HWP,

$$_{+2}B_{\ell s}^{\mathbb{P}} \equiv {}_2 b_{\ell s}^{\mathbb{P}} e^{4i\phi}, \quad (5.3)$$

where $\phi$ is the angle between HWP and telescope coordinates (equation 20 of [101]). In the non-ideal case, instead

$$_{+2}B_{\ell s}^{\mathbb{P}} = \sqrt{2} \left( \mathcal{C}_{IP} b_{\ell s+2}^{\mathbb{I}} + \mathcal{C}_{VP} b_{\ell s+2}^{\mathbb{V}} \right) e^{-2i\phi} + \mathcal{C}_{P*P} {}_{-2} b_{\ell s+4}^{\mathbb{P}} e^{-4i\phi} + \mathcal{C}_{PP} {}_2 b_{\ell s}^{\mathbb{P}}, \quad (5.4)$$

where the $\mathcal{C}$ matrix is a complex representation of the standard HWP Mueller matrix:

$$\mathcal{C} \equiv \mathcal{T} \mathcal{M}_{\text{HWP}} \mathcal{T}^\dagger, \quad (5.5)$$

where the dagger denotes the transpose conjugate, $\mathcal{M}_{\text{HWP}}$ is the HWP Mueller matrix in its coordinate system, and $\mathcal{T}$ is given by

$$\mathcal{T} = \begin{pmatrix} 1 & 0 & 0 & 0 \\ 0 & \frac{1}{\sqrt{2}} & \frac{i}{\sqrt{2}} & 0 \\ 0 & \frac{1}{\sqrt{2}} & -\frac{i}{\sqrt{2}} & 0 \\ 0 & 0 & 0 & 1 \end{pmatrix}. \quad (5.6)$$

The generalized data model of eq. (5.2) has been implemented in `beamconv` in 2020 [115]. The code needs information about the HWP Mueller matrix, which should be specified after the beam is initialized, and about its rotation, which should be provided after the beam is added to the focal plane. For example, to specify an ideal HWP that rotates continuously completing two rotations per second, we can run

```
4  single_beam.hwp_mueller = np.diag([1, 1, -1, -1])
5  S.add_to_focal_plane(single_beam)
6  S.set_hwp_mod(mode='continuous', freq=2)
7  S.scan_instrument_mpi(slm, ...)
```

In the following we only discuss the modifications we made to the code to tailor it to a LiteBIRD-like mission. The modified code is available in a GitHub repo[3], along with the `LiteBIRD-like.ipynb` notebook which can reproduce all the plots shown in this chapter.

---

[2]Given multiple processors, the code distributes the boresight pointing computations and assigns a subset of the detectors to each processor.

[3]https://github.com/martamonelli/beamconv_again/tree/LiteBIRD-like.

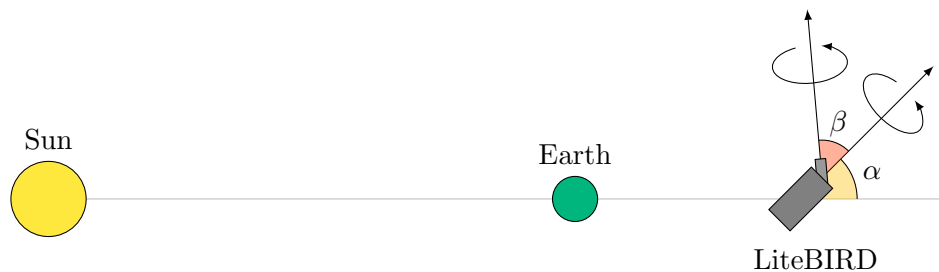

Figure 5.2: Schematic representation of LiteBIRD's set up. The antisun direction (in gray) forms an angle $\alpha = 45°$ with the spin axis which, in turn, forms an angle $\beta = 50°$ with the boresight. The spacecraft completes a precession every 192.348 minutes and a rotation around the spin axis every 20 minutes [29].

## 5.3 LiteBIRD's pointings

Although `beamconv` can read pointings in input, we preferred defining a new `scanning.py` module that could compute them on the fly for a LiteBIRD-like experiment. We also implemented the possibility of reading quaternion offsets in input, which makes it easier to interface with LiteBIRD's Instrument MOdule database (IMO).

### 5.3.1 Boresight pointings

LiteBIRD will observe the sky from L2, where it will remain in (Lissajous) orbit for the whole duration of the mission. As sketched in Figure 5.2, the spacecraft precesses around the anti-Sun direction, $\hat{n}_{as}$, while spinning around an axis that forms an angle $\alpha = 45°$ with $\hat{n}_{as}$. LiteBIRD's boresight is tilted by an angle $\beta = 50°$ with respect to the spin axis. This choice of the precession and boresight angles ensures that the main beam scans the entire sky without leaving any gaps, and without ever pointing at the Sun, the Earth or the Moon.

We implemented a new module in `beamconv` to compute the boresight pointings on the fly, given $\alpha$, $\beta$ and their respective angular frequencies (or periods). We used the same algorithm as `pyscan`[4]: we first compute the anti-Sun direction at a given time, and then apply some appropriate rotations to transform it into the boresight direction. In the current version of `beamconv` (this has been merged in the parent GitHub repository), this can be done on the fly by setting `use_l2_scan` to `True` when calling `scan_instrument_mpi`:

```
7  S.scan_instrument_mpi(slm, use_l2_scan=True,
8                        ctime_kwargs=dict(), q_bore_kwargs=dict())
```

This returns a LiteBIRD-like scanning strategy as the one shown in Figure 5.3.

---

[4] https://github.com/tmatsumu/LB_SYSPL_v4.2.

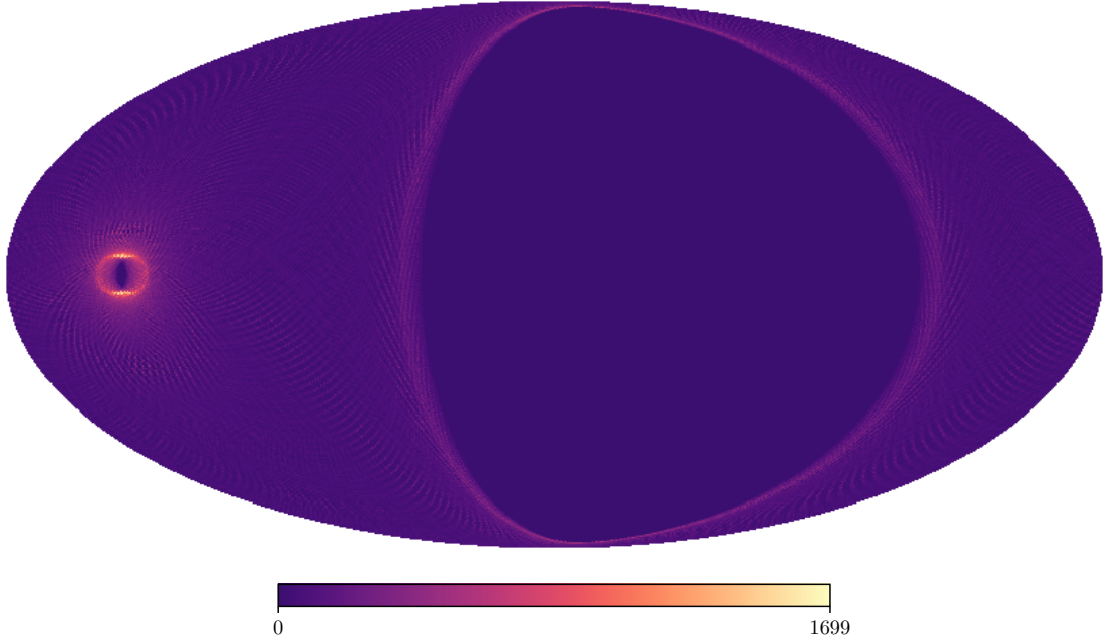

Figure 5.3: Hit map for a one detector 7-days LiteBIRD-like scanning strategy obtained by setting `ctime0 = 1510000000` and `use_l2_scan = True`.

## 5.3.2   Detectors' pointings

Knowing the boresight pointings is not the same as knowing the pointings of a given detector: detectors that are offset (i.e. not located in the center of the telescope) point in different directions than the boresight. In our version of `beamconv`, we introduced the possibility of feeding the detectors offsets directly to the `input_focal_plane` function, in the form of quaternions[5].

In Figure 5.4, we show a similar hit map to Figure 5.3, but with four detectors, one at boresight and the other three with non-zero offsets. By comparing the two figures, we can clearly see the difference in the detector pointings.

---

[5] A quaternion $q$ is composed of a scalar component $q_0$ and a vector component $\mathbf{q} = (q_1, q_2, q_3)$. Quaternions are generalized complex numbers in three dimensions, where the scalar component is real and the vector components are orthogonal imaginary quantities: $q = (q_0, \mathbf{q}) = q_0 + q_1 i + q_2 j + q_3 k$. The imaginary axes, $i$, $j$ and $k$, satisfy $i^2 = j^2 = k^2 = -1$, $ij = -ji = k$, $jk = -kj = i$, and $ki = -ik = j$. A unitary quaternion, $q$, provides a compact representation of a rotation around some arbitrary axis. To apply the rotation associated to $q$ to the vector $\mathbf{v}$, we treat the vector as a purely imaginary quaternion, $v = (0, \mathbf{v})$, and calculate $\mathbf{v}' = \mathbf{q}\mathbf{v}\mathbf{q}^{-1}$. This is equivalent to multiplying $\mathbf{v}$ by the matrix

$$M_q = \begin{pmatrix} q_0^2 + q_1^2 - q_2^2 - q_3^2 & 2(q_1 q_2 - q_0 q_3) & 2(q_1 q_3 + q_0 q_2) \\ 2(q_1 q_2 + q_0 q_3) & q_0^2 - q_1^2 + q_2^2 - q_3^2 & 2(q_2 q_3 - q_0 q_1) \\ 2(q_1 q_3 - q_0 q_2) & 2(q_2 q_3 + q_0 q_1) & q_0^2 - q_1^2 - q_2^2 + q_3^2 \end{pmatrix}.$$

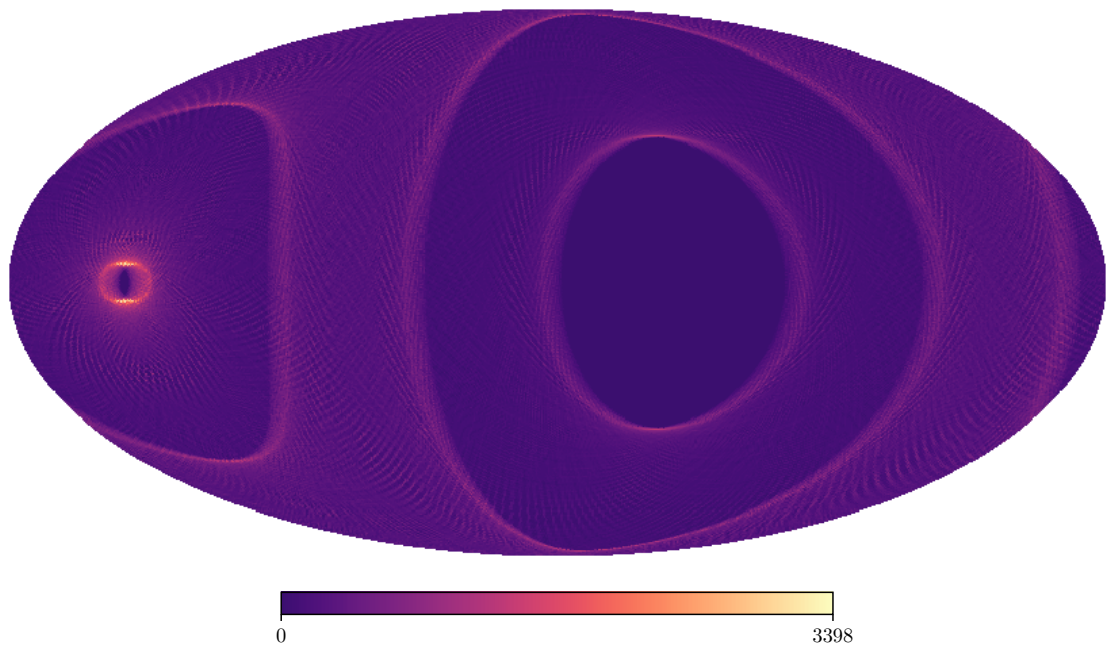

Figure 5.4: Same as Figure 5.3, but with one boresight and three offset detectors.

## 5.4   Adding noise and dipole signal

Another functionality we have added to `beamconv` is the possibility to generate noise- and dipole-only TODs on the fly and add them to the beam-convolved sky signal.

**Noise term**   If the `noise_tod` parameter is set to `True` when initializing the `ScanStrategy` class, a noise-only component is added to the TOD, chunk by chunk[6]. To generate it, we used the `OofaNoise` function from `ducc0`[7], that returns a realization of the noise TOD given the frequency power spectrum parameters of eq. (3.7): `sigma`, `f_min`, `f_knee` and `slope`= $2\alpha$. In Figure 5.5, we plot the noisy TOD (lighter teal), together with the noiseless signal (darker teal). To check if the noise term is doing what it is supposed to do, we can look at Figure 5.6, where we plot the frequency power spectrum $P(f)$ of the simulated noise together with its theoretical expectation. As expected, the agreement between the two increases with the number of realizations.

**Dipole term**   LiteBIRD's reference frame is not at rest with respect to the CMB, but it will have a non-zero velocity due to the motion of the spacecraft around L2, the motion of the L2 point in the Ecliptic plane, the motion of the Solar System around the Galactic Centre, and the motion of the Milky Way.

---

[6]By using the `partition_mission` function of the `ScanStrategy` class, it is possible to divide up the mission in equal-sized chunks. This can be helpful when simulating long missions.

[7]https://gitlab.mpcdf.mpg.de/mtr/ducc.

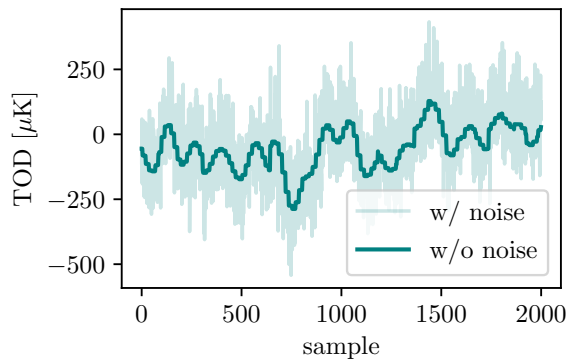

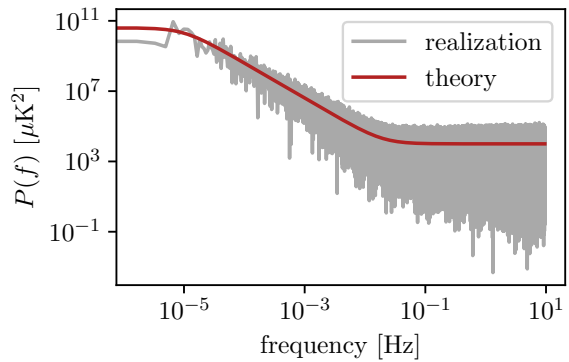

Figure 5.5: Simulated TOD with and without noise (lighter and darker teal lines, respectively) for 2000 observations.

Figure 5.6: Frequency power spectrum of the simulated noise (gray line) compared with the theoretical expectation (red line).

If, at some time $t$ the spacecraft is moving with a total velocity $\mathbf{v}(t)$ with respect to the CMB rest frame, the radiation coming from $\hat{n} = \hat{v}(t)$ will be blueshifted, while the radiation coming from $\hat{n} = -\hat{v}(t)$ will be redshifted due to the Doppler effect. As a consequence, the CMB temperature monopole will translate into an observed dipole anisotropy. Because of its apparent nature, the dipole signal is usually removed before any cosmological analysis is performed. However, the fact that it is bright and well known, makes this signal the most important photometric calibrator for balloon-borne and space-based missions.

In our version of beamconv, the dipole signal can be generated on the fly by setting dipole_tod = True when initializing the ScanStrategy class. What this does is to run a slightly modified version of the dipole module of LiteBIRD-sim framework[8], which returns the dipole TOD, given the time and direction of observation. The velocity is computed per chunk and a warning is issued whenever the chunk is too long (more than one day). In Figure 5.7 we show the simulated TOD, with and without the dipole signal, for 20000 observations.

---

[8] https://github.com/litebird/litebird_sim.

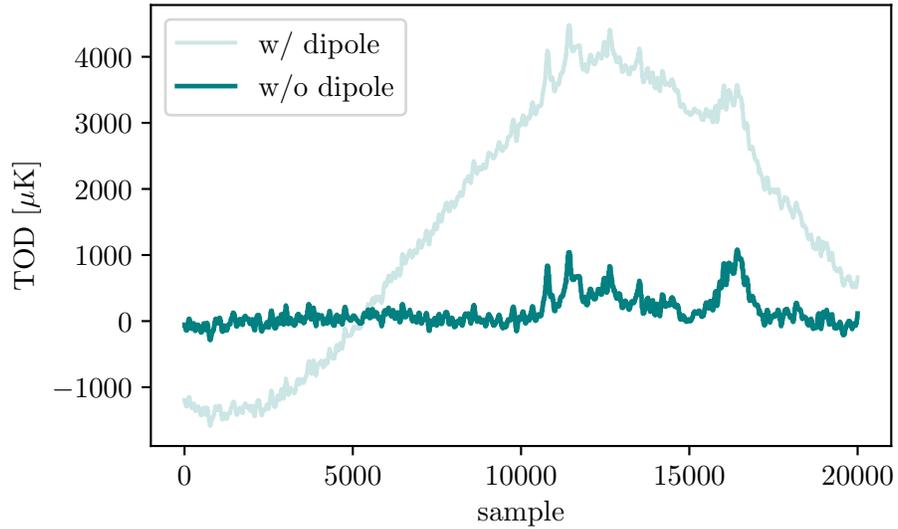

Figure 5.7: Simulated TOD with and without the dipole component (darker and lighter teal lines, respectively) for 20000 observations.

# Chapter 6

# HWP impact on cosmic birefringence

**Abstract:** Polarization of the CMB can probe new parity-violating physics such as cosmic birefringence (CB), which requires exquisite control over instrumental systematics. The non-idealities of the HWP represent a source of systematics when used as a polarization modulator. We study their impact on the CMB angular power spectra, which is partially degenerate with CB and miscalibration of the polarization angle. We use full-sky beam convolution simulations including HWP to generate mock noiseless time-ordered data, process them through a bin averaging map-maker, and calculate the power spectra including $TB$ and $EB$ correlations. We also derive analytical formulae which accurately model the observed spectra. For our choice of HWP parameters, the HWP-induced miscalibration angle amounts to a few degrees, which could be misinterpreted as CB. Accurate knowledge of the HWP is required to mitigate this. Our simulation and analytical formulae will be useful for deriving requirements for the accuracy of HWP calibration.

*This chapter is an adaptation of [78].*

## 6.1 Introduction

Temperature anisotropies in the cosmic microwave background (CMB) are an invaluable source of cosmological information [16, 19, 20]. Polarization anisotropies also contain a great wealth of complementary information [15, 17, 21, 22, 25, 27, 28, 116, 117], which has yet to be fully explored. A promising opportunity driving the development of a major experimental effort, involving both ground-based observatories (Simons Observatory [44], South Pole Observatory [45] and CMB Stage-4 [46]) and space missions (LiteBIRD [29] and PICO [47]), is to probe cosmic inflation [11–13]. Inflationary models predict the existence of a stochastic background of gravitational waves [35, 36] which would leave a distinctive $B$-mode signature on the CMB polarization [37–40].

The CMB polarization can also probe new parity-violating physics [49]. For example, in the presence of a time-dependent parity-violating pseudoscalar field, the linear polarization plane of CMB photons would rotate while they travel toward us [52–54]. Because of its similarity with photon propagation through a birefringent material, this phenomenon is referred to as cosmic birefringence (CB). The so-called CB angle, $\beta$, denotes the overall rotation angle from last scattering to present times. Although the effect of $\beta$ on the observed CMB angular power spectra is degenerate with an instrumental miscalibration of the polarization angle [55–58], the methodology proposed in [59–61], which relies on the polarized Galactic foreground emission to determine miscalibration angles, allowed to infer $\beta = 0.35 \pm 0.14°$ at 68% C.L. [62] from nearly full-sky *Planck* polarization data [63]. Subsequent works [64–66] reported more precise measurements for $\beta$. The statistical significance of $\beta$ is expected to improve with the next generation of CMB experiments, given the high precision at which they aim to calibrate the absolute position angle of linear polarization. This will make it unnecessary to rely on the Galactic foreground to calibrate angles and measure $\beta$ [49], hence avoiding the potential complications highlighted in [67].

The unprecedented sensitivity goals of future surveys, aiming to detect faint primordial $B$ modes, can only be achieved if systematics are kept under control. To this end, a promising strategy is to employ a rotating half-wave plate (HWP) as a polarization modulator. As shown by the previous analyses [68–75], a rotating HWP can both mitigate the $1/f$ noise component [68] and reduce a potential temperature-to-polarization ($I \rightarrow P$) leakage due to the pair differencing of orthogonal detectors [76, 77]. Because of these advantages, HWPs are used in the design of some next-generation experiments, including LiteBIRD [29]. However, non-idealities in realistic HWPs induce additional systematics which should be well understood in order for future experiments to meet their sensitivity requirements. This necessity motivated a number of recent works, from descriptions of HWP non-idealities [70, 113, 118, 119] and their impact on measured angular power spectra [120] to mitigation strategies [121–124].

In this chapter we study how HWP non-idealities can affect the estimated CMB angular power spectra if overlooked in the map-making step. We employ a modified version of the publicly available beam convolution code `beamconv`[1] [101, 115] and simulate two sets of noiseless time-ordered data (TOD). The two simulations make different assumptions on the HWP behavior. In the first case the HWP is assumed to be ideal, while non-idealities are included in the second case. We then process the two TOD sets with a map-maker assuming the ideal HWP and compare the output power spectra. We also derive a set of analytic expressions for the estimated angular power spectra as functions of the input spectra and the elements of the HWP Mueller matrix. These formulae accurately model the output power spectra. Finally, we show that neglecting the non-idealities in the map-maker affects the observed spectra in a way that is partially degenerate with the CB and instrumental miscalibration of the polarization angle. This effect is evident in the simulations and the

---

analytical formulae.

The rest of this chapter is organized as follows. In section 6.2 we present a simple data model for the signal measured by a single detector; generalize it to a larger focal plane and a longer observation time; and introduce the bin averaging map-making method we employ to convert the TOD to maps. In section 6.3 we discuss the instrument specifics we have implemented in the simulation and show the output angular power spectra. The interpretation of the result is the topic of section 6.4, where we derive some analytical formulae modeling it with good precision. In section 6.5 we show how the effect of the HWP non-idealities is partially degenerate with an instrumental miscalibration of the polarization angle, and can therefore be misinterpreted as CB. We quantify the HWP-induced miscalibration angle, which amounts to a few degrees for our choice of the HWP parameters. Conclusions and outlook are presented in section 6.6.

## 6.2 Data model and map-maker

**Data model for a single detector**    Polarized radiation can be described by the Stokes $I$, $Q$, $U$ and $V$ parameters or, more compactly, by a Stokes vector, $\mathbf{S} \equiv (I, Q, U, V)$. Here, we use the "CMB convention" for the sign of Stokes $U$ [125] and define the Stokes parameters in right-handed coordinates with the $z$ axis taken in the direction of the observer's line of sight (telescope boresight). The Stokes vector is transformed as $\mathbf{S} \to \mathbf{S}' = \mathcal{R}_\varphi \mathbf{S}$ by rotating the coordinates by an angle $\varphi$, where

$$\mathcal{R}_\varphi = \begin{pmatrix} 1 & 0 & 0 & 0 \\ 0 & \cos 2\varphi & \sin 2\varphi & 0 \\ 0 & -\sin 2\varphi & \cos 2\varphi & 0 \\ 0 & 0 & 0 & 1 \end{pmatrix}. \tag{6.1}$$

Defining the position angle of the plane of linear polarization, $\theta$, by $Q \pm iU = Pe^{\pm 2i\theta}$ with $P = \sqrt{Q^2 + U^2}$ and $2\theta = \arctan(U/Q)$, the rotation of coordinates shifts the position angle as $\theta \to \theta' = \theta - \varphi$.

The action of any polarization-altering device on $\mathbf{S}$ can be encoded in a Mueller matrix $\mathcal{M}$, so that the outgoing Stokes vector reads $\mathbf{S}' = \mathcal{M}\mathbf{S}$ [107]. In our case of interest, $\mathbf{S}$ represents the incoming CMB radiation and $\mathcal{M}$ the Mueller matrix of a telescope that employs a rotating HWP as a polarization modulator, i.e.

$$\mathbf{S}' = \mathcal{M}_{\mathrm{det}} \mathcal{R}_{\xi - \phi} \mathcal{M}_{\mathrm{HWP}} \mathcal{R}_{\phi + \psi} \mathbf{S}, \tag{6.2}$$

where $\mathcal{R}_\varphi$ is given in eq. (6.1). The meaning of each angle appearing in eq. (6.2) is clarified in Figure 6.1. For example, $\mathcal{R}_{\phi + \psi}$ rotates the sky coordinates by an angle $\psi$ to the telescope coordinates (the left panel) and further rotates by $\phi$ to the HWP coordinates (the middle panel). Here, $\mathcal{M}_{\mathrm{det}}$ and $\mathcal{M}_{\mathrm{HWP}}$ are the Mueller matrices of a detector along $x_{\mathrm{det}}$ and of a general HWP:

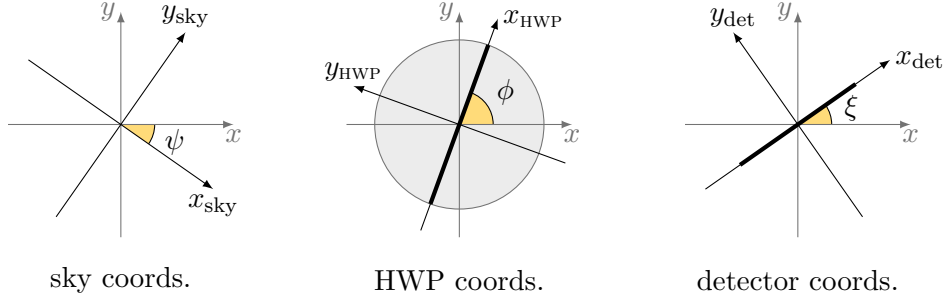

Figure 6.1: The **S** vector is defined in sky coordinates, forming an angle $\psi$ with the telescope ones (left panel). The HWP optical axis and the detector's sensitive direction are rotated with respect to the telescope coordinates by angles $\phi$ and $\xi$, respectively (center and right panels). The angles are defined in right-handed coordinates with the $z$ axis taken in the direction of the telescope boresight.

$$\mathcal{M}_{\mathrm{det}} = \frac{1}{2}\begin{pmatrix} 1 & 1 & 0 & 0 \\ 1 & 1 & 0 & 0 \\ 0 & 0 & 0 & 0 \\ 0 & 0 & 0 & 0 \end{pmatrix}, \quad \mathcal{M}_{\mathrm{HWP}} = \begin{pmatrix} m_{\mathrm{II}} & m_{\mathrm{IQ}} & m_{\mathrm{IU}} & m_{\mathrm{IV}} \\ m_{\mathrm{QI}} & m_{\mathrm{QQ}} & m_{\mathrm{QU}} & m_{\mathrm{QV}} \\ m_{\mathrm{UI}} & m_{\mathrm{UQ}} & m_{\mathrm{UU}} & m_{\mathrm{UV}} \\ m_{\mathrm{VI}} & m_{\mathrm{VQ}} & m_{\mathrm{VU}} & m_{\mathrm{VV}} \end{pmatrix}. \tag{6.3}$$

We can then model the signal $d$ measured by one detector as

$$d = \mathbf{a}^T \mathcal{M}_{\mathrm{det}} \mathcal{R}_{\xi-\phi} \mathcal{M}_{\mathrm{HWP}} \mathcal{R}_{\phi+\psi} \mathbf{S} + n, \quad \text{with} \quad \mathbf{a}^T = \begin{pmatrix} 1 & 0 & 0 & 0 \end{pmatrix}, \tag{6.4}$$

where $n$ represents an additional noise term.

**Modeling the TOD**   In a realistic CMB experiment, $n_{\mathrm{det}}$ detectors collect data by scanning the sky for an extended period of time, resulting in $n_{\mathrm{obs}}$ observations for each detector. All together, these $n_{\mathrm{det}} \times n_{\mathrm{obs}}$ measurements constitute the TOD. We represent the TOD as a vector $\mathbf{d}$ given by

$$\mathbf{d} = A\mathbf{m} + \mathbf{n}, \tag{6.5}$$

where $\mathbf{m}$ denotes the $\{I, Q, U, V\}$ pixelized sky maps, $A$ the response matrix, and $\mathbf{n}$ the noise component. Eq. (6.5) generalizes eq. (6.4) to larger $n_{\mathrm{obs}}$ and $n_{\mathrm{det}}$.

**Bin averaging map-maker**   To extract physical information from the TOD, we convert them to the map domain via some map-making procedure. A simple method often employed in the CMB analysis is the bin averaging [102], that estimates the sky map as

$$\widehat{\mathbf{m}} = \left(\widehat{A}^T \widehat{A}\right)^{-1} \widehat{A}^T \mathbf{d}, \tag{6.6}$$

where $\widehat{A}$ is the response matrix assumed by the map-maker. As long as the beam is axisymmetric and purely co-polarized, and the correlated component of the noise, such as $1/f$, is negligible, the bin averaging can, in principle, recover the input $\{I, Q, U, V\}$ maps. Whether the reconstructed maps actually reproduce the sky signal or not depends on how well the instrument specifics are encoded in the map-maker or, in other words, how close $\widehat{A}$ is to $A$. When $\widehat{A} = A$ and **n** is uncorrelated in time, $\widehat{\mathbf{m}}$ is the optimal (unbiased and minimum-variance) estimator of **m**.

| Scanning strategy parameters | | Instrument properties | |
| --- | --- | --- | --- |
| Precession angle | 45° | Number of detectors | 160 |
| Boresight angle | 50° | MFT frequency channel | 140 GHz |
| Precession period | 192.348 min | Sampling frequency | 19 Hz |
| Spin rate | 0.05 rpm | HWP rotation rate | 39 rpm |
| | | Beam FWHM | 30.8 arcmin |

Table 6.1: Simulation parameters used in this work. All values are taken from [29], except for the number of detectors and the central frequency, which we choose arbitrarily.

## 6.3   Simulation setup and output

We generate statistically isotropic random Gaussian $\{I, Q, U\}$ CMB maps with `HEALPix`[2] [126] resolution of $n_{\text{side}} = 512$ (high enough to avoid aliasing effects) by feeding the best-fit 2018 *Planck* power spectra [19] to the `synfast` function of `healpy`[3] (the Python implementation of `HEALPix`). We choose to neglect $V$ here[4].

The observation of the input maps is simulated by a modified version of the publicly available library `beamconv`. This choice is motivated by `beamconv`'s ability to simulate TOD with realistic HWPs, scanning strategies and beams, which makes it a promising framework to develop simulations for, among others, LiteBIRD-like experiments. The changes we have implemented to the library all aim to better tailor the simulations to LiteBIRD-like specifics. In particular:

**Scanning strategy** We implement a LiteBIRD-like scanning strategy by mimicking the relevant functionalities of `pyScan`[5]. The values of the telescope boresight and precession angles, together with their rotation parameters, are specified in Table 6.1. We simulate one year of observations to cover the full sky (see Figure 6.2).

**Instrument** We work with 160 detectors from the 140 GHz channel of LiteBIRD's Medium Frequency Telescope (MFT) and read the relevant parameters from [29]: the HWP rotation rate, the full-width-at-half-maximum (FWHM) of the (Gaussian and co-polarized) beam, the instrument sampling frequency and the detectors' pointing offsets. See Table 6.1 for their numerical values.

---

[2] http://healpix.sf.net

[3] https://github.com/healpy/healpy

[4] In the standard cosmological model, no circular polarization can be produced at last scattering. A number of models that could source $V$ have been proposed (see for instance [127–135]), but none of them predicts a strong signal, making $V \equiv 0$ a good first approximation.

[5] https://github.com/tmatsumu/LB_SYSPL_v4.2

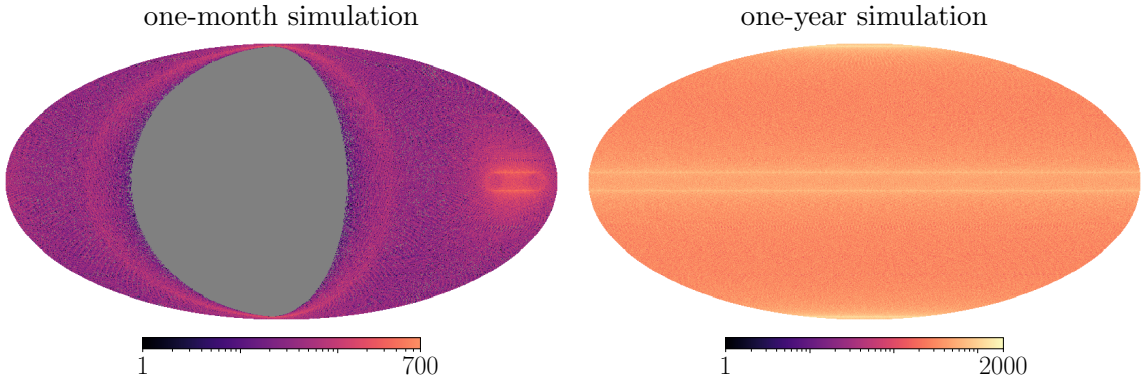

Figure 6.2: Simulated boresight hit maps for one-month (left) and one-year (right) observations. The two panels share the same logarithmic color map, although the range shown is truncated for the one-month case. Unobserved pixels appear gray.

**HWP Mueller matrix** The Mueller matrix elements for the MFT's HWP at 140 GHz are taken from [120], up to a coordinate change from International Astronomical Union (IAU) to CMB standards that flips the sign of the $m_{\text{IU}}$, $m_{\text{QU}}$, $m_{\text{UI}}$ and $m_{\text{UQ}}$ elements (see Section C.1):

$$\mathcal{M}_{\text{HWP}} = \begin{pmatrix} 9.80 \times 10^{-1} & 1.81 \times 10^{-2} & -9.81 \times 10^{-3} \\ 1.81 \times 10^{-2} & 9.71 \times 10^{-1} & -1.21 \times 10^{-1} \\ -9.81 \times 10^{-3} & -1.21 \times 10^{-1} & -8.40 \times 10^{-1} \end{pmatrix}. \tag{6.7}$$

This is the HWP Mueller matrix we assume when including non-idealities[6]. Since the elements of $\mathcal{M}_{\text{HWP}}$ are frequency-dependent, choosing a different frequency would result in slightly different output spectra.

We run two simulations for one-year observations. Noise is not included in either simulation to isolate the effect of HWP non-idealities in the signal; thus, using a different $n_{\text{det}}$ is almost free from consequences and our results do not change using fewer detectors. In the first simulation we assume the ideal HWP by setting $\mathcal{M}_{\text{HWP}} = \mathcal{M}_{\text{ideal}} \equiv \text{diag}(1, 1, -1)$, while we account for non-idealities in the second one. We convert both TODs to $\{I, Q, U\}$ maps by the bin averaging map-maker (see eq. (6.6)) whose response matrix $\widehat{A}$ assumes the ideal HWP described by $\mathcal{M}_{\text{ideal}}$. We then calculate two sets of full-sky angular power spectra using the `anafast` function of `healpy`. We denote the first (second) set of output spectra with $C_{\ell,\text{ideal}}^{XY}$ ($C_{\ell,\text{HWP}}^{XY}$), where $X, Y = \{T, E, B\}$. The rescaled $D_{\ell,\text{ideal}}^{XY} \equiv \ell(\ell+1) C_{\ell,\text{ideal}}^{XY}/2\pi$ and $D_{\ell,\text{HWP}}^{XY}$ spectra are plotted in Figure 6.3, together with the input spectra multiplied by the Gaussian beam transfer functions, $D_{\ell,\text{in}}^{XY}$. The simple map-maker recovers the input spectra with average deviations less than 0.1% in the plotted range when processing the TOD generated with $\mathcal{M}_{\text{ideal}}$, while important discrepancies arise for the non-ideal case.

---

[6]Doing so, we neglect the dependence of the HWP properties on the angle of incidence. The consequences of such approximation have not been tested yet.

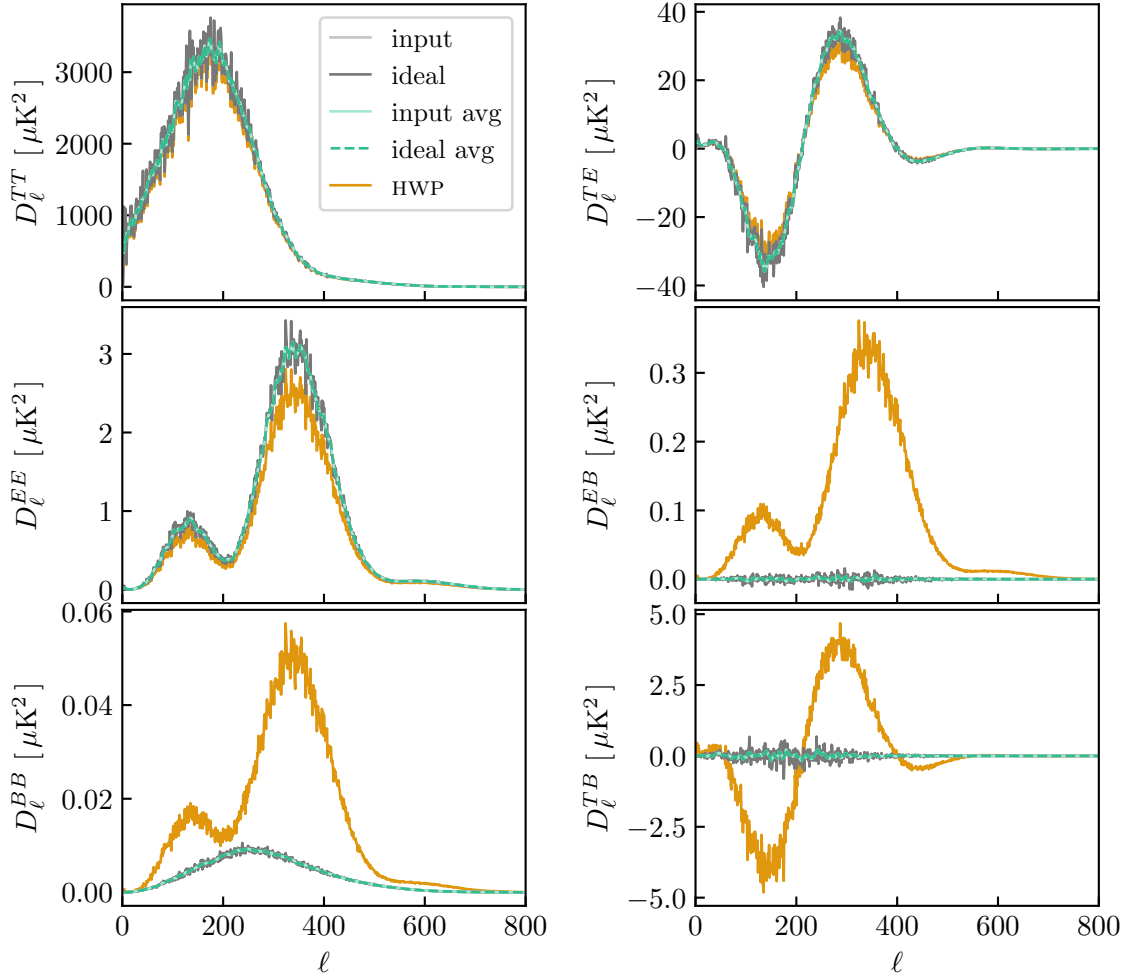

Figure 6.3: Comparison of the input angular power spectra $D_{\ell,\text{in}}^{XY}$ (light gray) with the ones computed from the outputs of the TOD simulations with ideal ($D_{\ell,\text{ideal}}^{XY}$, in dark gray) and non-ideal HWP ($D_{\ell,\text{HWP}}^{XY}$, in orange). The inputs are hard to see, since they almost perfectly overlap with the $D_{\ell,\text{ideal}}^{XY}$, while the $D_{\ell,\text{HWP}}^{XY}$ show clear deviations from the inputs. For clarity, we also show the simple moving average over 7 multipoles of $D_{\ell,\text{in}}^{XY}$ (lighter teal) and $D_{\ell,\text{ideal}}^{XY}$ (darker teal, dashed): $D_{\ell,\text{avg}}^{XY} \equiv \frac{1}{7} \sum_{\ell'=\ell-3}^{\ell+3} D_{\ell'}^{XY}$. The beam transfer function is not deconvolved.

We do not account for photometric calibration, although it represents a crucial step in any CMB analysis pipeline. Gain calibration, if perfect, would ensure intensity to be recovered exactly, hence compensating the lack of power in $D_{\ell}^{TT}$ visible in Figure 6.3. The discrepancies in $D_{\ell}^{TE}$ and $D_{\ell}^{TB}$ would also be reduced, although not removed. The discussion and results presented in the following would however not change, reason why we omit the step.

## 6.4   Analytical estimate of the output spectra

To understand the simulation results, we derive approximate analytical formulae for the angular power spectra affected by HWP non-idealities. Since we are neglecting any circularly polarized component, the Stokes vector is given by $\mathbf{S} = (I, Q, U)$. To obtain analytical formulae we apply the bin averaging map-maker of eq. (6.6) to a minimal TOD consisting of the signals measured by four detectors with different polarization sensitivity directions[7] (with 0°, 90°, 45° and 135° offsets) observing the same sky pixel. By expressing the signals observed by each of the four detectors as functions of the input Stokes parameters according to eq. (6.4), we obtain (see Section C.2 for the derivation)

$$\widehat{I} = m_{II}I_{in} + (m_{IQ}Q_{in} + m_{IU}U_{in})\cos(2\alpha) + (m_{IQ}U_{in} - m_{IU}Q_{in})\sin(2\alpha)\,, \tag{6.8a}$$

$$\begin{aligned}
\widehat{Q} = \frac{1}{2}\Big\{ &(m_{QQ} - m_{UU})Q_{in} + (m_{QU} + m_{UQ})U_{in} + 2m_{QI}I_{in}\cos(2\alpha) + 2m_{UI}I_{in}\sin(2\alpha) \\
&+ \big[(m_{QQ} + m_{UU})Q_{in} + (m_{QU} - m_{UQ})U_{in}\big]\cos(4\alpha) \\
&+ \big[-(m_{QU} - m_{UQ})Q_{in} + (m_{QQ} + m_{UU})U_{in}\big]\sin(4\alpha)\Big\}\,,
\end{aligned} \tag{6.8b}$$

$$\begin{aligned}
\widehat{U} = \frac{1}{2}\Big\{ &(m_{QQ} - m_{UU})U_{in} - (m_{QU} + m_{UQ})Q_{in} - 2m_{UI}I_{in}\cos(2\alpha) + 2m_{QI}I_{in}\sin(2\alpha) \\
&+ \big[-(m_{QQ} + m_{UU})U_{in} + (m_{QU} - m_{UQ})Q_{in}\big]\cos(4\alpha) \\
&+ \big[(m_{QU} - m_{UQ})U_{in} + (m_{QQ} + m_{UU})Q_{in}\big]\sin(4\alpha)\Big\}\,,
\end{aligned} \tag{6.8c}$$

where $m_{ss'}$ ($s, s' = I, Q, U$) are the elements of non-ideal $\mathcal{M}_{HWP}$ and $\alpha$ denotes the sum of the HWP's ($\phi$) and the telescope's ($\psi$) angles[8]: $\alpha \equiv \phi + \psi$. The quantities with the subscript "in" on the right hand side denote the sky signals, whereas $\widehat{\mathbf{S}} = (\widehat{I}, \widehat{Q}, \widehat{U})$ on the left hand side are maps recovered by the map-maker. These formulae are applicable to our case as long as eq. (6.4) accurately models the TOD simulated by `beamconv`, which is the case for an axisymmetric and purely co-polarized beam.

Eqs. (6.8) model $\widehat{\mathbf{S}}_p$ reconstructed from four observations of the pixel $p$, one for each detector. If each of those 4 detectors were to observe that same pixel $n_p$ times, the change in eqs. (6.8) would amount to substituting

$$\cos(n\alpha) \rightarrow \frac{1}{n_p}\sum_{t=t_1}^{t_{n_p}}\cos(n\alpha_t)\,, \quad \sin(n\alpha) \rightarrow \frac{1}{n_p}\sum_{t=t_1}^{t_{n_p}}\sin(n\alpha_t)\,, \tag{6.9}$$

for $n = \{2, 4\}$. If $p$ is observed with a uniform enough sample of $\alpha_t$ values and $n_p$ is large

---

[7]This is the minimal configuration that can reconstruct linearly polarized radiation.

[8]For the simple 4-detector configuration we are considering, the response matrix can be expressed as $A = B\mathcal{R}_{\xi-\phi}\mathcal{M}_{HWP}\mathcal{R}_{\phi+\psi}$, where $B$ accounts for the different $\xi$ angles of the four detectors and happens to satisfy $B^T B = \mathrm{diag}(1, 1/2, 1/2)$. As for the map-maker response matrix, $\widehat{A} = B\mathcal{R}_{\xi}\mathcal{M}_{ideal}\mathcal{R}_{\phi+\psi}$. All $B\mathcal{R}_{\xi-\phi}$ terms cancel out in eq. (6.6) and we are left with $\widehat{\mathbf{S}} = \mathcal{R}_{\phi+\psi}^T\mathcal{M}_{ideal}\mathcal{M}_{HWP}\mathcal{R}_{\phi+\psi}\mathbf{S}_{in}$. The discrepancies between $\widehat{\mathbf{S}}$ and $\mathbf{S}_{in}$ can therefore only depend on $\phi + \psi$.

enough, these terms can be neglected, resulting in

$$
\widehat{\mathbf{S}} \simeq \begin{pmatrix} m_{\mathrm{II}} I_{\mathrm{in}} \\ [(m_{\mathrm{QQ}} - m_{\mathrm{UU}})Q_{\mathrm{in}} + (m_{\mathrm{QU}} + m_{\mathrm{UQ}})U_{\mathrm{in}}]/2 \\ [(m_{\mathrm{QQ}} - m_{\mathrm{UU}})U_{\mathrm{in}} - (m_{\mathrm{QU}} + m_{\mathrm{UQ}})Q_{\mathrm{in}}]/2 \end{pmatrix}. \tag{6.10}
$$

We expect this to be a good approximation, given the presence of a rapidly spinning HWP and the good coverage of the simulation (see Figure 6.2). Note that eq. (6.10) can be derived even without assuming the four-detector configuration discussed here (see the alternative derivation in Section C.3).

By expanding eq. (6.10) in spherical harmonics, we write the corresponding angular power spectra as a mixing of the input ones weighted by combinations of the non-ideal HWP's Mueller matrix elements:

$$
\widehat{C}_\ell^{TT} \simeq m_{\mathrm{II}}^2 C_{\ell,\mathrm{in}}^{TT}, \tag{6.11a}
$$

$$
\widehat{C}_\ell^{EE} \simeq \frac{(m_{\mathrm{QQ}} - m_{\mathrm{UU}})^2}{4} C_{\ell,\mathrm{in}}^{EE} + \frac{(m_{\mathrm{QU}} + m_{\mathrm{UQ}})^2}{4} C_{\ell,\mathrm{in}}^{BB} + \frac{(m_{\mathrm{QQ}} - m_{\mathrm{UU}})(m_{\mathrm{QU}} + m_{\mathrm{UQ}})}{2} C_{\ell,\mathrm{in}}^{EB}, \tag{6.11b}
$$

$$
\widehat{C}_\ell^{BB} \simeq \frac{(m_{\mathrm{QQ}} - m_{\mathrm{UU}})^2}{4} C_{\ell,\mathrm{in}}^{BB} + \frac{(m_{\mathrm{QU}} + m_{\mathrm{UQ}})^2}{4} C_{\ell,\mathrm{in}}^{EE} - \frac{(m_{\mathrm{QQ}} - m_{\mathrm{UU}})(m_{\mathrm{QU}} + m_{\mathrm{UQ}})}{2} C_{\ell,\mathrm{in}}^{EB}, \tag{6.11c}
$$

$$
\widehat{C}_\ell^{TE} \simeq \frac{m_{\mathrm{II}}(m_{\mathrm{QQ}} - m_{\mathrm{UU}})}{2} C_{\ell,\mathrm{in}}^{TE} + \frac{m_{\mathrm{II}}(m_{\mathrm{QU}} + m_{\mathrm{UQ}})}{2} C_{\ell,\mathrm{in}}^{TB}, \tag{6.11d}
$$

$$
\widehat{C}_\ell^{EB} \simeq \frac{(m_{\mathrm{QQ}} - m_{\mathrm{UU}})^2 - (m_{\mathrm{QU}} + m_{\mathrm{UQ}})^2}{4} C_{\ell,\mathrm{in}}^{EB} - \frac{(m_{\mathrm{QQ}} - m_{\mathrm{UU}})(m_{\mathrm{QU}} + m_{\mathrm{UQ}})}{4} (C_{\ell,\mathrm{in}}^{EE} - C_{\ell,\mathrm{in}}^{BB}), \tag{6.11e}
$$

$$
\widehat{C}_\ell^{TB} \simeq \frac{m_{\mathrm{II}}(m_{\mathrm{QQ}} - m_{\mathrm{UU}})}{2} C_{\ell,\mathrm{in}}^{TB} - \frac{m_{\mathrm{II}}(m_{\mathrm{QU}} + m_{\mathrm{UQ}})}{2} C_{\ell,\mathrm{in}}^{TE}. \tag{6.11f}
$$

These analytical formulae explain quite well the non-ideal output spectra $C_{\ell,\mathrm{HWP}}^{XY}$ (see Figure 6.4). They are especially accurate on large scales, $\ell \lesssim 500$, where average deviations between $C_{\ell,\mathrm{HWP}}^{XY}$ and $\widehat{C}_\ell$ are less than 0.1%. Larger deviations on smaller scales are due to the approximate nature of eq. (6.10). Cosine and sine terms do not average out exactly, resulting in pixel-by-pixel fluctuations on smaller scales.

## 6.5 Impact on cosmic birefringence

Next generation CMB experiments are expected to measure the CMB polarization with unprecedented sensitivity and improve the constraints on the CB angle, $\beta$, recently obtained from the *Planck* data [62, 64–66]. Here we discuss how HWP non-idealities can impact such constraints in the particular case of a LiteBIRD-like mission discussed so far.

First, we recall that the sign of $\beta$ reported in the literature is also chosen to follow the CMB convention and a positive $\beta$ corresponds to a clockwise rotation on the sky [49]. The isotropic CB angle, $\beta$, and a miscalibration of the instrument polarization angle, $\Delta\alpha$, affect the observed spectra identically, since both rotate the observed Stokes parameters

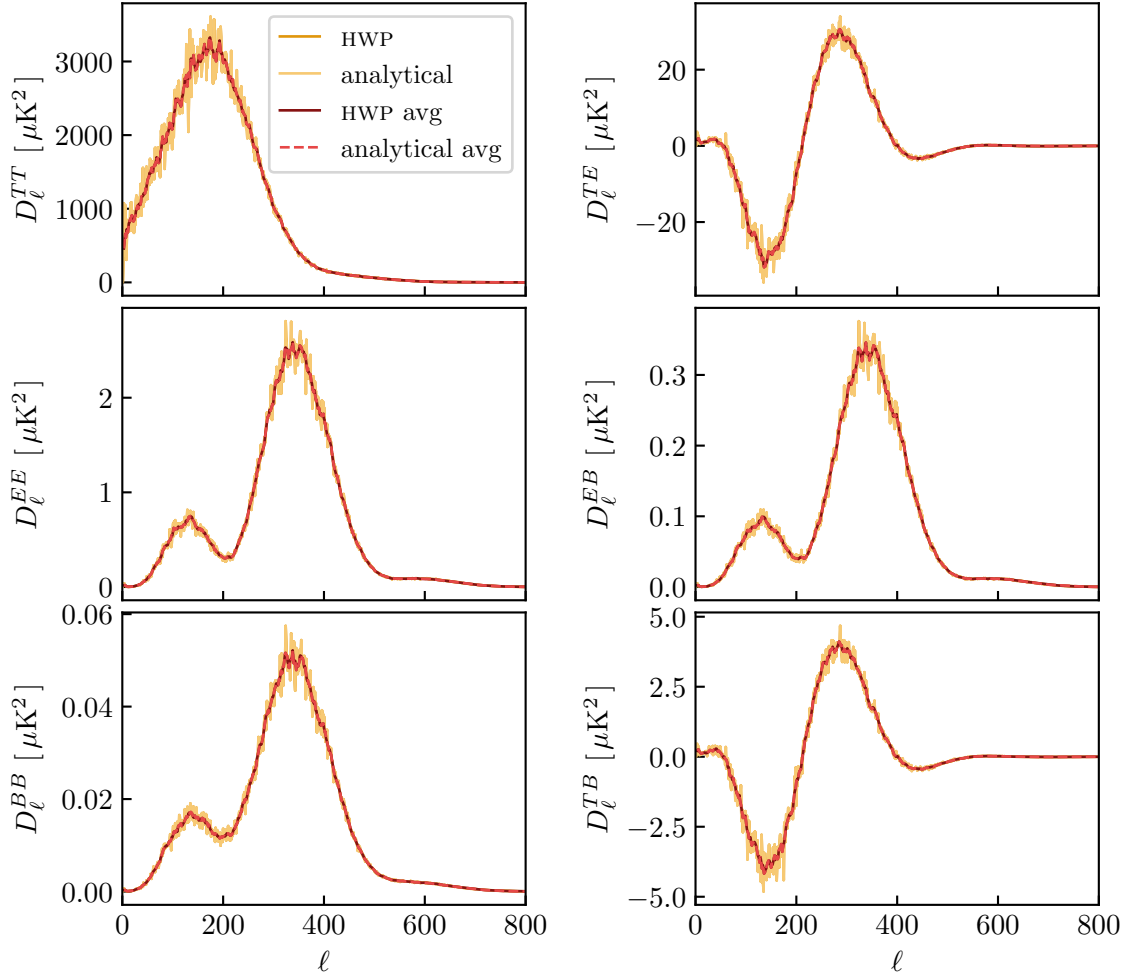

Figure 6.4: Comparison of the spectra computed from the output of the TOD simulation with non-ideal HWP, $D_{\ell,\text{HWP}}^{XY}$ (dark orange), with the $\widehat{D}_{\ell}^{XY}$ from the analytical formulae given in eqs. (6.11) (light orange). The non-ideal outputs are hard to see, since they almost perfectly overlap with the analytical curves. For clarity, we also show the simple moving average over 7 multipoles of $D_{\ell,\text{HWP}}^{XY}$ (dark red) and $\widehat{D}_{\ell}^{XY}$ (light red, dashed): $D_{\ell,\text{avg}}^{XY} \equiv \frac{1}{7} \sum_{\ell'=\ell-3}^{\ell+3} D_{\ell'}^{XY}$. The beam transfer function is not deconvolved.

in the same way. The observed spectra then satisfy the equations [81, 136]

$$C_{\ell,\text{obs}}^{EB} = \frac{\tan(4\theta)}{2}\left(C_{\ell,\text{obs}}^{EE} - C_{\ell,\text{obs}}^{BB}\right), \qquad C_{\ell,\text{obs}}^{TB} = \tan(2\theta)C_{\ell,\text{obs}}^{TE}, \qquad (6.12)$$

where $\theta$ represents rotation in the position angle of the plane of linear polarization including $\beta$, $\Delta\alpha$, or their sum. Not accounting for the HWP non-idealities in the map-maker step is degenerate with $\theta$, as it is evident from both our simulations and the analytical formulae given in eq. (6.11). We will refer to this additional rotation of the polarization plane as the HWP-induced miscalibration.

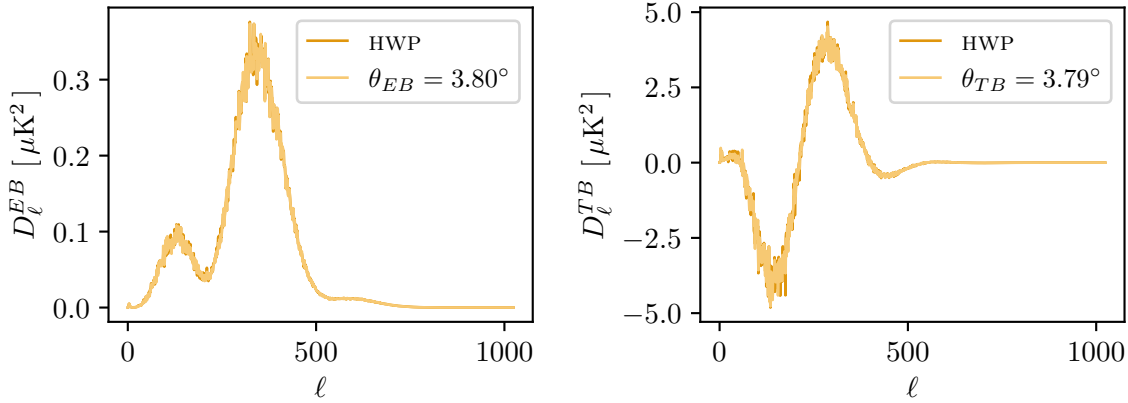

Figure 6.5: Comparison of $D_{\ell,\mathrm{HWP}}^{EB}$ and $D_{\ell,\mathrm{HWP}}^{TB}$ computed from the outputs of the TOD simulation with non-ideal HWP (dark orange) with the best-fit estimates of $\tan(4\theta_{EB})(D_{\ell,\mathrm{HWP}}^{EE} - D_{\ell,\mathrm{HWP}}^{BB})/2$ and $\tan(2\theta_{TB})D_{\ell,\mathrm{HWP}}^{TE}$, respectively.

**HWP-induced miscalibration from the simulated output spectra** We separately fit the simulated $C_{\ell,\mathrm{HWP}}^{EB}$ and $C_{\ell,\mathrm{HWP}}^{TB}$ for the angles $\theta_{EB}$ and $\theta_{TB}$, respectively, using the least-squares method with variance given by (see Section C.4 for the derivation)

$$\mathrm{Var}\left(C_{\ell,\mathrm{HWP}}^{XY}\right) = \frac{1}{2\ell+1}\left[C_{\ell,\mathrm{HWP}}^{XX}C_{\ell,\mathrm{HWP}}^{YY} + \left(C_{\ell,\mathrm{HWP}}^{XY}\right)^2\right], \tag{6.13}$$

for $XY = \{EB, TB\}$, respectively. The best-fit values, $\theta_{EB} = 3.800° \pm 0.007°$ and $\theta_{TB} = 3.79° \pm 0.02°$, are compatible with each other in agreement with eqs. (6.12). The observed and best-fit spectra are plotted in Figure 6.5 and are in good agreement.

**HWP-induced miscalibration from the analytical formulae** Using the fact that both $C_{\ell_{\mathrm{in}}}^{EB}$ and $C_{\ell_{\mathrm{in}}}^{TB}$ simply fluctuate around zero, eqs. (6.11) can be rearranged to express $\widehat{C}_\ell^{EB}$ and $\widehat{C}_\ell^{TB}$ similarly to the $C_{\ell,\mathrm{obs}}^{XY}$ of eqs. (6.12):

$$\widehat{C}_\ell^{EB} \simeq \frac{\tan(4\widehat{\theta})}{2}\left(\widehat{C}_\ell^{EE} - \widehat{C}_\ell^{BB}\right), \qquad \widehat{C}_\ell^{TB} \simeq \tan(2\widehat{\theta})\widehat{C}_\ell^{TE}, \tag{6.14}$$

where

$$\widehat{\theta} = -\frac{1}{2}\arctan\left(\frac{m_{\mathrm{QU}} + m_{\mathrm{UQ}}}{m_{\mathrm{QQ}} - m_{\mathrm{UU}}}\right) \simeq 3.8°, \tag{6.15}$$

in agreement with the best-fit values reported above.

If we were to relax all the underlying assumptions at once, we could not write $\widehat{\theta}$ this compactly. However, controlled generalizations do not necessarily spoil the simplicity of the analytical formulae. For instance, accounting for the frequency dependence of both the HWP Mueller matrix elements and the CMB signal, $\widehat{\theta}$ can be expressed as (see appendix

C.5 for the derivation):

$$\widehat{\theta} = -\frac{1}{2} \arctan\left(\frac{\int d\nu\, S_{\mathrm{CMB}}(\nu)\left[m_{\mathrm{QU}} + m_{\mathrm{UQ}}\right](\nu)}{\int d\nu\, S_{\mathrm{CMB}}(\nu)\left[m_{\mathrm{QQ}} - m_{\mathrm{UU}}\right](\nu)}\right), \tag{6.16}$$

where $S_{\mathrm{CMB}}(\nu)$ denotes the CMB spectral energy distribution (SED).

Another assumption that can be relaxed without spoiling the simplicity of the analytical formulae is the absence of miscalibration angles in the map-maker. When the telescope, HWP, and detector angles are not exactly known, $\psi = \widehat{\psi} + \delta\psi$, $\phi = \widehat{\phi} + \delta\phi$, and $\xi = \widehat{\xi} + \delta\xi$, where the hat denotes the values assumed by the map-maker. As long as we neglect the frequency dependence of $\delta\psi$, $\delta\phi$ and $\delta\xi$, we find (see appendix C.6 for the derivation)

$$\widehat{\theta} = -\frac{1}{2} \arctan\left(\frac{\int d\nu\, S_{\mathrm{CMB}}(\nu)\left[m_{\mathrm{QU}} + m_{\mathrm{UQ}}\right](\nu)}{\int d\nu\, S_{\mathrm{CMB}}(\nu)\left[m_{\mathrm{QQ}} - m_{\mathrm{UU}}\right](\nu)}\right) + \delta\xi - \delta\psi - 2\delta\phi\,. \tag{6.17}$$

The sign difference between the contributions from $\delta\xi$ and $\delta\psi + 2\delta\phi$ is due to the presence of the HWP. Ideally, the HWP acts on a polarization vector by reflecting it over its fast axis. This causes counterclockwise rotations applied before the HWP to look clockwise after, meaning that $\delta\phi + \delta\psi$ should be subtracted from $\delta\xi - \delta\phi$ (see eq. (6.2)).

## 6.6   Conclusions and outlook

In this work, we studied how overlooking HWP non-idealities during map-making can affect the reconstructed angular power spectra of CMB temperature and polarization fields. We focused on the impact of non-idealities on the measurement of the CB angle, $\beta$.

As a concrete working case, we considered a single frequency channel (140 GHz) of a space CMB mission with LiteBIRD-like specifics: scanning strategy, sampling frequency, detectors' pointing offsets and their polarization sensitivity directions, FWHM of the Gaussian beam and HWP specifics (rotation frequency and Mueller matrix elements). We employed the publicly available beam-convolution code `beamconv` to simulate the noiseless TOD for the above instrument and scanning specifications. We ran two different simulations: the HWP has been assumed to be ideal in the first simulation, while a realistic Mueller matrix has been employed in the second. We then converted both TODs to maps by a bin averaging map-maker that neglects the HWP non-idealities. As expected, the output spectra computed from the ideal simulation faithfully recovered the input spectra, while the spectra of the non-ideal maps showed a very different behavior (Figure 6.3). We also derived a set of analytical formulae (see eq. (6.11)) that accurately model the reconstructed angular power spectra as functions of the input spectra and the HWP Mueller matrix elements.

We studied the impact of the HWP non-idealities on $\beta$. We found that neglecting them in the map-making step induces an additional miscalibration of the polarization angle which might be erroneously interpreted as CB. For the concrete case we studied, the miscalibration angle induced by the HWP non-idealities amounts to $\theta \simeq 3.8°$. This value, obtained by fitting the output angular power spectra from the simulation, is compatible with the prediction from the analytical formulae (see eq. (6.15)).

Definitive confirmation of the current hint of CB [62, 64–66] requires the systematic uncertainty in the absolute position angle of linear polarization to be well below $0.1°$ [49]. We must therefore acquire accurate knowledge of the Mueller matrix elements via calibration, so that the systematic uncertainty in $\theta$ due to HWP non-idealities is well below $0.1°$. With such knowledge, one can take into account HWP non-idealities either during the map-making step or when interpreting the angular power spectra. As one cannot know the Mueller matrix elements perfectly, any remaining mismatch between the true Mueller matrix and the matrix assumed by the map-maker still affects the power spectra. Our simulation and analytical formulae will be useful for deriving the required accuracy of HWP calibration to meet specific science goals.

The situation we considered in this analysis is still simplistic: we simulated a single frequency channel in the absence of noise, and we used a Gaussian beam and a simple bin averaging map-maker. However, a similar analysis can be carried out for more complex cases. It is of utmost importance to make better predictions about how HWP non-idealities realistically affect the data collected by CMB experiments and, therefore, the cosmological information extracted from them. In this direction, we plan to carry on the following steps: i) drop the single frequency approximation, generalizing the results discussed here to a finite frequency bandwidth; ii) add a noise component to the TOD; iii) study the combined effect of beam asymmetries and HWP non-idealities; iv) include non-idealities in the map-maker and study how the uncertainties in our knowledge of non-idealities might propagate to the observed angular power spectra; and v) derive requirements for the accuracy of HWP calibration. We leave these topics for future work.

# Chapter 7

# HWP impact on $B$-mode polarization

**Abstract:** Polarization of the cosmic microwave background (CMB) can help probe the fundamental physics behind cosmic inflation via the measurement of primordial $B$ modes. As this requires exquisite control over instrumental systematics, some next-generation CMB experiments plan to use a rotating half-wave plate (HWP) as polarization modulator. However, the HWP non-idealities, if not properly treated in the analysis, can result in additional systematics. In this chapter, we present a simple, semi-analytical end-to-end model to propagate the HWP non-idealities through the macro-steps that make up any CMB experiment (observation of multi-frequency maps, foreground cleaning, and power spectra estimation) and compute the HWP-induced bias on the estimated tensor-to-scalar ratio, $r$. We find that the effective polarization efficiency of the HWP suppresses the polarization signal, leading to an underestimation of $r$. Laboratory measurements of the properties of the HWP can be used to calibrate this effect, but we show how gain calibration of the CMB temperature can also be used to partially mitigate it. On the basis of our findings, we present a set of recommendations for the HWP design that can help maximize the benefits of gain calibration.

*This chapter is an adaptation of [79].*

## 7.1   Introduction

Observations of temperature anisotropies in the cosmic microwave background (CMB) have been crucial in shaping our current understanding of cosmology [16, 19, 20]. Valuable complementary information is encoded in polarization anisotropies, which have only been partially explored [15, 17, 21, 22, 25, 27, 28, 116, 117]. The main goal of the next generation of CMB experiments, involving both ground-based (Simons Observatory [44], South Pole Observatory [45] and CMB Stage-4 [46]) and spaceborne (LiteBIRD [29] and PICO [47]) missions, is to probe the fundamental physics behind cosmic inflation [11–13] by measuring primordial $B$-mode polarization [48, 49].

Inflation sources initial conditions for cosmological perturbations via primordial vacuum quantum fluctuations [30–33]. The relative amplitude of the resulting scalar and tensor perturbations is quantified in terms of the tensor-to-scalar ratio, $r$. Since tensor perturbations [35, 36] would leave a distinct $B$-mode signature on the CMB polarization [37–40], $r$ can be inferred from the angular power spectrum of the primordial $B$ modes. To date, CMB observations have only placed upper bounds on $r$, the tightest being $r < 0.032$ (95% CL) [41] (see also [28, 42, 43]). Future surveys aim for unprecedentedly low overall uncertainties, which, depending on the true value of $r$, would lead to a detection or a tightening of the upper bounds, both of which would allow us to place strong constraints on inflationary models.

Such an ambitious goal can only be achieved through an exquisite control over systematics. To this end, some next-generation CMB experiments, including LiteBIRD, are planning to employ a rapidly spinning half-wave plate (HWP) as a polarization modulator, which can mitigate $1/f$ noise and reduce temperature-to-polarization leakage [68–77]. However, any realistic HWP is characterized by non-idealities [113, 118, 119] that can induce additional systematics if not properly accounted for in the analysis [78, 120–124, 137].

In this chapter, we present a simple framework to propagate the HWP non-idealities through the three macro-steps that characterize any CMB experiment: observation of multi-frequency maps, foreground cleaning, and power spectra estimation. We exploit the simplicity of the harmonic internal linear combination (HILC) foreground cleaning method [138] to keep the treatment semi-analytical. This choice, along with our working assumptions, makes the analysis computationally inexpensive[1] and reflects our intention to develop an intuitive understanding of how the HWP affects the observed CMB.

The remainder of this chapter is organized as follows. In section 7.2 we generalize the arguments presented in [78] and provide a simple model for multi-frequency maps observed through a rapidly spinning HWP. We then introduce the HILC foreground cleaning method and present the procedure we will use to infer $r$. In section 7.3, we discuss the specific choices we make to model sky, noise, and beams, and present the results of the analysis in two cases. First, we assume that the HWP is ideal and verify that the pipeline recovers the input CMB signal. Second, we consider LiteBIRD-like instrument specifics and assume realistic HWPs. We find that, for our choice of HWPs and $r_{\text{true}} = 0.00461$ in input, the HWP non-idealities introduce an effective polarization efficiency that suppresses the polarization signal, resulting in $\hat{r} = 0.0043 \pm 0.0005$. We also show how including gain calibration of the CMB temperature in the map model can partially mitigate this effect. In section 7.4, we derive a set of design recommendations that can help maximize the benefits of the gain calibration step. We also review the simplifying assumptions underlying the model and briefly discuss how they might be relaxed. Conclusions and perspectives are presented in section 7.5.

---

[1] The analysis presented here takes less than three minutes to run on a 32 GB RAM laptop computer.

## 7.2 Mathematical framework

In this section we present a simple model for multi-frequency maps observed through a rapidly spinning HWP. We also introduce the HILC foreground cleaning method and derive an explicit expression for the $B$-mode angular power spectrum of its solution, $C_{\ell,\mathrm{HILC}}^{BB}$, given the modeled multi-frequency maps. Finally, we present the methodology we use to estimate the tensor-to-scalar ratio parameter, $r$, from $C_{\ell,\mathrm{HILC}}^{BB}$.

### 7.2.1 Modeling the observed maps

We describe linearly polarized radiation[2] by the Stokes $I$, $Q$ and $U$ parameters defined in right-handed coordinates with the $z$ axis taken in the direction of the observer's line of sight (telescope boresight), according to the "CMB convention" [125]. Given an incoming Stokes vector $\mathbf{S} \equiv (I, Q, U)$, the effect of a polarization-altering device on $\mathbf{S}$ can be described by a Mueller matrix $\mathcal{M}$, so that $\mathbf{S}' = \mathcal{M}\mathbf{S}$ [107]. Assuming azimuthally symmetric and purely co-polarized beams, we can approximate the entire telescope's optical chain by means of a Mueller matrix acting on appropriately smoothed input Stokes parameters.

This setup allows us to write the telescope response matrix[3], $A$, analytically, and to obtain simple expressions for both time-ordered data (TOD), $\mathbf{d}$, and binned maps, $\widehat{\mathbf{m}}$ [102]:

$$\mathbf{d} = A\overline{\mathbf{m}} + \mathbf{n}, \qquad \widehat{\mathbf{m}} = \left(\widehat{A}^T \widehat{A}\right)^{-1} \widehat{A}^T \mathbf{d}, \qquad (7.1)$$

where $\overline{\mathbf{m}}$ denotes the pixelized $\{I, Q, U\}$ sky maps smoothed to the resolution of the instrument, $\mathbf{n}$ the noise contribution to the TOD, and $\widehat{A}$ the response matrix assumed by the map-maker.

If the telescope's first optical element is a rapidly rotating HWP with Mueller matrix

$$\mathcal{M}_{\mathrm{HWP}} = \begin{pmatrix} m_{\mathrm{II}} & m_{\mathrm{IQ}} & m_{\mathrm{IU}} \\ m_{\mathrm{QI}} & m_{\mathrm{QQ}} & m_{\mathrm{QU}} \\ m_{\mathrm{UI}} & m_{\mathrm{UQ}} & m_{\mathrm{UU}} \end{pmatrix}, \qquad (7.2)$$

---

[2]The standard cosmological model predicts that no circular polarization is produced at the surface of last scattering. Even beyond standard cosmology, none of the models that have been proposed to source circular polarization (see, for instance, [127–135]) allows for a significant signal. We therefore consider only linear polarization.

[3]The response matrix, $A$, relates the sky maps to the time-ordered data, i.e. the collection of signals observed by all the instrument's detectors. $A$ encodes information about the telescope's pointings and the instrument specifics, such as the HWP Mueller matrix and the detectors' orientations.

the maps reconstructed from the TOD of the $i$ channel's detectors by an ideal binning map-maker that assumes $\widehat{\mathcal{M}}_{\text{HWP}} = \text{diag}(1, 1, -1)$ read[4]

$$
\widehat{\mathbf{m}}^i \simeq \sum_\lambda \int_{\nu^i_{\min}}^{\nu^i_{\max}} \frac{\mathrm{d}\nu}{\Delta\nu^i} \begin{pmatrix} m_{\text{II}}(\nu) & 0 & 0 \\ 0 & [m_{\text{QQ}}(\nu) - m_{\text{UU}}(\nu)]/2 & [m_{\text{QU}}(\nu) + m_{\text{UQ}}(\nu)]/2 \\ 0 & -[m_{\text{QU}}(\nu) + m_{\text{UQ}}(\nu)]/2 & [m_{\text{QQ}}(\nu) - m_{\text{UU}}(\nu)]/2 \end{pmatrix} \overline{\mathbf{m}}^i_\lambda(\nu) + \mathbf{n}^i,
\tag{7.3}
$$

where the sum over $\lambda$ spans different sky components (CMB, dust, and synchrotron emission), the integral represents a top-hat bandpass with a bandwidth of $\Delta\nu^i \equiv \nu^i_{\min} - \nu^i_{\max}$, the superscript $i$ in $\overline{\mathbf{m}}^i_\lambda$ stresses that the input map is smoothed with the beam of the frequency channel $i$, and $\mathbf{n}^i$ denotes the noise maps.

Eq. (7.3) approximates the observed maps well when the cross-linking is good, that is, when each sky pixel is observed with a variety of scan angles. This condition is ensured by the rapid HWP rotation and the good LiteBIRD sky coverage, which guarantee that the scan angles are sampled uniformly enough for each pixel [78]. As a consequence, our model neglects intensity-to-polarization leakage, the effects of which have been shown to be correctable [137].

If we also make the simplifying assumption that the spectral energy distribution (SED) of each component is uniform throughout the sky, we can rewrite each sky map as $\overline{\mathbf{m}}_\lambda(\nu) \equiv a_\lambda(\nu)\overline{\mathbf{m}}_\lambda(\nu_\star)$, where $\nu_\star$ is some reference frequency. This is equivalent to using the *s0d0* option in the Python Sky Model (`PySM`) package [139], which has often been used in the literature for the study of systematics (e.g., [58, 140]). The reason for this assumption is twofold. First, it is often useful to separate the effects of systematics from the complexity of the foreground emission. Second, as shown in [58], the study of systematics is strongly influenced by the specific *class* of component separation methods, that is, whether it is a blind method, such as HILC [138], or a parametric method, such as `FGbuster` [141]. Here, we use HILC and leave the study based on a parametric method for future work.

The factorization, $\overline{\mathbf{m}}_\lambda(\nu) = a_\lambda(\nu)\overline{\mathbf{m}}_\lambda(\nu_\star)$, allows us to rewrite eq. (7.3) as

$$
\widehat{\mathbf{m}}^i \simeq \sum_\lambda \begin{pmatrix} g^i_\lambda & 0 & 0 \\ 0 & \rho^i_\lambda & \eta^i_\lambda \\ 0 & -\eta^i_\lambda & \rho^i_\lambda \end{pmatrix} \overline{\mathbf{m}}^i_\lambda + \mathbf{n}^i,
\tag{7.4}
$$

where we have dropped the $\nu_\star$ dependence for the sake of simplicity and defined

$$
g^i_\lambda \equiv \int_{\nu^i_{\min}}^{\nu^i_{\max}} \frac{\mathrm{d}\nu}{\Delta\nu^i} a_\lambda(\nu) m_{\text{II}}(\nu),
\tag{7.5a}
$$

$$
\rho^i_\lambda \equiv \frac{1}{2} \int_{\nu^i_{\min}}^{\nu^i_{\max}} \frac{\mathrm{d}\nu}{\Delta\nu^i} a_\lambda(\nu) \left[ m_{\text{QQ}}(\nu) - m_{\text{UU}}(\nu) \right],
\tag{7.5b}
$$

$$
\eta^i_\lambda \equiv \frac{1}{2} \int_{\nu^i_{\min}}^{\nu^i_{\max}} \frac{\mathrm{d}\nu}{\Delta\nu^i} a_\lambda(\nu) \left[ m_{\text{QU}}(\nu) + m_{\text{UQ}}(\nu) \right].
\tag{7.5c}
$$

---

[4]Eq. (7.3) follows from eq. (4.3) of [78] [i.e. eq. (6.10)] by relaxing the single frequency, CMB only, and no-noise assumptions.

The coefficients in these equations have a clear physical interpretation: $g_\lambda^i$ is an effective gain for the temperature data, $\rho_\lambda^i$ and $\eta_\lambda^i$ are effective polarization gain (or polarization efficiency) and cross-polarization coupling, respectively, caused by the non-idealities of the HWP.

**Including photometric calibration** Photometric calibration is a crucial step in any CMB analysis pipeline that allows us to map the instrumental output to the incoming physical signal [142]. Here, we assume that the CMB temperature dipole [143, 144] is used as a calibrator, as is commonly done in CMB experiments, and we neglect any imperfections in calibration. In other words, we assume to know $\tilde{g}^i = g_{\mathrm{CMB}}^i$ exactly after calibration. The photometrically calibrated counterpart of eq. (7.4) reads

$$\widehat{\mathbf{m}}^i \simeq \frac{1}{g_{\mathrm{CMB}}^i} \left[ \sum_\lambda \begin{pmatrix} g_\lambda^i & 0 & 0 \\ 0 & \rho_\lambda^i & \eta_\lambda^i \\ 0 & -\eta_\lambda^i & \rho_\lambda^i \end{pmatrix} \overline{\mathbf{m}}_\lambda^i + \mathbf{n}^i \right]. \tag{7.6}$$

**Spherical harmonics coefficients** To apply the HILC method to the modeled maps, we expand eq. (7.6) in spin-0 and spin-2 spherical harmonics and write the corresponding $B$-mode spherical harmonics coefficients as

$$\widehat{a}_{\ell m}^{B,i} = \frac{1}{g_{\mathrm{CMB}}^i} \left[ \sum_\lambda B_\ell^i \left( \rho_\lambda^i a_{\ell m}^{B,i} - \eta_\lambda^i a_{\ell m}^{E,i} \right) + n_{\ell m}^{B,i} \right], \tag{7.7}$$

where $a_{\ell m,\lambda}^E$ and $a_{\ell m,\lambda}^B$ are the $E$- and $B$-mode coefficients of the unsmoothed maps at some reference frequency $\nu_*$ (implicit here), and $B_\ell^i$ is the beam transfer function of the channel $i$.

## 7.2.2 Harmonic internal linear combination

The internal linear combination (ILC) [145] is a blind foreground cleaning method. It can be implemented in both map and multipole space, the latter case being referred to as HILC [138]. Given the spherical harmonics coefficients, $a_{\ell m}^{X,i}$ with $X = (T, E, B)$ and $i \in \{1, \ldots, n_{\mathrm{chan}}\}$, of the maps observed by each of the $n_{\mathrm{chan}}$ frequency channels, the HILC solution is given by [138]

$$a_{\ell m,\mathrm{HILC}}^X = \sum_{i=1}^{n_{\mathrm{chan}}} w_\ell^i a_{\ell m}^{X,i}, \quad \text{with weights} \quad \mathbf{w}_\ell = \frac{\mathbb{C}_\ell^{-1} \mathbf{e}}{\mathbf{e}^T \mathbb{C}_\ell^{-1} \mathbf{e}}, \tag{7.8}$$

where $\mathbb{C}_\ell$ is the $n_{\mathrm{chan}} \times n_{\mathrm{chan}}$ covariance matrix of the observed maps: $\mathbb{C}_\ell^{ij} = \langle a_{\ell m}^{i*} a_{\ell m}^j \rangle$.

By construction, the weights minimize the variance of the final map and add to unity, $\sum_i w_\ell^i = 1$, preserving the frequency independence of the CMB black-body spectrum. However, the frequency dependence of $g_{\mathrm{CMB}}^i$, $\rho_{\mathrm{CMB}}^i$, and $\eta_{\mathrm{CMB}}^i$ can violate this sum rule. This is the main point we study in this analysis.

**Modeling the HILC solution**   To apply the HILC to the analytical predictions discussed in section 7.2.1, we could simply use eq. (7.7); however, since different channels are characterized by different beams, it is preferable to perform the HILC on unsmoothed spherical harmonic coefficients, $a_{\ell m}^i \equiv \widehat{a}_{\ell m}^{B,i}/B_\ell^i$ and write the covariance matrix as

$$\mathbb{C}_\ell^{B,ij} = \frac{1}{g_{\mathrm{CMB}}^i g_{\mathrm{CMB}}^j} \left\{ \sum_\lambda \left[ \rho_\lambda^i \rho_\lambda^j C_{\ell,\lambda}^{BB} + \eta_\lambda^i \eta_\lambda^j C_{\ell,\lambda}^{EE} - \left( \rho_\lambda^i \eta_\lambda^j + \eta_\lambda^i \rho_\lambda^j \right) C_{\ell,\lambda}^{EB} \right] + \frac{\mathbb{N}_\ell^{BB,ij}}{B_\ell^i B_\ell^j} \right\}. \quad (7.9)$$

We use eq. (7.9) to compute the HILC weights, $\mathbf{w}_\ell$, and the spherical harmonics coefficients of the HILC solution according to eq. (7.8). The corresponding angular power spectrum reads

$$C_{\ell,\mathrm{HILC}}^{BB} = \sum_{i,j=1}^{n_{\mathrm{chan}}} \frac{w_\ell^i w_\ell^j}{g_{\mathrm{CMB}}^i g_{\mathrm{CMB}}^j} \left\{ \sum_\lambda \left[ \rho_\lambda^i \rho_\lambda^j C_{\ell,\lambda}^{BB} + \eta_\lambda^i \eta_\lambda^j C_{\ell,\lambda}^{EE} - \left( \rho_\lambda^i \eta_\lambda^j + \eta_\lambda^i \rho_\lambda^j \right) C_{\ell,\lambda}^{EB} \right] + \frac{\mathbb{N}_\ell^{BB,ij}}{B_\ell^i B_\ell^j} \right\}. \quad (7.10)$$

This is the main equation from which we derive all of our results.

Even at this early stage, we can make some educated guesses about which terms will contribute the most to the final angular power spectrum. By construction, the HILC tries to select the component $\lambda$ whose $\rho_\lambda^i$ and/or $\eta_\lambda^i$ are nearly constant across all frequency channels, i.e., a black-body spectrum. For example, if $m_{\mathrm{QQ}}(\nu) - m_{\mathrm{UU}}(\nu)$ or $m_{\mathrm{QU}}(\nu) + m_{\mathrm{UQ}}(\nu)$ depended on frequency as the inverse of the SED of the foreground emission, the foreground would leak into the HILC solution. However, the Mueller matrix elements of realistic HWPs do not exhibit such behavior. We therefore expect foreground-to-CMB leakage to be small in the final angular power spectrum.

Focusing on the CMB, eq. (7.10) tells us that there are two potential contaminations: $E$-to-$B$ leakage, which can occur if the effective cross-polarization coupling, $\eta_{\mathrm{CMB}}^i$, is nearly constant across the frequency channels, and suppression of the $B$ modes, which is instead driven by the effective polarization efficiency, $\rho_{\mathrm{CMB}}^i$. The relative importance of these effects depends on the specific design choice of the HWP.

### 7.2.3   Maximum likelihood estimate of the tensor-to-scalar ratio

The modeled angular power spectrum is

$$C_\ell^{BB}(r, A_{\mathrm{lens}}) = r C_\ell^{\mathrm{GW}} + A_{\mathrm{lens}} C_\ell^{\mathrm{lens}} + N_\ell^{BB}, \quad (7.11)$$

where $C_\ell^{\mathrm{GW}}$ is the primordial $B$-mode power spectrum with $r = 1$ [39, 40], $C_\ell^{\mathrm{lens}}$ is the lensed $B$-mode power spectrum [82], $A_{\mathrm{lens}}$ is its amplitude with $A_{\mathrm{lens}} = 1$ being the fiducial value, and $N_\ell^{BB}$ is the HILC solution for the total noise power spectrum [the last term in eq. (7.10)].

The probability density function (p.d.f.) of the observed $B$-mode power spectrum for a given value of $r$ and $A_{\mathrm{lens}}$, $P(C_{l,\mathrm{obs}}^{BB} \,|\, r, A_{\mathrm{lens}})$, is given by [see eq. (3.28) and, e.g., [146]]

$$\log P(C_{\ell,\mathrm{obs}}^{BB} \,|\, r, A_{\mathrm{lens}}) = \ - \ f_{\mathrm{sky}} \frac{2\ell+1}{2} \left[ \frac{C_{\ell,\mathrm{obs}}^{BB}}{C_\ell^{BB}(r, A_{\mathrm{lens}})} + \log C_\ell^{BB}(r, A_{\mathrm{lens}}) - \frac{2\ell-1}{2\ell+1} \log C_{\ell,\mathrm{obs}}^{BB} \right]$$
$$+ \ \mathrm{const.}, \quad (7.12)$$

where $f_{\text{sky}}$ is the sky fraction used to evaluate $C_{\ell,\text{obs}}^{BB}$. We use $f_{\text{sky}} = 0.78$, for which our sky model is defined (see Table 7.1 for details). Given the p.d.f., the likelihood function is

$$L(r, A_{\text{lens}}) \propto \prod_{\ell=\ell_{\text{min}}}^{\ell_{\text{max}}} P(C_{\ell,\text{obs}}^{BB} \,|\, r, A_{\text{lens}}) \,. \tag{7.13}$$

We use $\ell_{\text{max}} = 200$, which is the fiducial value for LiteBIRD [29]. Using Bayes' theorem, the posterior p.d.f. of $r$ with $A_{\text{lens}}$ marginalized over a flat prior is

$$L_{\text{m}}(r) \propto \int \mathrm{d}A_{\text{lens}} \, L(r, A_{\text{lens}}) \,. \tag{7.14a}$$

The frequentist profile likelihood is given instead by maximizing the bidimensional likelihood with respect to $A_{\text{lens}}$ for a set of values $\{r_0, \ldots, r_n\}$

$$L_{\text{p}}(r_i) \propto \max[L(r_i, A_{\text{lens}})] \,. \tag{7.14b}$$

Regardless of whether $L(r) \equiv L_{\text{m}}(r)$ or $L(r) \equiv L_{\text{p}}(r)$ is chosen, we define $\hat{r}$ as the maximum-likelihood estimate (MLE), i.e., the value of $r$ that maximizes $L(r)$. We compute the corresponding uncertainty as [146]

$$\sigma_r^2 = \int_0^\infty \mathrm{d}r \, L(r) r^2 - \left[ \int_0^\infty \mathrm{d}r \, L(r) r \right]^2 \,, \tag{7.15}$$

where $L(r)$ is normalized as $\int_0^\infty \mathrm{d}r \, L(r) = 1$. Eq. (7.15) defines the variance associated with a Gaussian random variable. We use eq. (7.15) whenever we compute $\sigma_r$, but we have also compared it with asymmetric 68% CL intervals. In our case, they are equal to the first significant digit.

## 7.3 Analysis

We apply the framework presented in section 7.2 to extract the bias on $r$ caused by a particular choice of HWP design. Given $\mathcal{M}_{\text{HWP}}$, our code[5] performs the following steps:

1. Compute the covariance matrix, $\mathbb{C}_\ell^{B,ij}$, as in eq. (7.9),

2. Invert $\mathbb{C}_\ell^{B,ij}$ to obtain the HILC weights, $w_\ell^i$, as in eq. (7.8),

3. Use $w_\ell^i$ to compute the $BB$ spectrum of the HILC solution, $C_{\ell,\text{HILC}}^{BB}$, as in eq. (7.10),

4. Compute the two-dimensional likelihood $L(r, A_{\text{lens}})$ from $C_{\ell,\text{HILC}}^{BB}$, using eq. (7.13),

5. Obtain the one-dimensional posterior p.d.f., $L_{\text{m}}(r)$, by marginalizing over $A_{\text{lens}}$, and the profile likelihood, $L_{\text{p}}(r)$, by maximization,

6. Return $\hat{r}$ and $\sigma_r$, defined as in eq. (7.15), computed from $L_{\text{m}}(r)$ and $L_{\text{p}}(r)$.

---

[5] https://github.com/martamonelli/HWP_end2end.

| Spectral parameters | | $C_\ell^{XX}$ parameters | $q\,[\mu\mathrm{K}^2]$ | $\alpha$ |
|---|---|---|---|---|
| CMB temperature $T_0$ | 2.725 K | Dust $EE$ | 323 | $-0.40$ |
| Dust temperature $T_{\mathrm{dust}}$ | 19.6 K | Dust $BB$ | 119 | $-0.50$ |
| Dust spectral index $\beta_{\mathrm{dust}}$ | 1.55 | Synchrotron $EE$ | 2.3 | $-0.84$ |
| Dust reference frequency $\nu_\star$ | 353 GHz | Synchrotron $BB$ | 0.8 | $-0.76$ |
| Synchrotron spectral index $\beta_{\mathrm{sync}}$ | $-3.1$ | | | |
| Synchtrotron reference frequency $\nu_\ast$ | 30 GHz | | | |

Table 7.1: Left panel: SED parameters entering in eqs. (7.16) for each component as reported in [104]. Right panel: The power-law parameters for the angular power spectra of synchrotron and thermal dust emission entering in eq. (7.17) as reported in [104] for the `Commander` [147] analysis with $f_{\mathrm{sky}} = 0.78$.

To validate our end-to-end model and code, we first perform the analysis for an ideal HWP and then move on to more realistic cases. However, before presenting our results, we review the additional assumptions that go into the explicit computation of the HILC covariance matrix $\mathbb{C}_\ell^B$, with the exception of the HWP choice.

**CMB, dust and synchtrotron spectral responses**  For maps in thermodynamic units, the $a_\lambda(\nu)$ functions entering in eqs. (7.5) read (see appendix A.2.2 for a complete derivation)

$$a_{\mathrm{CMB}}(\nu) = 1\,, \tag{7.16a}$$

$$a_{\mathrm{dust}}(\nu) = \left(\frac{\nu}{\nu_\star}\right)^{\beta_{\mathrm{dust}}} \frac{B_\nu(T_{\mathrm{dust}})}{B_{\nu_\star}(T_{\mathrm{dust}})} \frac{\nu_\star^2}{\nu^2} \frac{x_\star^2 e^{x_\star}}{x^2 e^x} \frac{(e^x - 1)^2}{(e^{x_\star} - 1)^2}\,, \tag{7.16b}$$

$$a_{\mathrm{sync}}(\nu) = \left(\frac{\nu}{\nu_\ast}\right)^{\beta_{\mathrm{sync}}} \frac{\nu_\ast^2}{\nu^2} \frac{x_\ast^2 e^{x_\ast}}{x^2 e^x} \frac{(e^x - 1)^2}{(e^{x_\ast} - 1)^2}\,, \tag{7.16c}$$

where $B_\nu(T)$ denotes a black-body spectrum at temperature $T$, $x \equiv h\nu/(k_B T_0)$ and $T_0 = 2.725$ K is the average temperature of the CMB [6]. The values of the remaining parameters entering in eqs. (7.16) are specified in Table 7.1.

**CMB, dust and synchtrotron angular power spectra**  The CMB angular power spectrum is computed with `CAMB` [148] assuming the best-fit 2018 *Planck* values for the cosmological parameters [19], except for the tensor-to-scalar ratio, which is set to $r_{\mathrm{true}} = 0.00461$. This is the same fiducial value as assumed in [29], and corresponds to Starobinsky's $R^2$ inflationary model [149] with the $e$-folding value of $N_* = 51$.

As for the polarized foreground emission, we parameterize their angular power spectra as a power law [104]

$$D_\ell \equiv \frac{\ell(\ell + 1)C_\ell}{2\pi} = q\left(\frac{\ell}{80}\right)^\alpha\,. \tag{7.17}$$

Specific values of the parameters are reported in Table 7.1 for both dust and synchrotron. Note that we neglect any intrinsic $EB$ correlation in the input, which is inaccurate (polarized dust emission has been observed to have non-zero $TB$ correlation [150, 151], which implies the presence of a $EB$ correlation [67, 152], and cosmic birefringence [49] would also result in a non-zero $EB$). When presenting our results in section 7.3.2, we comment on this assumption and argue that allowing non-zero $EB$ in input would not dramatically affect the analysis.

**Instrument specifics** To simulate LiteBIRD's design, we consider an instrument that mounts three different telescopes at low (LFT), medium (MFT), and high frequency (HFT). The specific frequency ranges of each telescope and frequency channel are taken from [29].

**Noise covariance matrix** Using a rotating HWP as polarization modulator suppresses the polarized $1/f$ noise component [68]. Being left with white noise only, we parameterize $N_\ell^{BB,i}$ as [146]

$$N_\ell^{BB,i} = \left[ \frac{\pi}{10800} \frac{n_p^i}{\mu\mathrm{K\,arcmin}} \right]^2 \mu\mathrm{K}^2\,\mathrm{str}\,, \tag{7.18}$$

where $n_p^i$ is the noise in Stokes parameters $Q$ or $U$ per pixel with solid angle $\Omega_{\mathrm{pix}} = 1$ arcmin$^2$. The specific values assumed for each $n_p^i$ are taken from [29].

**Beams** Since we assume the beams to be Gaussian and perfectly co-polarized, the $B_\ell^i$ coefficients only depend on the beam's full width at half maximum (FWHM). Specific FWHM values for each channel are taken from [29].

### 7.3.1 Validation: ideal HWP

An ideal HWP is described by a frequency-independent Mueller matrix with elements

$$\mathcal{M}_{\mathrm{ideal}} = \mathrm{diag}(1, 1, -1)\,. \tag{7.19}$$

In this case, the coefficients $g_\lambda^i$ and $\rho_\lambda^i$ reduce to the average of the correspondent $a_\lambda(\nu)$ function over the band $i$ [eq. (7.5)], which we will denote $a_\lambda^i$. The $\eta_\lambda^i$ coefficients go instead to zero. According to eq. (7.6), the multi-frequency maps reduce to

$$\widehat{\mathbf{m}}^i \simeq \overline{\mathbf{m}}_{\mathrm{CMB}}^i + \frac{1}{a_{\mathrm{CMB}}^i} \left[ \sum_{\lambda \neq \mathrm{CMB}} a_\lambda^i \, \overline{\mathbf{m}}_\lambda^i + \mathbf{n}^i \right]\,. \tag{7.20}$$

While the CMB component is not affected by the presence of the ideal HWP, the foreground emission suffers from a color correction, and the noise term is rescaled channel-by-channel. In this simple situation, the HILC should perform well and recover the CMB signal plus some noise bias given by

$$N_{\ell,\mathrm{HILC}}^{BB} = \sum_{i=1}^{n_{\mathrm{chan}}} \left( \frac{w_\ell^i}{a_{\mathrm{CMB}}^i B_\ell^i} \right)^2 \mathbb{N}_\ell^{BB,ii}\,. \tag{7.21}$$

We should therefore check that, for $\mathcal{M}_{\mathrm{ideal}} = \mathrm{diag}(1, 1, -1)$, the HILC output is in good agreement with the input CMB angular power spectrum, once the noise bias is removed.

In Figure 7.1, we show the angular $B$-mode power spectrum of the HILC solution, together with the input angular power spectra of CMB, dust, and synchrotron. For completeness, we also show the foreground residual and the noise bias. The noise bias has been removed from both the HILC solution and the foreground residual. The agreement between the HILC solution and the input CMB power spectrum is excellent up to $\ell \simeq 325$, roughly corresponding to LiteBIRD's beam resolution.

In Figure 7.2 we show the HILC weights for the three telescopes. All MFT channels have positive weights, consistent with them being CMB channels. On the other hand, some of LFT and HFT channels (at very low and very high frequencies, respectively) have negative weights, resulting in foreground subtraction.

The code returns the MLE $\hat{r} = 0.0047 \pm 0.0005$, which is compatible with the fiducial value of $r_{\mathrm{true}} = 0.00461$. This is what we expect, given the good agreement between the debiased HILC solution and the input CMB shown in Figure 7.1.

## 7.3.2  More realistic HWPs

For this analysis, we consider more realistic HWPs for each telescope. For LFT, we consider the Pancharatnam-type multi-layer sapphire symmetric stack design described in [153], provided with an anti-reflection coating (ARC) as presented in [112]. For the metal-mesh HWPs of MFT and HFT, we use the same input simulations and working assumptions as in [120].

We manipulate each set of Mueller matrices by performing a rotation of the angle $\theta_{\mathrm{T}}$ that minimizes the integral

$$\int_{\mathrm{T}} \mathrm{d}\nu \left\{ \left[ m_{\mathrm{QQ}}(\nu) - m_{\mathrm{UU}}(\nu) \right] \cos(4\theta_{\mathrm{T}}) + \left[ m_{\mathrm{QU}}(\nu) + m_{\mathrm{UQ}}(\nu) \right] \sin(4\theta_{\mathrm{T}}) \right\}^2 , \qquad (7.22)$$

over the entire frequency band of each telescope, specified by $\mathrm{T} = \{\mathrm{L}, \mathrm{M}, \mathrm{H}\}$. This choice is ultimately motivated by the specific design we assume for LFT, since there is no unique way to determine the position of the HWP's optical axes for a symmetric stack. Rotating $\mathcal{M}_{\mathrm{HWP,L}}$ of $\theta_{\mathrm{L}}$ then amounts to calibrate the HWP Mueller matrix and express it in a coordinate system aligned with the optical axes. Instead, the HWPs of MFT and HFT employ mesh-filter technology [154], for which optical axes can be more easily identified. However, for the sake of consistency, we choose to perform analogous rotations on the Mueller matrices of MFT and HFT metal-mesh HWPs. Rotation angles that minimize eq. (7.22) are 55.02° for LFT and 0.29° for M-HFT. The rotated Mueller matrix elements of each HWP are shown as a function of frequency in Figure 7.3.

Given the elements of the Mueller matrix, we compute the coefficients $\rho_{\lambda}^i$ and $\eta_{\lambda}^i$ according to eq. (7.5) and repeat all the steps outlined at the beginning of section 7.3. The HILC solution, $D_{\ell,\mathrm{HILC}}^{BB}$, is shown in Figure 7.4. Although the foreground residual (red dotted line) shows more features than in the ideal case of Figure 7.1, its contribution to $D_{\ell,\mathrm{HILC}}^{BB}$ is still subdominant. This confirms our intuition that reasonably optimized HWPs do not cause strong foreground leakage in the HILC solution [see the discussion below eq. (7.10)]. Note that, given the negligible foreground leakage, taking $C_{\ell,\mathrm{dust}}^{EB} = C_{\ell,\mathrm{synch}}^{EB} = 0$ in input is not such a strong assumption. Even if we allowed non-zero $EB$ correlations, they would not contribute significantly to the HILC solution.

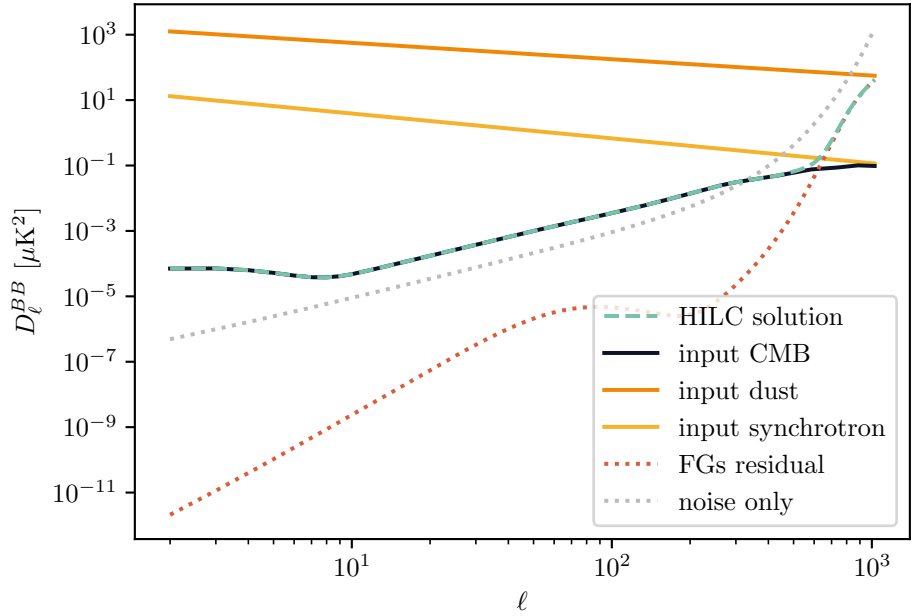

Figure 7.1: For an ideal HWP, the rescaled angular power spectrum, $D_\ell^{BB}$, of the HILC solution (dashed teal line) overlaps the input CMB spectrum (black solid line) for a wide range of multipoles. For large $\ell$, the two spectra begin to diverge as we approach the instrumental resolution. This can be seen by looking at the dotted gray line, representing the residual noise, which intersects the input spectrum at $\ell \sim 325$. For completeness, we also plot the input dust and synchrotron $D_\ell^{BB}$ (orange and yellow, respectively) and the foreground residual (red dotted line). The noise bias has been removed from both the HILC solution and the foreground residual spectra. The $w_\ell^i$ weights corresponding to the HILC solution are shown in Figure 7.2.

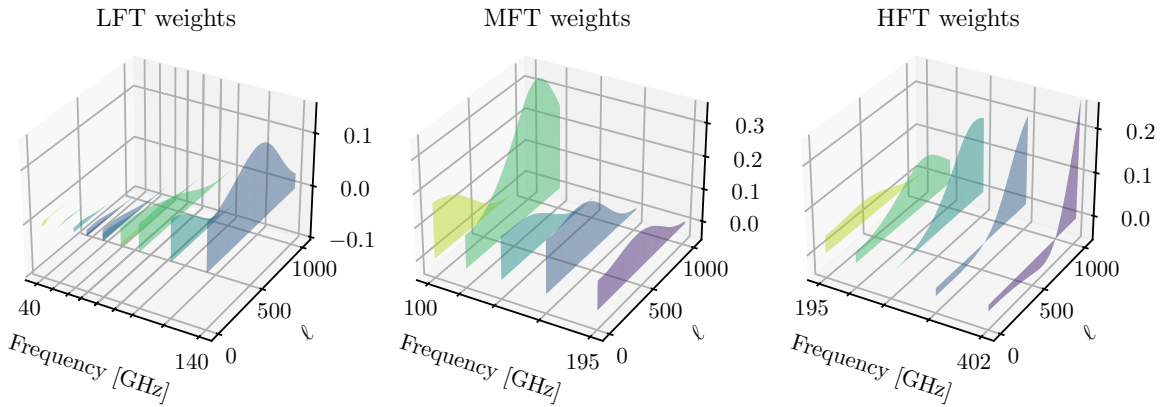

Figure 7.2: HILC weights, $w_\ell^i$, for each of the three telescopes with an ideal HWP. The corresponding $BB$ angular power spectrum is shown in Figure 7.1 (dashed teal line).

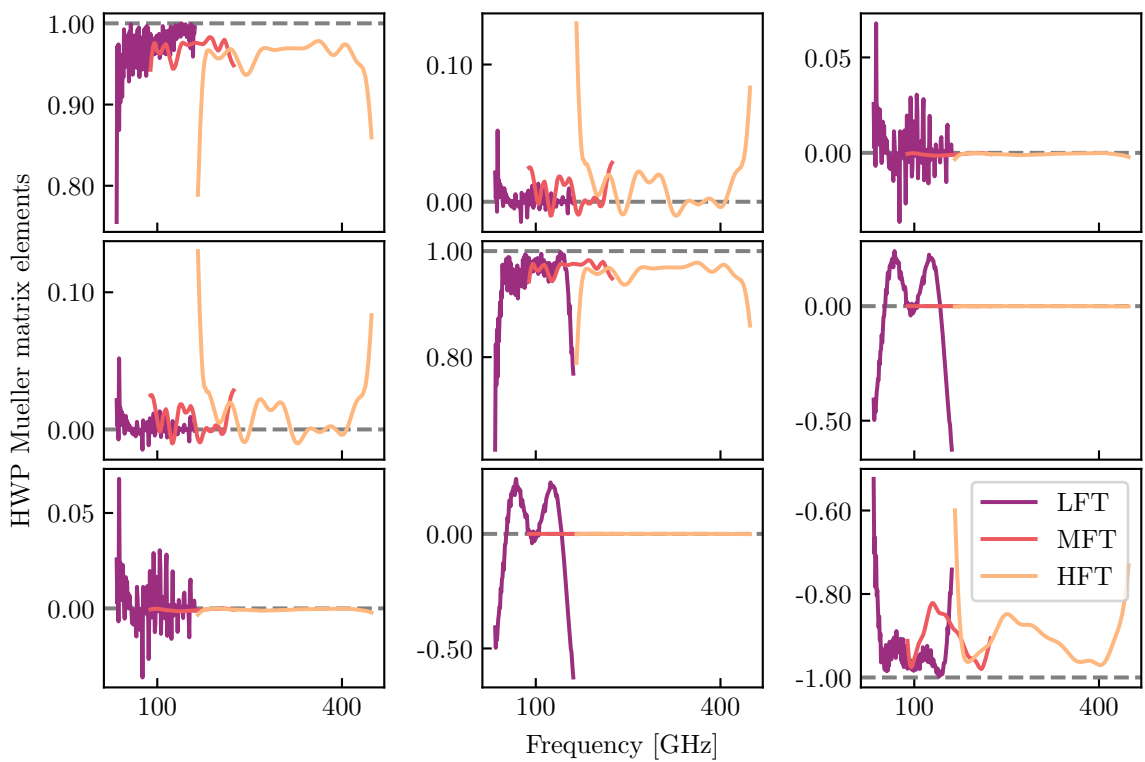

Figure 7.3: HWP Mueller matrix elements for LFT (purple), MFT (red) and HFT (orange) as function of frequency. For LFT, we consider a symmetric stack design [153] provided with ARC [112], compute its Mueller matrix elements, and rotate them of 55.02°, to express them in a reference frame with the $x$ axis parallel to the HWP optic axis. Instead, the Mueller matrix elements for MFT and HFT are obtained by following the same procedure and input simulations as done in [120], and rotating them of 0.29°. The dashed gray lines represent the ideal values of each element.

In Figure 7.5 we also show the HILC weights for the three telescopes. The weights look qualitatively similar to their ideal counterparts shown in Figure 7.2.

To give more precise considerations, Figure 7.6 shows the power spectra on large angular scales in more detail. We show the two independent terms that contribute to $D_{\ell,\mathrm{HILC}}^{BB}$ component-by-component: $\rho$-only (polarization efficiency) and $\eta$-only (cross-polarization coupling). These were obtained using the full covariance matrix $\mathbb{C}_\ell$ given in eq. (7.9) to compute the HILC weights, while neglecting some of the terms entering in eq. (7.10). For instance, the $\rho$-only dust contribution reads

$$C_{\ell,\mathrm{HILC}}^{BB,\mathrm{dust},\rho} = \sum_{i,j=1}^{n_{\mathrm{chan}}} \frac{w_\ell^i w_\ell^j}{g_{\mathrm{CMB}}^i g_{\mathrm{CMB}}^j} \rho_{\mathrm{dust}}^i \rho_{\mathrm{dust}}^j C_{\ell,\mathrm{dust}}^{BB} \,. \tag{7.23}$$

Intuitively, it makes sense for the effective polarization efficiency component to dominate in the CMB contribution. While $\eta_{\mathrm{CMB}}^i$ can be both positive and negative, all $\rho_{\mathrm{CMB}}^i$ are constrained to be smaller than 1. This means that, while the average $\langle \eta_{\mathrm{CMB}}^i \rangle$ across all frequency channels can be close to zero, $\langle \rho_{\mathrm{CMB}}^i \rangle$ cannot be arbitrarily close to 1. The HILC, which looks for the solution that minimizes the variance, may then be able to get rid of all cross-polarization coupling, while it cannot undo the average suppression due to the polarization efficiency. As a consequence of the smallness of the cross-polarization coupling component relative to the polarization efficiency, we argue that relaxing the $C_{\ell,\mathrm{CMB}}^{EB} = 0$ assumption for the input spectra would not significantly change our results.

Interestingly, the HILC solution approximately satisfies

$$\widehat{C}_{\ell,\mathrm{HILC}}^{BB} \simeq \frac{1}{n_{\mathrm{chan}}} \sum_{i=1}^{n_{\mathrm{chan}}} \left[ \frac{\rho_{\mathrm{CMB}}^i}{g_{\mathrm{CMB}}^i} \right]^2 \cdot C_{\ell,\mathrm{CMB}}^{BB} \,, \tag{7.24}$$

with $10^{-5}$ relative tolerance and $10^{-8}$ absolute tolerance for a wide range of multipoles, $25 \leq \ell \leq 372$. The upper limit has a simple interpretation: it roughly corresponds to the instrumental resolution.

**Bias on the tensor-to-scalar ratio** We finally employ the methodology introduced in section 7.2.3 to propagate the small discrepancy between the input CMB and the HILC solution shown in Figure 7.4 into a bias on $r$. We compare the marginalized posterior p.d.f., $L_{\mathrm{m}}(r)$, with the profile likelihood, $L_{\mathrm{p}}(r)$ [as defined in eqs. (7.14a) and (7.14b), respectively], and find that they are identical up to relative discrepancies of $\lesssim 10^{-3}$.

We show $L(r) = L_{\mathrm{p}}(r)$ in Figure 7.7 (teal solid line), together with a red vertical line corresponding to the input value, $r_{\mathrm{true}} = 0.00461$. The MLE is $\hat{r} = 0.0043 \pm 0.0005$. This bias is caused by the HWP polarization efficiency being lower than one. The $B$-mode signal is suppressed and $r$ is underestimated.

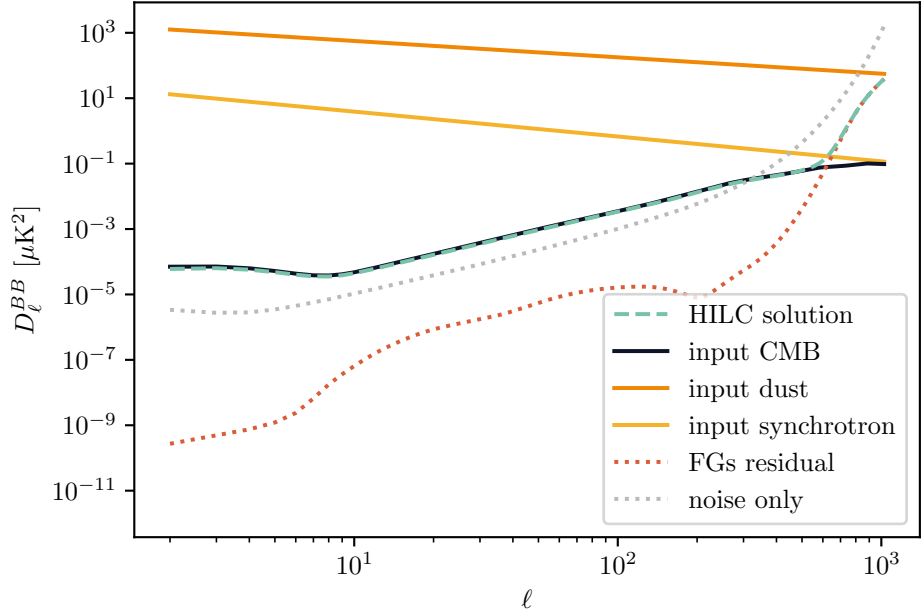

Figure 7.4: Same as Figure 7.1 but for the realistic HWP discussed in section 7.3.2 (dashed teal line). Compared to the ideal HWP case shown in Figure 7.1, the non-ideal HILC solution slightly differs from the input CMB at low multipoles. For comparison, we also show the residual noise bias (dotted gray line) and the foreground residual (red dotted line). They both show more features than their counterparts in Figure 7.1. The $w_\ell^i$ weights corresponding to the HILC solution are shown in Figure 7.5.

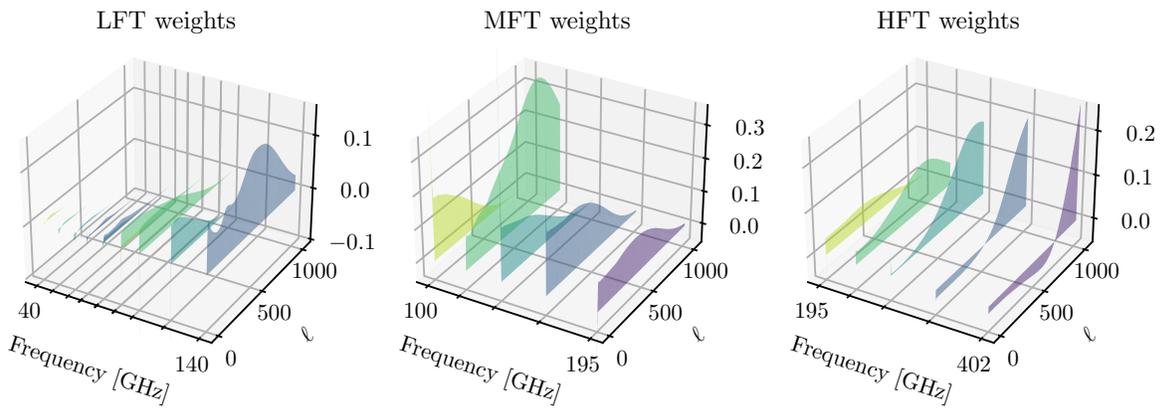

Figure 7.5: Same as Figure 7.2 but for the Mueller matrix elements given in Figure 7.3. The corresponding *BB* angular power spectrum is shown in Figure 7.4 (dashed teal line).

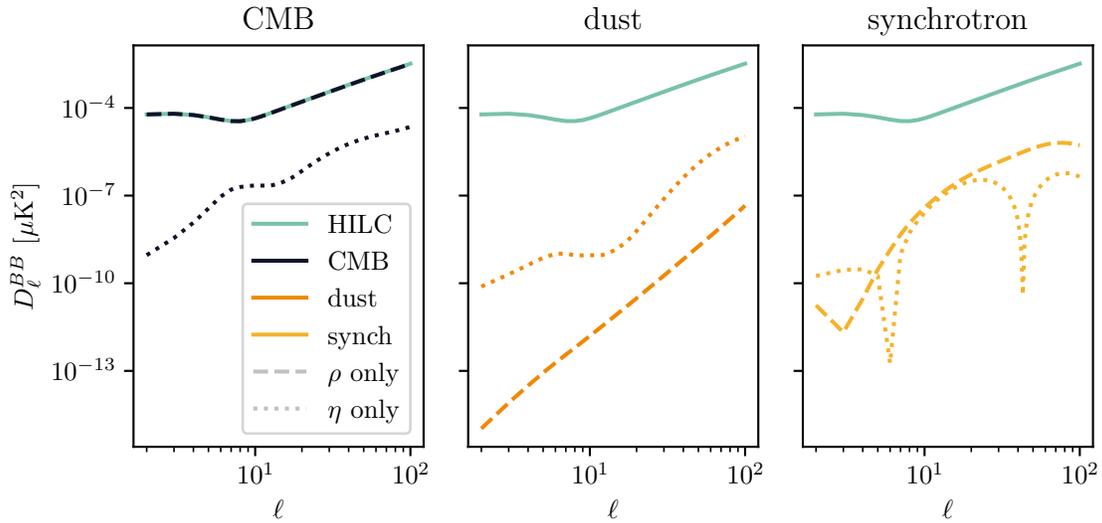

Figure 7.6: Different contributions to the *B*-mode power spectrum of the HILC solution (teal solid line). We focus on a different component (CMB, dust, and synchrotron) in each of the panels. The effective polarization efficiency and cross-polarization coupling components are shown in dashed and dotted, respectively. The largest contribution comes from the polarization efficiency component of the CMB.

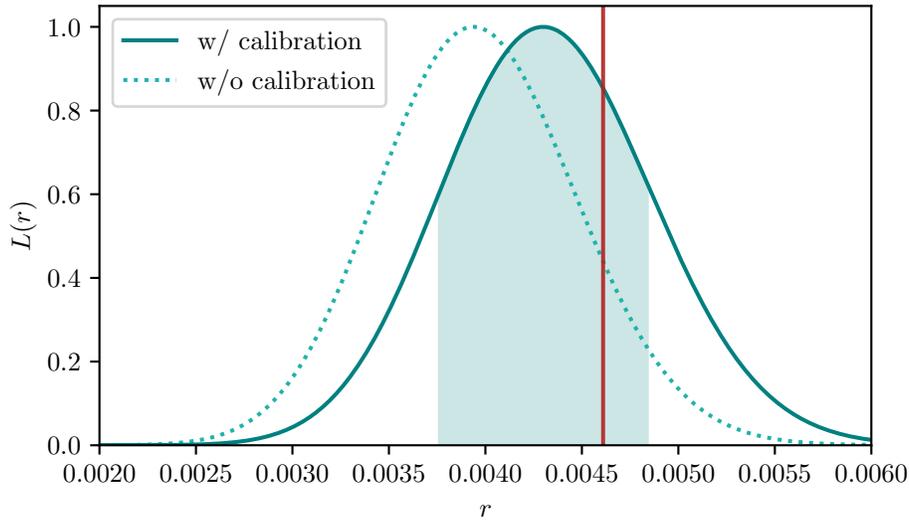

Figure 7.7: Normalized profile likelihood, $L(r) = L_p(r)$, obtained from the HILC solution, $\widehat{C}^{BB}_{\ell,\text{HILC}}$, given the HWP specifics presented in section 7.3.2 (teal solid line). The likelihood has a maximum at $\hat{r} = 0.0043$. The shaded region goes from $\hat{r} - \sigma_r$ to $\hat{r} + \sigma_r$ where $\sigma_r = 0.0005$. The solid red line represents the input tensor-to-scalar ratio parameter, $r_{\text{true}} = 0.00461$. The dotted light teal line shows the normalized profile likelihood obtained from the HILC solution when the gain calibration for the CMB temperature is not included.

**The weight of gain calibration**    The inclusion of the gain calibration for the CMB temperature in the modeling of multi-frequency maps may seem inconsequential, but it has strong implications. We repeat the analysis of section 7.3.2, except that we now skip the gain calibration, i.e., we model the $\widehat{\mathbf{m}}^i$ as in eq. (7.4) instead of eq. (7.6). The corresponding spherical harmonic coefficients read

$$\widehat{a}^{B,i}_{\ell m,\mathrm{w/o}} = \sum_\lambda B^i_\ell \left( \rho^i_\lambda a^{B,i}_{\ell m} - \eta^i_\lambda a^{E,i}_{\ell m} \right) + n^{B,i}_{\ell m} \,, \tag{7.25}$$

where the w/o subscript stresses that we are not calibrating the maps. By retracing the same steps as presented in section 7.2.2, we end up with an expression for the $BB$ angular power spectrum of the HILC solution that reads

$$C^{BB}_{\ell,\mathrm{HILC}} = \sum_{i,j=1}^{n_{\mathrm{chan}}} w^i_{\ell,\mathrm{w/o}} w^j_{\ell,\mathrm{w/o}} \left\{ \sum_\lambda \left[ \rho^i_\lambda \rho^j_\lambda C^{BB}_{\ell,\lambda} + \eta^i_\lambda \eta^j_\lambda C^{EE}_{\ell,\lambda} - \left( \rho^i_\lambda \eta^j_\lambda + \eta^i_\lambda \rho^j_\lambda \right) C^{EB}_{\ell,\lambda} \right] + \frac{\mathbb{N}^{BB,ij}_\ell}{B^i_\ell B^j_\ell} \right\} , \tag{7.26}$$

where the $w^i_{\ell,\mathrm{w/o}}$ are the HILC weights corresponding to the spherical harmonic coefficients of eq. (7.25). The corresponding normalized profile likelihood is shown in Figure 7.7 (dotted light teal line). We now find a much lower MLE of the tensor-to-scalar ratio, $\hat{r} = 0.0039 \pm 0.0005$, which is incompatible with $r_{\mathrm{true}}$.

## 7.4    Discussion

Clearly, gain calibration can partially mitigate the suppression of primordial $B$ modes caused by the HWP. Of course, one can characterize the non-idealities in laboratory measurements and correct for them in the data. However, if HWPs are properly designed, gain calibration for the CMB temperature allows us to mitigate the effects of non-idealities on polarization *in-flight* for space missions. The ability to perform in-flight calibration is always valuable.

To this end, we derive some realistic recommendations that can help maximize its benefits. In section 7.4.2, we also discuss the assumptions underlying our end-to-end model and comment on the possibility of relaxing some of them.

### 7.4.1    HWP design recommendations

We express the relevant combinations of Mueller matrix elements in terms of a set of 7 independent values that uniquely determine the components of $\mathcal{M}_{\mathrm{HWP}}$: the HWP Jones parameters, $h_{1,2}$, $\beta$, $\zeta_{1,2}$ and $\xi_{1,2}$ (see appendix B.4 for their definitions). The loss parameters $h_{1,2}$ describe the deviation from the unitary transmission of $E_{x,y}$; $\beta$ parametrizes the deviation from $\pi$ of the phase shift between $E_x$ and $E_y$; $\zeta_{1,2}$ and $\xi_{1,2}$ describe the amplitude and phase of the cross-polarization coupling. We write $g(\nu) \equiv m_{\mathrm{II}}(\nu)$, $\rho(\nu) \equiv [m_{\mathrm{QQ}}(\nu) - m_{\mathrm{UU}}(\nu)]/2$, and $\eta(\nu) \equiv [m_{\mathrm{QU}}(\nu) + m_{\mathrm{UQ}}(\nu)]/2$ as [120]

$$g = \frac{1}{2}\left[(1+h_1)^2 + (1+h_2)^2 + \zeta_1^2 + \zeta_2^2\right], \tag{7.27a}$$

$$\rho = \frac{1}{2}\left\{\frac{1}{2}\left[(1+h_1)^2 + (1+h_2)^2 - \zeta_1^2 - \zeta_2^2\right] + (1+h_1)(1+h_2)\cos\beta - \zeta_1\zeta_2\cos(\chi_1-\chi_2)\right\}, \tag{7.27b}$$

$$\eta = \frac{1}{2}\left\{(1+h_1)(\zeta_1\cos\chi_1 + \zeta_2\cos\chi_2) + (1+h_2)\left[\zeta_2\cos(\beta-\chi_2) + \zeta_1\cos(\beta-\chi_1)\right]\right\}, \tag{7.27c}$$

where any dependence on $\nu$ is kept implicit for the sake of compactness. Designing a perfectly ideal HWP with identically vanishing Jones parameters is technically impossible. However, some parameters are easier to minimize than others.

For example, $\zeta_{1,2}(\nu) \sim 10^{-2}$ can be achieved for both metal-mesh and multi-layer HWPs. If that is the case, the Taylor expansion of the above expressions for small $\zeta_{1,2}(\nu)$ yields, up to first order,

$$g = \frac{1}{2}\left[(1+h_1)^2 + (1+h_2)^2\right] + \mathcal{O}(10^{-4}), \tag{7.28a}$$

$$\rho = \frac{1}{2}\left\{\frac{1}{2}\left[(1+h_1)^2 + (1+h_2)^2\right] + (1+h_1)(1+h_2)\cos\beta\right\} + \mathcal{O}(10^{-4}), \tag{7.28b}$$

$$\eta = \frac{1}{2}\left\{(1+h_1)(\zeta_1\cos\chi_1 + \zeta_2\cos\chi_2) + (1+h_2)\left[\zeta_2\cos(\beta-\chi_2) + \zeta_1\cos(\beta-\chi_1)\right]\right\}. \tag{7.28c}$$

We can further simplify these expressions by requiring $h_{1,2} \sim 10^{-2}$, which implies $\rho(\nu) = g(\nu)\cos^2[\beta(\nu)/2]$ up to relative corrections of $\mathcal{O}(10^{-4})$. Alternatively, by keeping $h_{1,2}$ free while requiring $|h_1 - h_2|$ to be small, we ensure that $\rho(\nu) = g(\nu)\cos^2[\beta(\nu)/2]$ still holds up to relative corrections of $\mathcal{O}(|h_1 - h_2|)$. On the other hand, we cannot require $\beta(\nu)$ to be arbitrarily small due to the limitation of current technology. Keeping $\beta(\nu)$ free, we have

$$g_{\mathrm{CMB}}^i \simeq \int_{\nu_{\min}^i}^{\nu_{\max}^i} \frac{\mathrm{d}\nu}{\Delta\nu^i}\left[1 + h_1(\nu) + h_2(\nu)\right], \tag{7.29a}$$

$$\rho_{\mathrm{CMB}}^i \simeq \int_{\nu_{\min}^i}^{\nu_{\max}^i} \frac{\mathrm{d}\nu}{\Delta\nu^i}\left[1 + h_1(\nu) + h_2(\nu)\right]\cos^2[\beta(\nu)/2]. \tag{7.29b}$$

If at least one of $h_1(\nu) + h_2(\nu)$ and $\cos^2[\beta(\nu)/2] = [1 + \cos\beta(\nu)]/2$ is slowly varying within the band, we find that $\rho_{\mathrm{CMB}}^i \simeq A^i g_{\mathrm{CMB}}^i$, where $A^i$ is an appropriate factor that depends on $\beta$. Then, if we know $A^i$ with good precision, its effect can be undone by multiplying each multi-frequency polarization map by $1/A^i$. In this way, the gain calibration for the CMB temperature can partially mitigate the impact of the HWP polarization efficiency.

Regarding cross-polarization coupling, we argue that there are two strategies to keep its effects under control. First, we could simply require $\eta(\nu) \lesssim 10^{-3}$ so that the $E \to B$ leakage is negligible. However, this might be technically challenging. Another strategy is to exploit the fact that the HILC weights minimize the variance. Even if $\eta(\nu)$ is not vanishing small, as long as the $\eta_{\mathrm{CMB}}^i$ fluctuate around zero, the HILC should be able to mitigate their effect.

**HWP angle miscalibration**   An imperfect calibration of the HWP angle can dramatically affect the considerations we have presented so far. If an HWP with $g^i_{\mathrm{CMB}} \simeq \rho^i_{\mathrm{CMB}}$ and $\langle \eta^i_{\mathrm{CMB}} \rangle \simeq 0$, is rotated by some angle $\theta$, its effective gain, polarization efficiency, and cross-polarization coupling are transformed as

$$g' = g\,, \qquad \rho' = \rho \cos 4\theta - \eta \sin 4\theta\,, \qquad \eta' = \eta \cos 4\theta + \rho \sin 4\theta\,. \qquad (7.30)$$

On the one hand, this causes the cross-polarization coupling coefficients to fluctuate around some non-zero value, making it impossible for the HILC to filter them out. On the other hand, the polarization efficiency and gain coefficients might strongly deviate from each other, reducing the benefits of gain calibration.

   Therefore, a good calibration of the HWP position angle, $\theta$, is crucial to ensure the validity of our considerations and recommendations. Derotating the polarization maps by $\theta$ prior to the foreground cleaning step, as suggested in [58], would allow us to account for potential differences in the miscalibration angles of the HWPs.

## 7.4.2   Reviewing the underlying assumptions

We derived the model for multi-frequency maps and their spherical harmonics coefficients [eqs. (7.6) and (7.7), respectively] under several assumptions. We list them in order of appearance:

1. We assumed axisymmetric and perfectly co-polarized beams,

2. We assumed the maps to be obtained from an ideal bin averaging map-maker,

3. We considered a top-hat bandpass,

4. We assumed the SED of each component to be uniform throughout the sky,

5. We assumed a perfect gain calibration for the CMB temperature.

Assumptions 1 and 2 cannot be relaxed while maintaining the semi-analytical treatment, since more complex beams and more refined map-makers can only be included in numerical simulations. On the other hand, assumptions 3 and 5 can be straightforwardly relaxed within our simple analytical model (given our focus on the HWP non-idealities, however, we chose not to play around with the bandpass shape or imperfect temperature gain calibration).

   Assumption 4 can also be relaxed easily, but allowed us to analytically model the foreground cleaning step. Indeed, as soon as the SED of the foreground emission becomes anisotropic, the simple implementation of the HILC presented in section 7.2.2 is no longer able to recover the CMB signal accurately, and more elaborate methods such as Needlet ILC [155] and its moment [156] and Multiclustering [157] extensions will be needed. Although our quantitative results may be affected, qualitative conclusions will remain valid as long as the method is still based on ILC.

   It would be interesting to relax some of these assumptions and check whether the recommendations presented in section 7.4.1 still ensure that gain calibration for the CMB temperature can mitigate polarization systematics due to the HWP non-idealities. We leave this analysis for future work.

## 7.5   Conclusions and perspectives

In this work, we presented a simple framework to propagate the HWP non-idealities through the three macro-steps of any CMB experiment: observation of multi-frequency maps, foreground cleaning, and power spectra estimation. We focused on the impact of non-idealities on the tensor-to-scalar ratio parameter, $r$.

We generalized the formalism presented in [78] to include the polarized Galactic foreground emission (dust and synchrotron), foreground cleaning using a blind method (HILC), bandpass integration, noise, beam smoothing, and gain calibration for the CMB temperature. As a concrete working case, we considered a full-sky CMB mission with LiteBIRD-like specifics [29].

We validated the code against an ideal HWP and confirmed that the MLE $\hat{r}$ was compatible with the input value, $r = 0.00461$, within the uncertainty. Then, we employed more realistic Mueller matrix elements for each of the three telescopes of LiteBIRD and found $\hat{r} = 0.0043 \pm 0.0005$. We showed how the suppression is mostly due to the effective polarization efficiency of the HWP, which averages to a value lower than 1. The effective cross-polarization coupling and the foreground residual are found to be subdominant in our output $B$-mode power spectrum.

We found that the bias in $r$ significantly worsens if gain calibration for the CMB temperature is not included in the modeled multi-frequency maps: $\hat{r} = 0.0039 \pm 0.0005$, which is incompatible with the input value. Gain calibration would perfectly remove the HWP effects if $\rho^i_{\mathrm{CMB}} = g^i_{\mathrm{CMB}}$ and $\eta^i_{\mathrm{CMB}} = 0$, which are, however, unrealistic requirements. Still, we showed that an effective mitigation can be achieved if we can factorize $\rho^i_{\mathrm{CMB}} \simeq A^i g^i_{\mathrm{CMB}}$, we have good knowledge of the $A^i$ coefficients, and $\langle \eta^i_{\mathrm{CMB}} \rangle \simeq 0$. These considerations helped us to formulate some recommendations on the HWP design in terms of the HWP Jones parameters:

▷ Cross-polarization coupling should be small, $\zeta_{1,2} \lesssim 10^{-2}$, which can be achieved for both metal-mesh and multi-layer HWPs;

▷ The loss parameters should also be small, $h_{1,2} \lesssim 10^{-2}$, or, alternatively, $|h_1 - h_2| \lesssim 10^{-3}$;

▷ At least one of $h_1(\nu) + h_2(\nu)$ and $[1 + \cos\beta(\nu)]/2$ should be slowly varying within the band, so that $\rho^i_{\mathrm{CMB}} \simeq A^i g^i_{\mathrm{CMB}}$;

▷ Cross-polarization coupling can be kept under control by requiring $\zeta_{1,2}$ to be even smaller, or alternatively, by ensuring that $\eta^i_{\mathrm{CMB}}$ fluctuates around zero.

One can characterize the non-idealities of the HWP in laboratory measurements, and a requirement for the smallness of a bias in $r$ gives a requirement for the accuracy of the calibration in the laboratory. However, if the above recommendations are implemented in the design of the HWP used for space missions, the in-flight gain calibration for the CMB temperature can also be used to check and correct for the effects of HWP non-idealities in the data, complementing the laboratory calibration.

Some of the recommendations above depend strongly on the class of foreground cleaning methods we used in our end-to-end model. We used a blind method (HILC), but if one were to use a parametric component separation method to derive design recommendations, they would likely be different from those listed above. This highlights the importance of developing analysis strategies together with hardware designs.

This work represents a first generalization of the model presented in [78] toward a more realistic account of how the HWP non-idealities affect the observed CMB. However, being semi-analytical, this framework still relies on several simplifying assumptions (see section 7.4.2). One of the most crucial is the isotropy of the foreground SED. It would be interesting to relax this assumption and repeat the analysis carried out here, using more elaborate ILC-based methods (e.g., [156, 157]). This would help us test the robustness of our recommendations for the design of HWPs in a more realistic context. We leave this study for future work.

# Chapter 8

# Conclusions

## 8.1 Summary

In this thesis, we studied systematic effects in CMB polarization experiments using realistic simulations and analytical methods. We focused on HWP non-idealities and studied how they propagate through the main analysis steps of any CMB experiment (data acquisition, reconstruction of multi-frequency maps, foreground cleaning, and power spectra estimation) and how they ultimately impact the observed CMB polarization.

In Chapter 2 we outlined the broader scientific context of this thesis. We briefly discussed CMB anisotropies in both temperature and polarization, and introduced $E$ and $B$ modes. We showed how inhomogeneities at the time of photon decoupling can source polarization via Thomson scattering, and provided an intuitive understanding of why scalar modes do not produce $B$ modes at linear order. We discussed the possibility of extracting new physics from CMB polarization: constraining inflationary theories from $B$ modes, and testing parity violation from $EB$ correlation.

In Chapter 3 we described the macro steps that make up any CMB experiment. First, we provided a simple data model for the TOD and then discussed data analysis. We introduced map-making (focusing on bin-averaging), foreground cleaning and parameter inference. All these concepts and concrete examples are used extensively in the rest of this thesis.

Chapter 4 is a brief introduction to the HWP, its advantages and disadvantages. We first considered the ideal limit, and presented a concrete example to show the HWP's ability to mitigate $1/f$ noise and reduce pair-differencing systematics. We then emphasized the inevitable presence of non-idealities in any real HWP and the importance of carefully studying them to ensure that they do not induce crucial systematic effects.

In Chapter 5, we presented a framework for simulating TOD and binned maps given realistic HWPs, arbitrary beams, and noise specifics (including $1/f$). We discussed the functionalities we added to `beamconv` [101, 115] to tailor its output to a LiteBIRD-like mission: the implementation of a new scanning strategy, the option to read detector offsets as quaternions, and the possibility to produce noise and dipole TOD at the chunks level.

We discussed a concrete application of the simulation framework in Chapter 6, where we simulated binned maps for 160 detectors of a single frequency channel (140 GHz) for a LiteBIRD-like experiment (under the CMB-only and no noise assumptions). We used a realistic Mueller matrix to simulate the TOD, but neglected the non-idealities in the map-maker. This allowed us to study how overlooking HWP non-idealities in the analysis can affect the reconstructed CMB angular power spectra. We focused on the measured cosmic birefringence angle, $\beta$, and found that the non-idealities induce a bias on $\beta$ of a few degrees. This large miscalibration highlights the importance of having a good knowledge of the non-idealities to mitigate their effect at the data analysis level. Another important new result presented in Chapter 6 is a simple analytical formula that models the effect of the HWP non-idealities on the observed maps. After testing it against the simulation output, we highlighted the importance of such analytical tools, which can help tremendously in gaining intuition about the problem at hand.

In Chapter 7, we generalized the analytical model to include the polarized Galactic foreground emission (dust and synchrotron), foreground cleaning using a blind method (HILC), bandpass integration, noise, beam smoothing, and gain calibration for the CMB temperature. We used this framework to study the impact of non-idealities on the observed tensor-to-scalar ratio, $r$, and found a suppression of $\sim 5\%$. We showed how the bias on $r$ significantly worsens when the gain calibration for the CMB temperature is not included in the modeled multi-frequency maps. We provided some recommendations on the HWP design to maximize the benefits of gain calibration, which would allow to mitigate the effects of non-idealities in flight.

## 8.2   Outlook and future perspectives

Given the number of new CMB experiments that will see their first light in the next few years, studying systematic effects is now as timely as vital. The work presented in this thesis lays the foundations in this direction and opens up to a few generalizations that will help refine the tools and results discussed here.

**Optimizing the simulation framework** The simulation framework presented in Chapter 5, based on a modified version of `beamconv`, could be extended and optimized even further. It would be interesting to include additional instrumental effects, such as the detector non-idealities and the polarization wobble of the sinuous antennas. It would be also important to develop an efficient strategy to perform integrations over the frequency bands (this is a crucial issue in CMB simulations, as both the sky maps and the instrumental response are frequency-dependent). Finally, the code should be optimized to run on a computer cluster, which would allow simulating full-scale missions. Implementing all these changes would allow us to run realistic CMB simulations, which is essential for studying systematic effects in their complexity.

**Refining HWP design recommendations** The semi-analytical end-to-end model presented in Chapter 7 allowed us to derive a set of recommendations for the HWP design that can help maximize the benefits of gain calibration. To keep the treatment semi-analytical, however, we took a number of simplifying assumptions. One of the most crucial was the isotropy of the foreground SEDs, which allowed us to use the HILC foreground cleaning method. However, foregrounds are known to have anisotropic SEDs, making this approximation invalid. It would be interesting to relax this assumption and study how this affects the HWP design recommendations.

As a concrete path forward, we plan to consider realistic sky models generated with `PySM` [139, 158], together with a more sophisticated foreground cleaning method: the multi-clustering needlet ILC (MCNILC) [157]. Our expectation is that the results obtained in Chapter 7 will remain (at least qualitatively) valid, since the MCNILC belongs to the same class of foreground cleaning methods as the ILC. Regardless of whether this analysis will confirm our previous results or not, this work will be crucial for optimizing the HWP design for LiteBIRD, as it will provide a concrete strategy to perform *in-flight* calibration of the non-idealities.

**Beyond HWPs** Although the work discussed in this thesis focused on HWP-induced systematics, the tools we have presented here (TOD simulation pipeline and semi-analytical framework) could also be used to study other instrumental effects. Specifically, we plan to investigate how the observed angular power spectra are affected by the polarization wobble of the sinuous antennas, which is a frequency-dependent rotation of the transmitted polarization vector due to the antennas' geometry. Although this effect can be removed by pairing the signal from sinuous antennas with opposite orientations, limitations of the focal plane could prevent us from achieving perfect cancellation. Also, one cannot exclude that one of the detector wafers could malfunction, nulling the signal from one orientation (but not the other). These scenarios might result in some $E \to B$ leakage, which should be absolutely avoided to be able to measure $B$ modes. It is therefore essential to fully understand the implications of the polarization wobble, and work to develop concrete mitigation strategy. This study, will be relevant for both LiteBIRD and SO, as sinuous antennas are used in the designs of both experiments.

Finally, we would like to emphasize the importance of the work presented in this thesis for the search for new physics from CMB polarization. Finding new physics requires new strategies for controlling systematics and instrumental effects, which must be carefully studied to achieve the ambitious observational goals of the next-generation CMB experiments. This work takes a step in that direction by providing two complementary tools for understanding the impact of HWP non-idealities on new physics.

# Appendices

# Appendix A

# On unit conversion

**Summary:** In this appendix, we introduce some of the units of measurement commonly used in the CMB literature and show how to convert between them. We give two concrete examples: we convert COBE (DMR), WMAP and *Planck* noise levels to the same units to compare them, and we derive dust and synchrotron spectral properties in thermodynamic units.

*Section A.1 is inspired by some personal notes that Eiichiro Komatsu shared with me at the beginning of my PhD. Section A.2.2 is adapted from the appendix of [79].*

## A.1  Definitions

For a given $\nu$, the specific intensity of the sky signal at that frequency, $I_\nu$, has units of $\mathrm{J\,s^{-1}\,m^{-2}\,str^{-1}\,Hz^{-1}}$ and can be decomposed as

$$I_\nu = B_\nu(T_0) + \delta I_\nu\,, \tag{A.1}$$

where $B_\nu(T) = 2h\nu^3/[c^2(e^x-1)]$ is a black-body spectrum, $x \equiv h\nu/(k_B T)$ and $T_0 = 2.725$ K is the average temperature of the CMB [6]. The term $\delta I_\nu$ represents any excess signal over the isotropic CMB, such as CMB temperature fluctuations or foreground emission.

In radio astronomy, the specific intensity $\delta I_\nu$ is often expressed in terms of some kind of temperature in units of K (or its multiples, often $\mu$K). Here we discuss a few examples.

**Brightness (or Rayleigh-Jeans) temperature fluctuations**  The brightness temperature (also called Rayleigh-Jeans temperature), $T_B$, is defined by

$$I_\nu \equiv \frac{2\nu^2}{c^2} k_B T_B(\nu)\,. \tag{A.2}$$

Similarly, the brightness temperature fluctuations, $\delta T_B$, can be defined as the brightness temperature associated to the excess intensity, $\delta I_\nu$:

$$\delta I_\nu = \frac{2\nu^2}{c^2} k_B \delta T_B(\nu)\,. \tag{A.3}$$

Working with brightness temperatures was convenient in the the past, when CMB experiments were only sensitive to relatively low frequencies, $\nu \ll 100\,\text{GHz}$, corresponding to $x \ll 1$. In this regime, $T_B$ for a black-body approximates the thermodynamic temperature[1], making it easy to interpret. However, this nice interpretation does not hold at higher frequencies. Also, most sources in the sky do not have a black-body spectrum, in which case $T_B$ has no thermodynamic interpretation.

**Temperature fluctuation in thermodynamic units**   As CMB experiments began to operate at higher frequencies, $\nu \gtrsim 100$ GHz, where $T_B \not\equiv T$, it became necessary to define the temperature fluctuation in thermodynamic units as

$$\delta I_\nu = \frac{dB_\nu(\bar{T})}{d\bar{T}} \delta T(\nu) = \frac{2\nu^2}{c^2} \frac{x^2 e^x}{(e^x - 1)^2} k_B \delta T(\nu)\,. \tag{A.4}$$

Since $\delta I_\nu$ can include non-CMB components (foreground emission), $\delta T(\nu)$ does not refer only to the CMB temperature anisotropies, but also to the other components. Note that $\delta T(\nu)$ is actually frequency independent for CMB anisotropies, while it depends on $\nu$ for the other components. By comparing eqs. (A.3) and (A.4), we obtain the formula to convert thermodynamic units to brightness temperature units and vice versa:

$$\delta T_B(\nu) = \frac{x^2 e^x}{(e^x - 1)^2} \delta T(\nu)\,. \tag{A.5}$$

**Antenna temperature**   The antenna temperature, $T_A$, measures the power, $P$, received by an antenna and is defined by

$$P \equiv k_B \int_{\nu - \Delta\nu/2}^{\nu + \Delta\nu/2} d\nu'\; T_A(\nu') \approx k_B T_A(\nu) \Delta\nu\,, \tag{A.6}$$

where $P$ has units of $\text{J}\,\text{s}^{-1}$ and $\Delta\nu$ is the bandwidth. This definition holds when we distinguish between two polarization states of incoming photons. If $P$ includes both polarization states, instead, the relationship is modified to $P \approx 2k_B T_A \Delta\nu$.

To see how the antenna temperature is related to the brightness temperature of the sky signal, we start by writing the power $P$ received by a telescope with aperture $A$ as[2]

$$P = \frac{A}{2} \int_{\nu - \Delta\nu/2}^{\nu + \Delta\nu/2} d\nu' \int d\Omega\; \text{Beam}(\theta, \varphi, \nu') I_{\nu'}(\theta, \varphi)\,, \tag{A.7}$$

where $d\Omega = d\cos\theta\,d\varphi$ is the solid angle element on the sphere and "Beam" represents the beam response of the antenna. We can rewrite eq. (A.7) in terms of the beam-averaged brightness temperature, $\langle T_B(\nu) \rangle_{\text{beam}}$:

$$\langle T_B(\nu) \rangle_{\text{beam}} = \frac{2\nu^2}{c^2} \langle I_\nu \rangle_{\text{beam}} \equiv \frac{2\nu^2}{c^2} \frac{\int d\Omega\; \text{Beam}(\theta, \varphi, \nu) I_{\nu'}(\theta, \varphi)}{\Omega_A}\,, \tag{A.8}$$

---

[1] By taking the low-frequency (Rayleigh-Jeans) limit of $B_\nu(T)$, i.e. $x \ll 1$, we get $B_\nu(T) \to \frac{2\nu^2}{c^2} k_B T$.

[2] Here, we distinguish between two polarization states of incoming photons.

where $\langle I_\nu \rangle_{\text{beam}}$ denotes the beam-averaged intensity, and $\Omega_A$ is the beam solid angle:

$$\Omega_A(\nu) \equiv \int d\Omega \; \text{Beam}(\theta, \varphi, \nu) \,. \tag{A.9}$$

By combining eqs. (A.7) and (A.8), we get

$$P = \frac{A}{2} \int_{\nu - \Delta\nu/2}^{\nu + \Delta\nu/2} d\nu' \; \Omega_A(\nu') \frac{2\nu'^2}{c^2} k_B \langle T_B(\nu') \rangle_{\text{beam}} \,. \tag{A.10}$$

Now, since an ideal antenna has the property that $\Omega_A = \lambda^2/A = c^2/(\nu^2 A)$, the above equation simplifies to

$$P = k_B \int_{\nu - \Delta\nu/2}^{\nu + \Delta\nu/2} d\nu' \; \langle T_B(\nu') \rangle_{\text{beam}} \approx k_B \langle T_B(\nu) \rangle_{\text{beam}} \Delta\nu \,. \tag{A.11}$$

We thus conclude that the antenna temperature is equal to the beam-averaged brightness temperature of the sky intensity:

$$T_A = \langle T_B \rangle_{\text{beam}} \,. \tag{A.12}$$

**Noise-equivalent power**   The noise-equivalent power (NEP) is a measure of noise power with 1 Hz bandwidth, or noise power with 0.5 second of integration time, in units of J s$^{-1}$ Hz$^{-1/2}$. When we observe noise power of $P$ (with no sky signal) over some bandwidth $\Delta\nu$, the NEP is defined by

$$\text{NEP} \equiv \frac{P}{\sqrt{\Delta\nu}} \,, \tag{A.13}$$

where $\Delta\nu$ is in units of Hz. Using the definition of the antenna temperature given in eq. (A.6), we obtain

$$\text{NEP} = k_B T_A^{\text{NEP}} \sqrt{\Delta\nu} \,, \tag{A.14}$$

which is the expression for the *equivalent* antenna temperature for noise power.

**Noise equivalent temperature**   The noise equivalent temperature (NET) is also a measure of noise power with 1 Hz bandwidth, in units of K Hz$^{-1/2}$. In the low-frequency limit (Rayleigh-Jeans limit), we define it as

$$k_B \text{NET}_{\text{RJ}} \equiv \text{NEP} = k_B T_A^{\text{NEP}} \sqrt{\Delta\nu} \,. \tag{A.15}$$

Often, the NET is expressed in units of CMB thermodynamic temperature. This can be done by using the relationships $\delta T_A = \langle \delta T_B \rangle_{\text{beam}} = x^2 e^x / (e^x - 1)^2 \langle \delta T \rangle_{\text{beam}}$. As noise does not care about antenna response, we can remove $\langle \ldots \rangle_{\text{beam}}$. We write

$$\text{NET}_{\text{RJ}} = \frac{x^2 e^x}{(e^x - 1)^2} \text{NET}_{\text{CMB}} \,. \tag{A.16}$$

If we write the band average explicitly,

$$\text{NET}_{\text{RJ}} = \text{NET}_{\text{CMB}} \times \int_{\nu - \Delta\nu/2}^{\nu + \Delta\nu/2} \frac{d\nu'}{\Delta\nu} \frac{x'^2 e^{x'}}{(e^{x'} - 1)^2} \,. \tag{A.17}$$

## A.2   Concrete examples

Here we discuss two concrete examples of unit conversion. In Section A.2.1, we show how to convert COBE, WMAP and *Planck* noise levels to the same units and compare them, while in Section A.2.2, we write the spectral properties of dust and synchtrotron in thermodynamic units.

### A.2.1   Comparing noise levels of COBE, WMAP and *Planck*

We choose to consider the channels with central frequency around $100\,\mathrm{GHz}$, i.e. the $90\,\mathrm{GHz}$ DMR channel for COBE, the W-band for WMAP (centered at $94\,\mathrm{GHz}$), and the HFI $100$ GHz channel for *Planck*.

***Planck***   The $\mathrm{NET_{CMB}}$ of the $100\,\mathrm{GHz}$ *Planck*-HFI frequency channel is $40.0\,\mu\mathrm{K}\sqrt{\mathrm{s}}$ [159].

**WMAP**   The W-band has 8 radiometers, and its sensitivity per radiometer is $1.48\,\mathrm{mK}\sqrt{\mathrm{s}}$, in brightness temperature[3]. The total channel sensitivity can be estimated by dividing the sensitivity per radiometer by the square root of the number of radiometers, obtaining $\simeq 0.52\,\mathrm{mK}\sqrt{\mathrm{s}}$. To compare this value with *Planck*'s, we must convert it from brightness temperature to thermodynamic units. Making use of eq. (A.5) with $\nu = 94\,\mathrm{GHz}$, we obtain $\mathrm{NET_{CMB}} \simeq 0.65\,\mathrm{mK}\sqrt{\mathrm{s}}$.

**COBE**   The noise levels in units of antenna temperature per 0.5 sec measurement of the 90A and 90B DMR channels, $\mathrm{NET_{RJ,90A}}$ and $\mathrm{NET_{RJ,90B}}$, amount to $39.10\,\mathrm{mK}\sqrt{\mathrm{s}}$ and $30.76\,\mathrm{mK}\sqrt{\mathrm{s}}$, respectively[4]. We compute the combined noise level of the 90A and 90B channels by taking their inverse-variance average:

$$\mathrm{NET_{RJ}} = \sqrt{\left(\frac{1}{\mathrm{NET^2_{RJ,90A}}} + \frac{1}{\mathrm{NET^2_{RJ,90A}}}\right)^{-1}} \simeq 24\,\mathrm{mK}\sqrt{\mathrm{s}}\,. \tag{A.18}$$

To convert those values in thermodynamic units, we can use again eq. (A.5) with $\nu = 90\,\mathrm{GHz}$ and obtain $\mathrm{NET_{CMB}} \simeq 30\,\mathrm{mK}\sqrt{\mathrm{s}}$.

### A.2.2   Spectral properties in thermodynamic units

As shown in eq. (3.21), the specific intensity of CMB anisotropies follows a differential black-body, while dust and synchrotron can be modeled as a modified black-body and a power law, respectively [104]. By making use of eq. (A.4), we can write the relation between the thermodynamic temperatures at $\nu$ and at some other reference frequency $\nu_*$:

$$\delta T(\nu) = \frac{\delta I_\nu}{\delta I_{\nu_*}} \frac{\nu_*^2}{\nu} \frac{x_*^2 e^{x_*}}{(e^{x_*}-1)^2} \frac{(e^x-1)^2}{x^2 e^x} \delta T(\nu_*)\,. \tag{A.19}$$

---

[3]See Table 1.3 of WMAP Nine–Year Explanatory Supplement available at https://lambda.gsfc.nasa.gov/product/wmap/dr5/pub_papers/nineyear/supplement/WMAP_supplement.pdf.

[4]See Table 1 of COBE-DMR Four-Year Explanatory Supplement available at https://lambda.gsfc.nasa.gov/data/cobe/dmr/doc4/dmr_explanatory_supplement_4yr.pdf.

By plugging these expressions in eq. (A.19), we obtain the SED of CMB, dust, and synchrotron in terms of the CMB thermodynamic temperature:

$$\delta T_{\text{CMB}}(\nu) = \delta T_{\text{CMB}}\,, \tag{A.20a}$$

$$\delta T_{\text{dust}}(\nu) = \left(\frac{\nu}{\nu_\star}\right)^{\beta_{\text{dust}}} \frac{B_\nu(T_{\text{dust}})}{B_{\nu_\star}(T_{\text{dust}})} \frac{\nu_\star^2}{\nu^2} \frac{x^2 e^{x_\star}}{x_\star^2 e^x} \frac{(e^x - 1)^2}{(e^{x_\star} - 1)^2}\, \delta T_{\text{dust}}(\nu_\star)\,, \tag{A.20b}$$

$$\delta T_{\text{sync}}(\nu) = \left(\frac{\nu}{\nu_*}\right)^{\beta_{\text{sync}}} \frac{\nu_*^2}{\nu^2} \frac{x^2 e^{x_*}}{x_*^2 e^x} \frac{(e^x - 1)^2}{(e^{x_*} - 1)^2}\, \delta T_{\text{sync}}(\nu_*)\,. \tag{A.20c}$$

# Appendix B

# Polarized light

**Summary:** In this appendix, we define polarization and provide a brief introduction to Jones and Mueller calculus.

*Section B.4 is adapted from the appendix of [79].*

## B.1   Definitions

Electromagnetic fields can be described by a pair of three-dimensional vector and pseudovector fields: the electric and magnetic fields, $\mathbf{E}(x, y, z, t)$ and $\mathbf{B}(x, y, z, t)$. These fields depend on charges, currents and each other according to the Maxwell's equations which, in SI units, read

$$\nabla \cdot \mathbf{E} = \frac{\rho}{\varepsilon} , \qquad \text{(B.1a)}$$

$$\nabla \cdot \mathbf{B} = 0 , \qquad \text{(B.1b)}$$

$$\nabla \times \mathbf{E} = -\frac{\partial \mathbf{B}}{\partial t} , \qquad \text{(B.1c)}$$

$$\nabla \times \mathbf{B} = \mu \mathbf{j} + \mu \varepsilon \frac{\partial \mathbf{E}}{\partial t} , \qquad \text{(B.1d)}$$

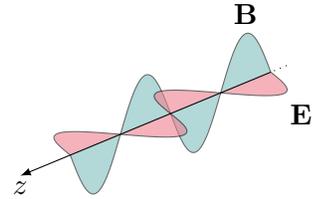

Figure B.1: Electric and magnetic fields for a wave propagating along $z$.

where $\rho$ is the charge density, $\varepsilon$ is the electric constant, $\mu$ is the magnetic constant, and $\mathbf{j}$ is the current density. In general, $\rho$ and $\mathbf{j}$ can depend on both time and position. Solutions to the Maxwell equations in vacuum can be expressed as a superposition of plane waves

$$\mathbf{E}(\mathbf{r}, t) = \mathbf{A}_{\text{EM}} \cos(\mathbf{k} \cdot \mathbf{r} - \omega t + \phi) , \qquad \text{(B.2a)}$$

$$\mathbf{B}(\mathbf{r}, t) = (\hat{k} \times \mathbf{A}_{\text{EM}}) \cos(\mathbf{k} \cdot \mathbf{r} - \omega t + \phi) , \qquad \text{(B.2b)}$$

with $\omega = kc \equiv k \sqrt{\mu_0 \varepsilon_0}$ and $\mathbf{A}_{\text{EM}} \cdot \mathbf{k} = 0$, over all possible values of amplitude $\mathbf{A}_{\text{EM}}$, phase $\phi$, and wave vector $\mathbf{k}$. These plane wave solutions consist of oscillating electric and magnetic fields perpendicular to each other and to $\mathbf{k}$ (see Fig. B.1).

Instead of the most general solution, consider the superposition of plane waves (B.2) with the same $\mathbf{k}$. Depending on the behavior of the resulting electric field $\mathbf{E}_{\text{tot}}$, defined as the sum of the electric fields of all the superposed waves, the radiation is said to be

*Unpolarized* if $\mathbf{E}_{\text{tot}}$ changes randomly in time at some point in space;

*Lineraly polarized* if $\mathbf{E}_{\text{tot}}$ is constant in time (this is the case for a single plane wave);

*Circularly polarized* if $\mathbf{E}_{\text{tot}}$ draws a circle on the plane perpendicular to $\mathbf{k}$;

*Elliptically polarized* if $\mathbf{E}_{\text{tot}}$ draws an ellipse on the plane perpendicular to $\mathbf{k}$.

These pure polarization states are often represented as in Figure B.2. If the radiation has a polarized and an unpolarized component, it is partially polarized.

In problems involving polarized light, it is often necessary to determine the effect of various types of polarizers (linear, circular, elliptical, etc.), rotators, retardation plates, and other optical elements on the state of polarization of a light beam. Rather than

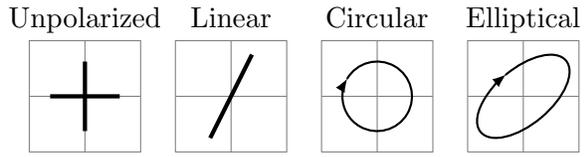

Unpolarized   Linear   Circular   Elliptical

Figure B.2: Two orthogonal rods denote no polarization. Linear polarization is represented by a rod directed as $\mathbf{E}_{\text{tot}}$. For circular and elliptical polarization, an arrow specifies if $\mathbf{E}_{\text{tot}}$ is rotating clock- or counterclock-wise.

using Maxwell's equations to study how the electromagnetic field propagates through an optical element, it is often more convenient to use matrix methods, which are based on the fact that the effect of a polarizer or retarder is to perform a linear transformation (represented by a matrix) on the vector representation of a polarized light beam. The advantage of these methods over conventional techniques is that problems are reduced to simple matrix operations.

## B.2   Jones calculus

The Jones calculus is one of the most common matrix methods. To define the Jones vector representation, consider a plane wave given by a superposition of solutions (B.2) with the same $\mathbf{k} \propto \hat{z}$. At a given $z$, the electric fields has components

$$E_x(t) = A_{\text{EM},x} \cos[kz - \omega t + \phi_x], \tag{B.3a}$$

$$E_y(t) = A_{\text{EM},y} \cos[kz - \omega t + \phi_y]. \tag{B.3b}$$

We define the complex electric field, $\boldsymbol{\mathcal{E}}(t)$, with components $\mathcal{E}_i(t) = A_{\text{EM},i}\, e^{i(kz - \omega t + \phi_i)}$, such that $E_i = \text{Re}\,\mathcal{E}_i$. The complex electric field can be factorized as

$$\boldsymbol{\mathcal{E}}(t) = \begin{pmatrix} \mathcal{E}_x(t) \\ \mathcal{E}_y(t) \end{pmatrix} = \begin{pmatrix} A_{\text{EM},x} e^{i(kz - \omega t + \phi_x)} \\ A_{\text{EM},y} e^{i(kz - \omega t + \phi_y)} \end{pmatrix} = \begin{pmatrix} A_{\text{EM},x} e^{i\phi_x} \\ A_{\text{EM},y} e^{i\phi_y} \end{pmatrix} e^{i(kz - \omega t)} \equiv \mathbf{J}\, e^{i(kz - \omega t)}, \tag{B.4}$$

where $\mathbf{J}$ is the Jones vector, representing the amplitude and phase of the electric field in the $x$ and $y$ directions.

The effect of optical elements on the Jones vector is described by a Jones matrix, $\mathcal{J}$. For instance, the Jones matrix for an ideal HWP is

$$\mathcal{J}_{\text{HWP}} = \begin{pmatrix} 1 & 0 \\ 0 & -1 \end{pmatrix}. \tag{B.5}$$

# B.3  Mueller calculus

Another widely used matrix method is the Mueller calculus, where the polarization state is described by a Stokes vector, $\mathbf{S}$, and the effect of an optical element is expressed as a Mueller matrix, $\mathcal{M}$.

## B.3.1  Stokes vectors

Again, consider a plane wave given by a superposition of solutions (B.2) with the same $\mathbf{k} \propto \hat{z}$. At a given $z$, the components of the complex electric can be written as

$$\mathcal{E}_x(t) = A_{\text{EM},x} e^{i(kz - \omega t + \phi_x)}, \tag{B.6a}$$

$$\mathcal{E}_y(t) = A_{\text{EM},y} e^{i(kz - \omega t + \phi_y)}. \tag{B.6b}$$

We define the $2 \times 2$ Hermitian tensor $\langle \mathcal{E}_i \mathcal{E}_j^* \rangle$, where the angle brackets denote a temporal average over some periods of the wave, and decompose it as a combination of the identity matrix and the Pauli matrices

$$\langle \mathcal{E}_i \mathcal{E}_j^* \rangle = \frac{1}{2} I \delta_{ij} + \frac{1}{2} (U \sigma_1 + V \sigma_2 + Q \sigma_3)_{ij} = \frac{1}{2} \begin{pmatrix} I + Q & U - iV \\ U + iV & I - Q \end{pmatrix}_{ij}, \tag{B.7}$$

where $I$, $Q$, $U$ and $V$ are the Stokes parameters:

$$\begin{aligned} I = \text{Tr}(\langle \mathcal{E}_i \mathcal{E}_j^* \rangle) &= \langle |\mathcal{E}_x|^2 \rangle + \langle |\mathcal{E}_y|^2 \rangle \\ &= A_{\text{EM},x}^2 + A_{\text{EM},y}^2, \end{aligned} \tag{B.8a}$$

$$\begin{aligned} Q = \text{Tr}(\langle \mathcal{E}_i \mathcal{E}_j^* \rangle \sigma_3) &= \langle |\mathcal{E}_x|^2 \rangle - \langle |\mathcal{E}_y|^2 \rangle \\ &= A_{\text{EM},x}^2 - A_{\text{EM},y}^2, \end{aligned} \tag{B.8b}$$

$$\begin{aligned} U = \text{Tr}(\langle \mathcal{E}_i \mathcal{E}_j^* \rangle \sigma_1) &= 2 \,\text{Re} \langle \mathcal{E}_x^* \mathcal{E}_y \rangle \\ &= 2 A_{\text{EM},x} A_{\text{EM},y} \cos(\theta_x - \theta_y), \end{aligned} \tag{B.8c}$$

$$\begin{aligned} V = \text{Tr}(\langle \mathcal{E}_i \mathcal{E}_j^* \rangle \sigma_2) &= 2 \,\text{Im} \langle \mathcal{E}_x^* \mathcal{E}_y \rangle \\ &= 2 A_{\text{EM},x} A_{\text{EM},y} \sin(\phi_x - \phi_y). \end{aligned} \tag{B.8d}$$

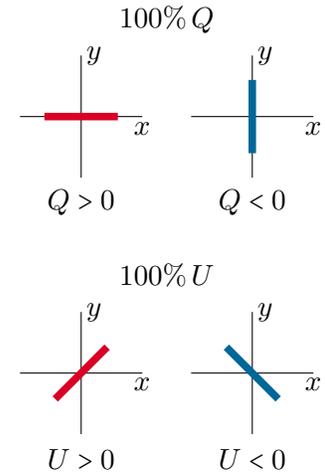

Figure B.3: Pure $Q$ and $U$ polarization states for a wave propagating along $z$.

$I$ is proportional to the intensity of the radiation, $Q$ and $U$ describe linear polarization as sketched in Figure B.3, and $V$ describes circular polarization.

The Stokes parameters can be organized in a vector

$$\mathbf{S} = \begin{pmatrix} I \\ Q \\ U \\ V \end{pmatrix}, \tag{B.9}$$

called Stokes vector. In Mueller calculus, Stokes vectors are used to describe the polarization state of radiation.

## B.3.2 Mueller matrices

The Mueller matrix $\mathcal{M}$ for a polarization-altering device is defined as the matrix which transforms an incident Stokes vector $\mathbf{S}$ into the outgoing Stokes vector $\mathbf{S}'$

$$\mathbf{S}' = \mathcal{M}\mathbf{S} = \begin{pmatrix} \mathcal{M}_{\mathrm{II}} & \mathcal{M}_{\mathrm{IQ}} & \mathcal{M}_{\mathrm{IU}} & \mathcal{M}_{\mathrm{IV}} \\ \mathcal{M}_{\mathrm{QI}} & \mathcal{M}_{\mathrm{QQ}} & \mathcal{M}_{\mathrm{QU}} & \mathcal{M}_{\mathrm{QV}} \\ \mathcal{M}_{\mathrm{UI}} & \mathcal{M}_{\mathrm{UQ}} & \mathcal{M}_{\mathrm{UU}} & \mathcal{M}_{\mathrm{UV}} \\ \mathcal{M}_{\mathrm{VI}} & \mathcal{M}_{\mathrm{VQ}} & \mathcal{M}_{\mathrm{VU}} & \mathcal{M}_{\mathrm{VV}} \end{pmatrix} \begin{pmatrix} I \\ Q \\ U \\ V \end{pmatrix}. \tag{B.10}$$

The Mueller matrix contains within its elements all of the polarization properties: diattenuation, retardance and depolarization, as well as their form, either linear, circular, or elliptical. When the Mueller matrix is known, the exiting polarization state is known for an arbitrary incident polarization state.

The Mueller matrix $\mathcal{M}$ associated with a cascade of polarization elements $q = 1, 2, \ldots, Q$ is the right-to-left product of the individual matrices $\mathcal{M}_q$:

$$\mathcal{M} = \mathcal{M}_Q \mathcal{M}_{Q-1} \ldots \mathcal{M}_2 \mathcal{M}_1 = \prod_{q=1}^{Q} \mathcal{M}_q. \tag{B.11}$$

Here we list a few common Mueller matrices.

**Rotations**  By rotating the $x$ and $y$ axes of an angle $\psi$ around $z$, the components of the complex electric field transform as

$$\begin{pmatrix} \mathcal{E}'_x \\ \mathcal{E}'_y \end{pmatrix} = \begin{pmatrix} \cos\psi & \sin\psi \\ -\sin\psi & \cos\psi \end{pmatrix} \begin{pmatrix} \mathcal{E}_x \\ \mathcal{E}_y \end{pmatrix}. \tag{B.12}$$

Substituting these new expressions into eq. (B.8), we have that

$$\begin{pmatrix} Q' \\ U' \end{pmatrix} = \begin{pmatrix} \cos 2\psi & \sin 2\psi \\ -\sin 2\psi & \cos 2\psi \end{pmatrix} \begin{pmatrix} Q \\ U \end{pmatrix}, \tag{B.13}$$

while $I$ and $V$ are invariant. The Mueller matrix for a coordinate rotation is then

$$\mathcal{R}_\psi = \begin{pmatrix} 1 & 0 & 0 & 0 \\ 0 & \cos 2\psi & \sin 2\psi & 0 \\ 0 & -\sin 2\psi & \cos 2\psi & 0 \\ 0 & 0 & 0 & 1 \end{pmatrix}. \tag{B.14}$$

**Linear polarizer** In general, a polarizer is an optical filter that lets light waves of a specific polarization pass through while blocking light waves of other polarizations. For instance, a linear polarizer with horizontal transmission only lets linear polarization along $x$ pass, and it is described by the Mueller matrix

$$\mathcal{M}_{\mathrm{pol}} = \frac{1}{2} \begin{pmatrix} 1 & 1 & 0 & 0 \\ 1 & 1 & 0 & 0 \\ 0 & 0 & 0 & 0 \\ 0 & 0 & 0 & 0 \end{pmatrix}.$$

(B.15)

**Half-wave plate** A waveplate or retarder is an optical device that alters the polarization state of a light wave traveling through it. In particular, the half-wave plate reflects the polarization vector with respect to the $y$ axis, and is described by the Mueller matrix

$$\mathcal{M}_{\mathrm{HWP}} = \begin{pmatrix} 1 & 0 & 0 & 0 \\ 0 & 1 & 0 & 0 \\ 0 & 0 & -1 & 0 \\ 0 & 0 & 0 & -1 \end{pmatrix}.$$

(B.16)

Eq. (B.16) describes the effect of an idealized HWP. In reality, there can be a number of effects that result in deviations from the ideal Mueller matrix elements, and we can write

$$\mathcal{M}_{\mathrm{HWP}} = \begin{pmatrix} m_{\mathrm{II}} & m_{\mathrm{IQ}} & m_{\mathrm{IU}} & m_{\mathrm{IV}} \\ m_{\mathrm{QI}} & m_{\mathrm{QQ}} & m_{\mathrm{QU}} & m_{\mathrm{QV}} \\ m_{\mathrm{UI}} & m_{\mathrm{UQ}} & m_{\mathrm{UU}} & m_{\mathrm{UV}} \\ m_{\mathrm{VI}} & m_{\mathrm{VQ}} & m_{\mathrm{VU}} & m_{\mathrm{VV}} \end{pmatrix}.$$

(B.17)

We refer to deviations $\mathcal{M}_{\mathrm{HWP}} - \mathrm{diag}(1, 1, -1, -1)$ as HWP non-idealities.

## B.4 Relating Mueller to Jones parameters

Mueller and Jones calculus are two different matrix methods to describe and manipulate polarized radiation. Mueller calculus works with intensities, while Jones calculus works directly with the $x$ and $y$ components of the electric field. Any Jones matrix, $J$, can be transformed into the corresponding Mueller–Jones matrix $\mathcal{M} = A \left( J \otimes J^* \right) A^{-1}$, where

$$A = \begin{pmatrix} 1 & 0 & 0 & 1 \\ 1 & 0 & 0 & -1 \\ 0 & 1 & 1 & 0 \\ 0 & i & -i & 0 \end{pmatrix}.$$

(B.18)

Here, $*$ denotes the complex conjugate and $\otimes$ is the Kronecker product. The Jones matrix for a non-ideal HWP is

$$J_{\mathrm{HWP}} = \begin{pmatrix} 1 + h_1 & \zeta_1 e^{i\chi_1} \\ \zeta_2 e^{i\chi_2} & -(1 + h_2) e^{i\beta} \end{pmatrix},$$

(B.19)

where $h_{1,2}$ are loss parameters describing the deviation from the unitary transmission of $E_{x,y}$; $\beta$ parametrizes the deviation from $\pi$ of the phase shift between $E_x$ and $E_y$; $\zeta_{1,2}$ and $\xi_{1,2}$ describe the amplitude and phase of the cross-polarization coupling. All Jones parameters tend to zero in the ideal limit.

# Appendix C

# Derivations for Chapter 6

**Summary:** In this appendix, we provide some explicit derivations of the results presented in Chapter 6.

*Sections C.5 and C.6 are adapted from the appendix of [78].*

Note that we neglect circular polarization here for the sake of compactness (and because we ignore it in Chapter 6), but these results can be easily generalized to include $V$.

## C.1   Mueller matrices in CMB and IAU conventions

Stokes vectors defined in CMB convention, $\mathbf{S}_{(\text{CMB})}$, can be transformed into their IAU counterpart, $\mathbf{S}_{(\text{IAU})}$, by flipping the sign of the $U$ parameter. In other words

$$\mathbf{S}_{(\text{IAU})} = \begin{pmatrix} 1 & 0 & 0 \\ 0 & 1 & 0 \\ 0 & 0 & -1 \end{pmatrix} \mathbf{S}_{(\text{CMB})}. \tag{C.1}$$

Now, assume to know a HWP Mueller matrix, $\mathcal{M}_{(\text{IAU})}$, that acts on Stokes vectors defined in IAU convention. Using eq. (C.1), the identity $\mathbf{S}'_{(\text{IAU})} = \mathcal{M}_{(\text{IAU})}\mathbf{S}_{(\text{IAU})}$ becomes

$$\begin{pmatrix} 1 & 0 & 0 \\ 0 & 1 & 0 \\ 0 & 0 & -1 \end{pmatrix} \mathbf{S}'_{(\text{CMB})} = \mathcal{M}_{(\text{IAU})} \begin{pmatrix} 1 & 0 & 0 \\ 0 & 1 & 0 \\ 0 & 0 & -1 \end{pmatrix} \mathbf{S}_{(\text{CMB})}. \tag{C.2}$$

By multiplying both sides by $\text{diag}(1, 1, -1)$, we get $\mathbf{S}'_{(\text{CMB})} = \mathcal{M}_{(\text{CMB})}\mathbf{S}_{(\text{CMB})}$, where

$$\mathcal{M}_{(\text{CMB})} \equiv \begin{pmatrix} 1 & 0 & 0 \\ 0 & 1 & 0 \\ 0 & 0 & -1 \end{pmatrix} \mathcal{M}_{(\text{IAU})} \begin{pmatrix} 1 & 0 & 0 \\ 0 & 1 & 0 \\ 0 & 0 & -1 \end{pmatrix}. \tag{C.3}$$

By denoting the elements of the Mueller matrix $\mathcal{M}_{(\mathrm{IAU})}$ in IAU convention as $m_{\mathrm{ss'}}$, i.e.

$$\mathcal{M}_{(\mathrm{IAU})} = \begin{pmatrix} m_{\mathrm{II}} & m_{\mathrm{IQ}} & m_{\mathrm{IU}} \\ m_{\mathrm{QI}} & m_{\mathrm{QQ}} & m_{\mathrm{QU}} \\ m_{\mathrm{UI}} & m_{\mathrm{UQ}} & m_{\mathrm{UU}} \end{pmatrix}, \tag{C.4}$$

and performing the scalar products of eq. (C.3), we get

$$\mathcal{M}_{(\mathrm{CMB})} = \begin{pmatrix} m_{\mathrm{II}} & m_{\mathrm{IQ}} & -m_{\mathrm{IU}} \\ m_{\mathrm{QI}} & m_{\mathrm{QQ}} & -m_{\mathrm{QU}} \\ -m_{\mathrm{UI}} & -m_{\mathrm{UQ}} & m_{\mathrm{UU}} \end{pmatrix}. \tag{C.5}$$

## C.2   Derivation of eq. (6.8)

Given an incoming Stokes vector, $\mathbf{S}$, the signal $d$ detected by an ideal detector sensitive to the polarization direction that forms an angle $\xi$ wuith the $x$ telescope axis can be modelled as $d = \mathbf{a}^T \mathcal{R}_\xi \mathcal{M} \mathbf{S}$, where $\mathbf{a}^T = \frac{1}{2} \begin{pmatrix} 1 & 1 & 0 & 0 \end{pmatrix}$, and $\mathcal{R}_\xi$ represents a rotation from the telescope to the detector coordinates. As for $\mathcal{M}$, its explicit form will depend on the telescope optics. For now, let us simply denote its elements as

$$\mathcal{M} = \begin{pmatrix} \mathcal{M}_{\mathrm{II}} & \mathcal{M}_{\mathrm{IQ}} & \mathcal{M}_{\mathrm{IU}} \\ \mathcal{M}_{\mathrm{QI}} & \mathcal{M}_{\mathrm{QQ}} & \mathcal{M}_{\mathrm{QU}} \\ \mathcal{M}_{\mathrm{UI}} & \mathcal{M}_{\mathrm{UQ}} & \mathcal{M}_{\mathrm{UU}} \end{pmatrix}. \tag{C.6}$$

Consider now a set of 4 detectors sensitive to different polarization directions, with 0, 90, 45 and 135 degrees offsets, and assume them to be observing the same pixel. The signals they measure can be modeled as

$$\begin{pmatrix} d^{(0)} \\ d^{(90)} \\ d^{(45)} \\ d^{(135)} \end{pmatrix} = \frac{1}{2} \begin{pmatrix} \mathcal{M}_{\mathrm{II}} + \mathcal{M}_{\mathrm{QI}} & \mathcal{M}_{\mathrm{IQ}} + \mathcal{M}_{\mathrm{QQ}} & \mathcal{M}_{\mathrm{IU}} + \mathcal{M}_{\mathrm{QU}} \\ \mathcal{M}_{\mathrm{II}} - \mathcal{M}_{\mathrm{QI}} & \mathcal{M}_{\mathrm{IQ}} - \mathcal{M}_{\mathrm{QQ}} & \mathcal{M}_{\mathrm{IU}} - \mathcal{M}_{\mathrm{QU}} \\ \mathcal{M}_{\mathrm{II}} + \mathcal{M}_{\mathrm{UI}} & \mathcal{M}_{\mathrm{IQ}} + \mathcal{M}_{\mathrm{UQ}} & \mathcal{M}_{\mathrm{IU}} + \mathcal{M}_{\mathrm{UU}} \\ \mathcal{M}_{\mathrm{II}} - \mathcal{M}_{\mathrm{UI}} & \mathcal{M}_{\mathrm{IQ}} - \mathcal{M}_{\mathrm{UQ}} & \mathcal{M}_{\mathrm{IU}} - \mathcal{M}_{\mathrm{UU}} \end{pmatrix} \mathbf{S}. \tag{C.7}$$

Eq. (C.7) models the minimal TOD constituted by the four detector readings, and the above matrix can be thought as a "response matrix". For the sake of compactness, we will denote it by $\mathcal{A}$ from now on.

As a concrete case, we assume an HWP to be the first optical element in the telescope chain, so that $\mathcal{M} \equiv \mathcal{R}_{-\phi} \mathcal{M}_{\mathrm{HWP}} \mathcal{R}_{\phi+\psi}$, where $\mathcal{M}_{\mathrm{HWP}}$ denotes the Mueller matrix of the HWP. We reconstruct the sky signal via a binning map-maker that neglects the HWP non-idealities. The reconstructed Stokes vector will read

$$\widehat{\mathbf{S}} = (\mathcal{A}_{\mathrm{ideal}}^T \mathcal{A}_{\mathrm{ideal}})^{-1} \mathcal{A}_{\mathrm{ideal}}^T \mathcal{A} \mathbf{S}, \tag{C.8}$$

By evaluating $\widehat{\mathbf{S}}$ explicitly with `Mathematica`[1], we get eq. (6.8).

---

[1] Although a bit tedious, the derivation can also be done with pen and paper. It helps noticing that

## C.3 Alternative derivation of eqs. (6.10)

Here we provide a derivation of eqs. (6.10) that does not rely on the four-detector configuration. We start from eq. (3.20) for the binning map-maker, and write explicitly the detected signal as $d_{it} = \mathbb{S}_{it}^T \mathbf{S}_p$:

$$
\begin{pmatrix} \widehat{I} \\ \widehat{Q} \\ \widehat{U} \end{pmatrix}_p = \left[ \sum_{i't' \in \{it\}_p} \begin{pmatrix} \widehat{\mathbb{I}}^2 & \widehat{\mathbb{I}}\widehat{\mathbb{Q}} & \widehat{\mathbb{I}}\widehat{\mathbb{U}} \\ \widehat{\mathbb{Q}}\widehat{\mathbb{I}} & \widehat{\mathbb{Q}}^2 & \widehat{\mathbb{Q}}\widehat{\mathbb{U}} \\ \widehat{\mathbb{U}}\widehat{\mathbb{I}} & \widehat{\mathbb{U}}\widehat{\mathbb{Q}} & \widehat{\mathbb{U}}^2 \end{pmatrix}_{i't'} \right]^{-1} \sum_{it \in \{it\}_p} \begin{pmatrix} \widehat{\mathbb{I}}\mathbb{I} & \widehat{\mathbb{I}}\mathbb{Q} & \widehat{\mathbb{I}}\mathbb{U} \\ \widehat{\mathbb{Q}}\mathbb{I} & \widehat{\mathbb{Q}}\mathbb{Q} & \widehat{\mathbb{Q}}\mathbb{U} \\ \widehat{\mathbb{U}}\mathbb{I} & \widehat{\mathbb{U}}\mathbb{Q} & \widehat{\mathbb{U}}\mathbb{U} \end{pmatrix}_{it} \begin{pmatrix} I \\ Q \\ U \end{pmatrix}_p . \tag{C.10}
$$

The Stokes vector $\mathbb{S}_{it}$ encodes the response of the detector $i$ at time $t$, while $\widehat{\mathbb{S}}_{it}$ is the response Stokes vector assumed by the binning map-maker.

We apply eq. (C.10) to the following concrete case: a telescope provided with a non-ideal HWP whose TOD is processed through an ideal binning map-maker. In this case, $\mathbb{S}_{it}$ and $\widehat{\mathbb{S}}_{it}$ read

$$
\mathbb{S}_{it} = \frac{1}{2} \begin{pmatrix} m_{\mathrm{II}} + m_{\mathrm{QI}} c_\beta + m_{\mathrm{UI}} s_\beta \\ m_{\mathrm{IQ}} c_\alpha - m_{\mathrm{IU}} s_\alpha + (m_{\mathrm{QQ}} c_\alpha - m_{\mathrm{QU}} s_\alpha) c_\beta + (m_{\mathrm{UQ}} c_\alpha - m_{\mathrm{UU}} s_\alpha) s_\beta \\ m_{\mathrm{IQ}} s_\alpha + m_{\mathrm{IU}} c_\alpha + (m_{\mathrm{QQ}} s_\alpha + m_{\mathrm{QU}} c_\alpha) c_\beta + (m_{\mathrm{UQ}} s_\alpha + m_{\mathrm{UU}} c_\alpha) s_\beta \end{pmatrix} , \tag{C.11a}
$$

$$
\widehat{\mathbb{S}}_{it} = \left[ \frac{1}{2} \begin{pmatrix} 1 & 1 & 0 \end{pmatrix} \mathcal{R}_{\xi_i - \phi_t} \mathcal{M}_{\mathrm{ideal}} \mathcal{R}_{\psi_t - \phi_t} \right]^T = \frac{1}{2} \begin{pmatrix} 1 \\ c_{\alpha+\beta} \\ s_{\alpha+\beta} \end{pmatrix} , \tag{C.11b}
$$

where we adopt the compact notation $c_\theta = \cos(2\theta)$ and $s_\theta = \sin(2\theta)$. By plugging these expression into eq. (C.10), assuming that the $\alpha$ and $\beta$ angles are sampled uniformly enough, and using the orthogonality of sine and cosine, we obtain

$$
\begin{pmatrix} \widehat{I} \\ \widehat{Q} \\ \widehat{U} \end{pmatrix}_p = \left[ \sum_{i't' \in \{it\}_p} \begin{pmatrix} 1 & c_{\alpha+\beta} & s_{\alpha+\beta} \\ c_{\alpha+\beta} & c_{\alpha+\beta}^2 & c_{\alpha+\beta} s_{\alpha+\beta} \\ s_{\alpha+\beta} & s_{\alpha+\beta} c_{\alpha+\beta} & s_{\alpha+\beta}^2 \end{pmatrix}_{i't'} \right]^{-1} \sum_{it \in \{it\}_p} \begin{pmatrix} \mathbb{I} & \mathbb{Q} & \mathbb{U} \\ c_{\alpha+\beta}\mathbb{I} & c_{\alpha+\beta}\mathbb{Q} & c_{\alpha+\beta}\mathbb{U} \\ s_{\alpha+\beta}\mathbb{I} & s_{\alpha+\beta}\mathbb{Q} & s_{\alpha+\beta}\mathbb{U} \end{pmatrix}_{it} \begin{pmatrix} I \\ Q \\ U \end{pmatrix}_p
$$

$$
\simeq \left[ \begin{pmatrix} \lfloor 1 \rfloor & 0 & 0 \\ 0 & \lfloor c_{\alpha+\beta}^2 \rfloor & 0 \\ 0 & 0 & \lfloor s_{\alpha+\beta}^2 \rfloor \end{pmatrix} \right]^{-1} \begin{pmatrix} \lfloor m_{\mathrm{II}} \rfloor & 0 & 0 \\ 0 & \lfloor c_{\alpha+\beta}^2 \rfloor (m_{\mathrm{QQ}} - m_{\mathrm{UU}})/2 & \lfloor c_{\alpha+\beta}^2 \rfloor (m_{\mathrm{QU}} + m_{\mathrm{UQ}})/2 \\ 0 & -\lfloor s_{\alpha+\beta}^2 \rfloor (m_{\mathrm{QU}} + m_{\mathrm{UQ}})/2 & \lfloor s_{\alpha+\beta}^2 \rfloor (m_{\mathrm{QQ}} - m_{\mathrm{UU}})/2 \end{pmatrix} \begin{pmatrix} I \\ Q \\ U \end{pmatrix}_p \tag{C.12}
$$

where $\lfloor \cdot \rfloor$ compactly denotes the sum over all the observations of the pixel $p$, i.e. $\sum_{it \in \{it\}_p}$. By evaluating eq. (C.12) explictly, it reduces to eqs. (6.10).

---

both response matrices satisfy $\mathcal{A} = A\mathcal{M}$, with

$$
A \equiv \frac{1}{2} \begin{pmatrix} 1 & 1 & 0 \\ 1 & -1 & 0 \\ 1 & 0 & 1 \\ 1 & 0 & -1 \end{pmatrix} , \tag{C.9}
$$

and that the $A^T A$ product reduces to $\mathrm{diag}(1, 1/2, 1/2)$.

## C.4   Derivation of eq. (6.13)

By definition, the variance associated to $C_{\ell,\mathrm{obs}}^{XY}$ is

$$\mathrm{Var}\!\left(C_{\ell,\mathrm{obs}}^{XY}\right) = \left\langle\left(C_{\ell,\mathrm{obs}}^{XY}\right)^2\right\rangle - \left\langle C_{\ell,\mathrm{obs}}^{XY}\right\rangle^2 . \tag{C.13}$$

The first term on the right-hand side can be expressed as

$$\left\langle\left(C_{\ell,\mathrm{obs}}^{XY}\right)^2\right\rangle = \frac{1}{(2\ell+1)^2}\sum_{mm'}\left\langle X_{\ell m}Y_{\ell m}^* X_{\ell m'}^* Y_{\ell m'}\right\rangle . \tag{C.14}$$

Under the assumption that the $X_{\ell m}$ are Gaussian random variables, the expectation value $\left\langle X_{\ell m}Y_{\ell m}^* X_{\ell m'}^* Y_{\ell m'}\right\rangle$ can be rewritten by making use of Isserlis' theorem [160] (also referred to in the literature as Wick's probability theorem):

$$\left\langle X_{\ell m}Y_{\ell m}^*\right\rangle\left\langle X_{\ell m'}^* Y_{\ell m'}\right\rangle + \left\langle X_{\ell m}X_{\ell m'}^*\right\rangle\left\langle Y_{\ell m}^* Y_{\ell m'}\right\rangle + \left\langle X_{\ell m}Y_{\ell m'}\right\rangle\left\langle Y_{\ell m}^* X_{\ell m'}^*\right\rangle , \tag{C.15}$$

which, when plugged back in (C.13), gives

$$\begin{aligned}
\mathrm{Var}\!\left(C_{\ell,\mathrm{obs}}^{XY}\right) &= \frac{1}{(2\ell+1)^2}\sum_{mm'}\left\langle X_{\ell m}X_{\ell m'}^*\right\rangle\left\langle Y_{\ell m}^* Y_{\ell m'}\right\rangle + \frac{1}{(2\ell+1)^2}\sum_{mm'}\left\langle X_{\ell m}Y_{\ell m'}\right\rangle\left\langle Y_{\ell m}^* X_{\ell m'}^*\right\rangle \\
&= \frac{1}{(2\ell+1)^2}\sum_{mm'}C_\ell^{XX}C_\ell^{YY}\delta_{mm'}\delta_{mm'} + \frac{1}{(2\ell+1)^2}\sum_{mm'}\left(C_\ell^{XY}\right)^2\delta_{m-m'}\delta_{m-m'}(-1)^{m-m'} \\
&= \frac{1}{2\ell+1}\left[C_\ell^{XX}C_\ell^{YY} + \left(C_\ell^{XY}\right)^2\right] ,
\end{aligned} \tag{C.16}$$

where $C_\ell^{XY}$ represents the theoretical angular power spectra. Estimating

$$\mathrm{Var}\!\left(C_{\ell,\mathrm{obs}}^{XY}\right) \simeq \frac{1}{2\ell+1}\left[C_{\ell,\mathrm{obs}}^{XX}C_{\ell,\mathrm{obs}}^{YY} + \left(C_{\ell,\mathrm{obs}}^{XY}\right)^2\right] , \tag{C.17}$$

is typically a good approximation for $\ell \gtrsim 10$, where the sample of $m$ is large enough.

## C.5   Derivation of eq. (6.16)

Taking into account the frequency dependence of both the HWP Mueller matrix elements and the CMB signal, we write the data model of eq. (6.4) as

$$d = \mathbf{a}^T \mathcal{M}_{\mathrm{det}}\mathcal{R}_{\xi-\phi}\int \mathrm{d}\nu\, \mathcal{M}_{\mathrm{HWP}}(\nu)\mathcal{R}_{\phi+\psi}\mathbf{S}(\nu) + n . \tag{C.18}$$

Repeating the analysis presented in section 6.4, eq. (6.11) reads

$$\widehat{C}_\ell^{TT} \simeq \langle m_{\text{II}} \rangle^2 \bar{C}_{\ell,\text{in}}^{TT}, \tag{C.19a}$$

$$\widehat{C}_\ell^{EE} \simeq \frac{\langle m_{\text{QQ}} - m_{\text{UU}} \rangle^2}{4} \bar{C}_{\ell,\text{in}}^{EE} + \frac{\langle m_{\text{QU}} + m_{\text{UQ}} \rangle^2}{4} \bar{C}_{\ell,\text{in}}^{BB} + \frac{\langle m_{\text{QQ}} - m_{\text{UU}} \rangle \langle m_{\text{QU}} + m_{\text{UQ}} \rangle}{2} \bar{C}_{\ell,\text{in}}^{EB}, \tag{C.19b}$$

$$\widehat{C}_\ell^{BB} \simeq \frac{\langle m_{\text{QQ}} - m_{\text{UU}} \rangle^2}{4} \bar{C}_{\ell,\text{in}}^{BB} + \frac{\langle m_{\text{QU}} + m_{\text{UQ}} \rangle^2}{4} \bar{C}_{\ell,\text{in}}^{EE} - \frac{\langle m_{\text{QQ}} - m_{\text{UU}} \rangle \langle m_{\text{QU}} + m_{\text{UQ}} \rangle}{2} \bar{C}_{\ell,\text{in}}^{EB}, \tag{C.19c}$$

$$\widehat{C}_\ell^{TE} \simeq \frac{\langle m_{\text{II}} \rangle \langle m_{\text{QQ}} - m_{\text{UU}} \rangle}{2} \bar{C}_{\ell,\text{in}}^{TE} + \frac{\langle m_{\text{II}} \rangle \langle m_{\text{QU}} + m_{\text{UQ}} \rangle}{2} \bar{C}_{\ell,\text{in}}^{TB}, \tag{C.19d}$$

$$\widehat{C}_\ell^{EB} \simeq \frac{\langle m_{\text{QQ}} - m_{\text{UU}} \rangle^2 - \langle m_{\text{QU}} + m_{\text{UQ}} \rangle^2}{4} \bar{C}_{\ell,\text{in}}^{EB} - \frac{\langle m_{\text{QQ}} - m_{\text{UU}} \rangle \langle m_{\text{QU}} + m_{\text{UQ}} \rangle}{4} (\bar{C}_{\ell,\text{in}}^{EE} - \bar{C}_{\ell,\text{in}}^{BB}), \tag{C.19e}$$

$$\widehat{C}_\ell^{TB} \simeq \frac{\langle m_{\text{II}} \rangle \langle m_{\text{QQ}} - m_{\text{UU}} \rangle}{2} \bar{C}_{\ell,\text{in}}^{TB} - \frac{\langle m_{\text{II}} \rangle \langle m_{\text{QU}} + m_{\text{UQ}} \rangle}{2} \bar{C}_{\ell,\text{in}}^{TE}, \tag{C.19f}$$

where the brackets denote frequency integrals weighted over the SED of the CMB,

$$\langle f \rangle \equiv \frac{\int \mathrm{d}\nu \, S_{\text{CMB}}(\nu) f(\nu)}{\int \mathrm{d}\nu \, S_{\text{CMB}}(\nu)}, \tag{C.20}$$

and $\bar{C}_{\ell,\text{in}}^{XY}$ the input angular power spectra at some reference frequency $\bar{\nu}$. This modifies eq. (6.15) to

$$\widehat{\theta} = -\frac{1}{2} \arctan\left( \frac{\int \mathrm{d}\nu \, S_{\text{CMB}}(\nu) \left[ m_{\text{QU}} + m_{\text{UQ}} \right](\nu)}{\int \mathrm{d}\nu \, S_{\text{CMB}}(\nu) \left[ m_{\text{QQ}} - m_{\text{UU}} \right](\nu)} \right). \tag{C.21}$$

# C.6   Derivation of eq. (6.17)

So far, we neglected any miscalibration angles in the map-maker, i.e. we assumed the response matrix $\widehat{A}$ to encode the true values of the telescope, HWP, and detector angles: $\widehat{\psi} \equiv \psi$, $\widehat{\phi} \equiv \phi$, and $\widehat{\xi} \equiv \xi$, where the hat denotes the values assumed by the map-maker. We now consider a more general case by allowing for deviations: $\psi = \widehat{\psi} + \delta\psi$, $\phi = \widehat{\phi} + \delta\phi$, and $\xi = \widehat{\xi} + \delta\xi$.

**Single frequency**   Repeating the analysis presented in section 6.4 with miscalibration angles, eq. (6.10) reads

$$\widehat{I} \simeq m_{\text{II}} I_{\text{in}}, \tag{C.22a}$$

$$\widehat{Q} \simeq \left[ \cos(2\delta\theta)(m_{\text{QQ}} - m_{\text{UU}}) + \sin(2\delta\theta)(m_{\text{QU}} + m_{\text{UQ}}) \right] Q_{\text{in}}/2$$
$$+ \left[ \cos(2\delta\theta)(m_{\text{QU}} + m_{\text{UQ}}) - \sin(2\delta\theta)(m_{\text{QQ}} - m_{\text{UU}}) \right] U_{\text{in}}/2, \tag{C.22b}$$

$$\widehat{U} \simeq \left[ \cos(2\delta\theta)(m_{\text{QQ}} - m_{\text{UU}}) + \sin(2\delta\theta)(m_{\text{QU}} + m_{\text{UQ}}) \right] U_{\text{in}}/2$$
$$- \left[ \cos(2\delta\theta)(m_{\text{QU}} + m_{\text{UQ}}) + \sin(2\delta\theta)(m_{\text{QQ}} - m_{\text{UU}}) \right] Q_{\text{in}}/2, \tag{C.22c}$$

where $\delta\theta \equiv \delta\xi - \delta\psi - 2\delta\phi$. This modifies eq. (6.15) to

$$
\begin{aligned}
\widehat{\theta} &= -\frac{1}{2}\arctan\left(\frac{\cos(2\delta\theta)(m_{\mathrm{QU}} + m_{\mathrm{UQ}}) - \sin(2\delta\theta)(m_{\mathrm{QQ}} - m_{\mathrm{UU}})}{\cos(2\delta\theta)(m_{\mathrm{QQ}} - m_{\mathrm{UU}}) + \sin(2\delta\theta)(m_{\mathrm{QU}} + m_{\mathrm{UQ}})}\right) \\
&= -\frac{1}{2}\arctan\left(\frac{m_{\mathrm{QU}} + m_{\mathrm{UQ}}}{m_{\mathrm{QQ}} - m_{\mathrm{UU}}}\right) + \delta\theta\,.
\end{aligned}
\tag{C.23}
$$

Therefore, the additional miscalibration angles simply shift $\widehat{\theta}$, as expected.

**Finite frequency bandwidth**   Taking into account a finite frequency bandwidth and miscalibration angles simultaneously is slightly more complicated, but does not spoil the analytic treatment as long as $\delta\theta$ is assumed to be frequency-independent. The generalization of eq. (6.15) in this case reads

$$
\widehat{\theta} = -\frac{1}{2}\arctan\left(\frac{\int \mathrm{d}\nu\, S_{\mathrm{CMB}}(\nu)\left[m_{\mathrm{QU}} + m_{\mathrm{UQ}}\right](\nu)}{\int \mathrm{d}\nu\, S_{\mathrm{CMB}}(\nu)\left[m_{\mathrm{QQ}} - m_{\mathrm{UU}}\right](\nu)}\right) + \delta\theta\,.
\tag{C.24}
$$

# Bibliography

[1] S. Weinberg, *Cosmology*, Oxford University Press (2008).

[2] S. Dodelson and F. Schmidt, *Modern Cosmology*, Elsevier Science (2020).

[3] V. Mukhanov, *Physical Foundations of Cosmology*, Cambridge University Press, Oxford (2005).

[4] A.A. Penzias and R.W. Wilson, *A Measurement of Excess Antenna Temperature at 4080 Mc/s.*, *Astrophys. J.* **142** (1965) 419.

[5] R.H. Dicke, P.J.E. Peebles, P.G. Roll and D.T. Wilkinson, *Cosmic Black-Body Radiation.*, *Astrophys. J.* **142** (1965) 414.

[6] D.J. Fixsen, *The Temperature of the Cosmic Microwave Background*, *Astrophys. J.* **707** (2009) 916 [0911.1955].

[7] R.K. Sachs and A.M. Wolfe, *Perturbations of a Cosmological Model and Angular Variations of the Microwave Background*, *Astrophys. J.* **147** (1967) 73.

[8] M.J. Rees and D.W. Sciama, *Large scale Density Inhomogeneities in the Universe*, *Nature* **217** (1968) 511.

[9] Y.B. Zeldovich and R.A. Sunyaev, *The Interaction of Matter and Radiation in a Hot-Model Universe*, *Astrophys. and Space Science* **4** (1969) 301.

[10] G.F. Smoot, C.L. Bennett, A. Kogut, E.L. Wright, J. Aymon, N.W. Boggess et al., *Structure in the COBE Differential Microwave Radiometer First-Year Maps*, *Astrophys. J. Lett.* **396** (1992) L1.

[11] A.H. Guth, *Inflationary universe: A possible solution to the horizon and flatness problems*, *Phys. Rev. D* **23** (1981) 347.

[12] K. Sato, *First Order Phase Transition of a Vacuum and Expansion of the Universe*, *Mon. Not. Roy. Astron. Soc.* **195** (1981) 467.

[13] A.D. Linde, *A New Inflationary Universe Scenario: A Possible Solution of the Horizon, Flatness, Homogeneity, Isotropy and Primordial Monopole Problems*, *Phys. Lett. B* **108** (1982) 389.

[14] WMAP collaboration, *Nine-Year Wilkinson Microwave Anisotropy Probe (WMAP) Observations: Cosmological Parameter Results*, *Astrophys. J. Suppl.* **208** (2013) 19 [1212.5226].

[15] WMAP collaboration, *Nine-Year Wilkinson Microwave Anisotropy Probe (WMAP) Observations: Final Maps and Results*, *Astrophys. J. Suppl.* **208** (2013) 20 [1212.5225].

[16] WMAP collaboration, *Results from the Wilkinson Microwave Anisotropy Probe*, *PTEP* **2014** (2014) 06B102 [1404.5415].

[17] PLANCK collaboration, *Planck 2018 results. I. Overview and the cosmological legacy of Planck*, *Astron. Astrophys.* **641** (2020) A1 [1807.06205].

[18] PLANCK collaboration, *Planck 2018 results. V. CMB power spectra and likelihoods*, *Astron. Astrophys.* **641** (2020) A5 [1907.12875].

[19] PLANCK collaboration, *Planck 2018 results. VI. Cosmological parameters*, *Astron. Astrophys.* **641** (2020) A6 [1807.06209].

[20] C.J. MacTavish, P.A.R. Ade, J.J. Bock, J.R. Bond, J. Borrill, A. Boscaleri et al., *Cosmological parameters from the 2003 flight of BOOMERANG*, *Astrophys. J.* **647** (2006) 799 [astro-ph/0507503].

[21] SPIDER collaboration, *A Constraint on Primordial B-modes from the First Flight of the Spider Balloon-borne Telescope*, *Astrophys. J.* **927** (2022) 174 [2103.13334].

[22] ACT collaboration, *The Atacama Cosmology Telescope: DR4 Maps and Cosmological Parameters*, *JCAP* **12** (2020) 047 [2007.07288].

[23] ACT collaboration, *The Atacama Cosmology Telescope: a measurement of the Cosmic Microwave Background power spectra at 98 and 150 GHz*, *JCAP* **12** (2020) 045 [2007.07289].

[24] SPT collaboration, *Measurements of the Temperature and E-Mode Polarization of the CMB from 500 Square Degrees of SPTpol Data*, *Astrophys. J.* **852** (2018) 97 [1707.09353].

[25] SPT collaboration, *Measurements of B-mode Polarization of the Cosmic Microwave Background from 500 Square Degrees of SPTpol Data*, *Phys. Rev. D* **101** (2020) 122003 [1910.05748].

[26] POLARBEAR collaboration, *A Measurement of the Cosmic Microwave Background B-Mode Polarization Power Spectrum at Sub-Degree Scales from 2 years of POLARBEAR Data*, *Astrophys. J.* **848** (2017) 121 [1705.02907].

[27] POLARBEAR collaboration, *A Measurement of the CMB E-mode Angular Power Spectrum at Subdegree Scales from670 Square Degrees of POLARBEAR Data*, *Astrophys. J.* **904** (2020) 65 [2005.06168].

[28] BICEP, KECK collaboration, *Improved Constraints on Primordial Gravitational Waves using Planck, WMAP, and BICEP/Keck Observations through the 2018 Observing Season*, *Phys. Rev. Lett.* **127** (2021) 151301 [2110.00483].

[29] LITEBIRD collaboration, *Probing Cosmic Inflation with the LiteBIRD Cosmic Microwave Background Polarization Survey*, *PTEP* **2023** (2023) 042F01 [2202.02773].

[30] V.F. Mukhanov and G.V. Chibisov, *Quantum Fluctuations and a Nonsingular Universe*, *JETP Lett.* **33** (1981) 532.

[31] A.A. Starobinsky, *Dynamics of Phase Transition in the New Inflationary Universe Scenario and Generation of Perturbations*, *Phys. Lett. B* **117** (1982) 175.

[32] S.W. Hawking, *The Development of Irregularities in a Single Bubble Inflationary Universe*, *Phys. Lett. B* **115** (1982) 295.

[33] A.H. Guth and S.Y. Pi, *Fluctuations in the New Inflationary Universe*, *Phys. Rev. Lett.* **49** (1982) 1110.

[34] J.M. Bardeen, P.J. Steinhardt and M.S. Turner, *Spontaneous creation of almost scale-free density perturbations in an inflationary universe*, *Phys. Rev. D* **28** (1983) 679.

[35] L.P. Grishchuk, *Amplification of gravitational waves in an istropic universe*, *Zh. Eksp. Teor. Fiz.* **67** (1974) 825.

[36] A.A. Starobinsky, *Spectrum of relict gravitational radiation and the early state of the universe*, *JETP Lett.* **30** (1979) 682.

[37] M. Zaldarriaga and U. Seljak, *An all sky analysis of polarization in the microwave background*, *Phys. Rev. D* **55** (1997) 1830 [astro-ph/9609170].

[38] M. Kamionkowski, A. Kosowsky and A. Stebbins, *Statistics of cosmic microwave background polarization*, *Phys. Rev. D* **55** (1997) 7368 [astro-ph/9611125].

[39] U. Seljak and M. Zaldarriaga, *Signature of gravity waves in polarization of the microwave background*, *Phys. Rev. Lett.* **78** (1997) 2054 [astro-ph/9609169].

[40] M. Kamionkowski, A. Kosowsky and A. Stebbins, *A Probe of primordial gravity waves and vorticity*, *Phys. Rev. Lett.* **78** (1997) 2058 [astro-ph/9609132].

[41] M. Tristram et al., *Improved limits on the tensor-to-scalar ratio using BICEP and Planck data*, *Phys. Rev. D* **105** (2022) 083524 [2112.07961].

[42] M. Tristram et al., *Planck constraints on the tensor-to-scalar ratio*, *Astron. Astrophys.* **647** (2021) A128 [2010.01139].

[43] P. Campeti and E. Komatsu, *New Constraint on the Tensor-to-scalar Ratio from the Planck and BICEP/Keck Array Data Using the Profile Likelihood*, *Astrophys. J.* **941** (2022) 110 [2205.05617].

[44] Simons Observatory collaboration, *The Simons Observatory: Science goals and forecasts*, *JCAP* **02** (2019) 056 [1808.07445].

[45] L. Moncelsi et al., *Receiver development for BICEP Array, a next-generation CMB polarimeter at the South Pole*, *Proc. SPIE Int. Soc. Opt. Eng.* **11453** (2020) 1145314 [2012.04047].

[46] K. Abazajian et al., *CMB-S4 Science Case, Reference Design, and Project Plan*, 1907.04473.

[47] NASA PICO collaboration, *PICO: Probe of Inflation and Cosmic Origins*, 1902.10541.

[48] M. Kamionkowski and E.D. Kovetz, *The Quest for B Modes from Inflationary Gravitational Waves*, *Ann. Rev. Astron. Astrophys.* **54** (2016) 227 [1510.06042].

[49] E. Komatsu, *New physics from the polarized light of the cosmic microwave background*, *Nature Rev. Phys.* **4** (2022) 452 [2202.13919].

[50] D.J.E. Marsh, *Axion Cosmology*, *Phys. Rept.* **643** (2016) 1 [1510.07633].

[51] E.G.M. Ferreira, *Ultra-light dark matter*, *Astron. Astrophys. Rev.* **29** (2021) 7 [2005.03254].

[52] S.M. Carroll, G.B. Field and R. Jackiw, *Limits on a Lorentz and Parity Violating Modification of Electrodynamics*, *Phys. Rev. D* **41** (1990) 1231.

[53] S.M. Carroll and G.B. Field, *Einstein equivalence principle and the polarization of radio galaxies*, *Phys. Rev. D* **43** (1991) 3789.

[54] D. Harari and P. Sikivie, *Effects of a Nambu-Goldstone boson on the polarization of radio galaxies and the cosmic microwave background*, *Phys. Lett. B* **289** (1992) 67.

[55] QUaD collaboration, *Parity Violation Constraints Using Cosmic Microwave Background Polarization Spectra from 2006 and 2007 Observations by the QUaD Polarimeter*, *Phys. Rev. Lett.* **102** (2009) 161302 [0811.0618].

[56] WMAP collaboration, *Seven-Year Wilkinson Microwave Anisotropy Probe (WMAP) Observations: Cosmological Interpretation*, *Astrophys. J. Suppl.* **192** (2011) 18 [1001.4538].

[57] B.G. Keating, M. Shimon and A.P.S. Yadav, *Self-Calibration of CMB Polarization Experiments*, *Astrophys. J. Lett.* **762** (2012) L23 [1211.5734].

[58] LiteBIRD collaboration, *In-flight polarization angle calibration for LiteBIRD: blind challenge and cosmological implications*, *JCAP* **01** (2022) 039 [2111.09140].

[59] Y. Minami, H. Ochi, K. Ichiki, N. Katayama, E. Komatsu and T. Matsumura, *Simultaneous determination of the cosmic birefringence and miscalibrated polarization angles from CMB experiments*, *PTEP* **2019** (2019) 083E02 [1904.12440].

[60] Y. Minami, *Determination of miscalibrated polarization angles from observed cosmic microwave background and foreground EB power spectra: Application to partial-sky observation*, *PTEP* **2020** (2020) 063E01 [2002.03572].

[61] Y. Minami and E. Komatsu, *Simultaneous determination of the cosmic birefringence and miscalibrated polarization angles II: Including cross frequency spectra*, *PTEP* **2020** (2020) 103E02 [2006.15982].

[62] Y. Minami and E. Komatsu, *New Extraction of the Cosmic Birefringence from the Planck 2018 Polarization Data*, *Phys. Rev. Lett.* **125** (2020) 221301 [2011.11254].

[63] Planck collaboration, *Planck 2018 results. III. High Frequency Instrument data processing and frequency maps*, *Astron. Astrophys.* **641** (2020) A3 [1807.06207].

[64] P. Diego-Palazuelos et al., *Cosmic Birefringence from the Planck Data Release 4*, *Phys. Rev. Lett.* **128** (2022) 091302 [2201.07682].

[65] J.R. Eskilt, *Frequency-Dependent Constraints on Cosmic Birefringence from the LFI and HFI Planck Data Release 4*, *Astron. Astrophys.* **662** (2022) A10 [2201.13347].

[66] J.R. Eskilt and E. Komatsu, *Improved constraints on cosmic birefringence from the WMAP and Planck cosmic microwave background polarization data*, *Phys. Rev. D* **106** (2022) 063503 [2205.13962].

[67] S.E. Clark, C.-G. Kim, J.C. Hill and B.S. Hensley, *The Origin of Parity Violation in Polarized Dust Emission and Implications for Cosmic Birefringence*, *Astrophys. J.* **919** (2021) 53 [2105.00120].

[68] B.R. Johnson et al., *MAXIPOL: Cosmic Microwave Background Polarimetry Using a Rotating Half-Wave Plate*, *Astrophys. J.* **665** (2007) 42 [astro-ph/0611394].

[69] B. Reichborn-Kjennerud et al., *EBEX: A balloon-borne CMB polarization experiment*, *Proc. SPIE Int. Soc. Opt. Eng.* **7741** (2010) 77411C [1007.3672].

[70] ABS collaboration, *Modulation of cosmic microwave background polarization with a warm rapidly rotating half-wave plate on the Atacama B-Mode Search instrument*, *Rev. Sci. Instrum.* **85** (2014) 024501 [`1310.3711`].

[71] A.S. Rahlin et al., *Pre-flight integration and characterization of the SPIDER balloon-borne telescope*, *Proc. SPIE Int. Soc. Opt. Eng.* **9153** (2014) 915313 [`1407.2906`].

[72] R. Misawa et al., *PILOT: a balloon-borne experiment to measure the polarized FIR emission of dust grains in the interstellar medium*, *Proc. SPIE Int. Soc. Opt. Eng.* **9153** (2014) 91531H [`1410.5760`].

[73] C.A. Hill et al., *Design and development of an ambient-temperature continuously-rotating achromatic half-wave plate for CMB polarization modulation on the POLARBEAR-2 experiment*, *Proc. SPIE Int. Soc. Opt. Eng.* **9914** (2016) 99142U [`1607.07399`].

[74] S. Takakura et al., *Performance of a continuously rotating half-wave plate on the POLARBEAR telescope*, *JCAP* **05** (2017) 008 [`1702.07111`].

[75] N. Galitzki, P. Ade, F.E. Angilè, P. Ashton, J. Austermann, T. Billings et al., *Instrumental performance and results from testing of the BLAST-TNG receiver, submillimeter optics, and MKID detector arrays*, in *Millimeter, Submillimeter, and Far-Infrared Detectors and Instrumentation for Astronomy VIII*, W.S. Holland and J. Zmuidzinas, eds., vol. 9914 of *Society of Photo-Optical Instrumentation Engineers (SPIE) Conference Series*, p. 99140J, July, 2016, DOI [`1608.05456`].

[76] S. Bryan et al., *A cryogenic rotation stage with a large clear aperture for the half-wave plates in the Spider instrument*, *Rev. Sci. Instrum.* **87** (2016) 014501 [`1510.01771`].

[77] ABS collaboration, *Systematic effects from an ambient-temperature, continuously rotating half-wave plate*, *Rev. Sci. Instrum.* **87** (2016) 094503 [`1601.05901`].

[78] M. Monelli, E. Komatsu, A.E. Adler, M. Billi, P. Campeti, N. Dachlythra et al., *Impact of half-wave plate systematics on the measurement of cosmic birefringence from CMB polarization*, *JCAP* **03** (2023) 034 [`2211.05685`].

[79] M. Monelli, E. Komatsu, T. Ghigna, T. Matsumura, G. Pisano and R. Takaku, *Impact of half-wave plate systematics on the measurement of CMB B-mode polarization*, `2311.07999`.

[80] M. Zaldarriaga, *Nature of the E B decomposition of CMB polarization*, *Phys. Rev. D* **64** (2001) 103001 [`astro-ph/0106174`].

[81] A. Lue, L. Wang and M. Kamionkowski, *Cosmological Signature of New Parity-Violating Interactions*, *Phys. Rev. Lett.* **83** (1999) 1506 [`astro-ph/9812088`].

[82] M. Zaldarriaga and U. Seljak, *Gravitational lensing effect on cosmic microwave background polarization*, *Phys. Rev. D* **58** (1998) 023003 [`astro-ph/9803150`].

[83] V.A. Rubakov, M.V. Sazhin and A.V. Veryaskin, *Graviton Creation in the Inflationary Universe and the Grand Unification Scale*, *Phys. Lett. B* **115** (1982) 189.

[84] R. Fabbri and M.d. Pollock, *The Effect of Primordially Produced Gravitons upon the Anisotropy of the Cosmological Microwave Background Radiation*, *Phys. Lett. B* **125** (1983) 445.

[85] L.F. Abbott and M.B. Wise, *Constraints on Generalized Inflationary Cosmologies*, *Nucl. Phys. B* **244** (1984) 541.

[86] A.A. Starobinsky, *Cosmic Background Anisotropy Induced by Isotropic Flat-Spectrum Gravitational-Wave Perturbations*, *Sov. Astron. Lett.* **11** (1985) 133.

[87] R. Crittenden, J.R. Bond, R.L. Davis, G. Efstathiou and P.J. Steinhardt, *The Imprint of gravitational waves on the cosmic microwave background*, *Phys. Rev. Lett.* **71** (1993) 324 [`astro-ph/9303014`].

[88] M.M. Basko and A.G. Polnarev, *Polarization and anisotropy of the RELICT radiation in an anisotropic universe*, *Mon. Not. Roy. Astron. Soc.* **191** (1980) 207.

[89] J.R. Bond and G. Efstathiou, *Cosmic background radiation anisotropies in universes dominated by nonbaryonic dark matter*, *Astrophys. J. Lett.* **285** (1984) L45.

[90] A.G. Polnarev, *Polarization and Anisotropy Induced in the Microwave Background by Cosmological Gravitational Waves*, *Sov. Astron.* **29** (1985) 607.

[91] R.G. Crittenden, D. Coulson and N.G. Turok, *Temperature-polarization correlations from tensor fluctuations*, *Phys. Rev. D* **52** (1995) R5402.

[92] D. Coulson, R.G. Crittenden and N.G. Turok, *Polarization and anisotropy of the microwave sky*, *Phys. Rev. Lett.* **73** (1994) 2390.

[93] A. Friedmann, *Über die Krümmung des Raumes*, *Zeitschrift fur Physik* **10** (1922) 377.

[94] H.P. Robertson, *Kinematics and World-Structure*, *Astrophys. J.* **82** (1935) 284.

[95] A.G. Walker, *On Milne's Theory of World-Structure*, *Proceedings of the London Mathematical Society* **42** (1937) 90.

[96] C. Dickinson, *CMB foregrounds - A brief review*, in *51st Rencontres de Moriond on Cosmology*, pp. 53–62, 6, 2016 [`1606.03606`].

[97] K. Ichiki, *CMB foreground: A concise review*, *PTEP* **2014** (2014) 06B109.

[98] G. Cowan, *Statistical data analysis*, Oxford University Press (1998).

[99] H.T. Ihle et al., *BEYONDPLANCK - VI. Noise characterization and modeling*, *Astron. Astrophys.* **675** (2023) A6 [2011.06650].

[100] E. Milotti, *1/f noise: a pedagogical review*, *arXiv e-prints* (2002) physics/0204033 [physics/0204033].

[101] A.J. Duivenvoorden, J.E. Gudmundsson and A.S. Rahlin, *Full-Sky Beam Convolution for Cosmic Microwave Background Applications*, *Mon. Not. Roy. Astron. Soc.* **486** (2019) 5448 [1809.05034].

[102] M. Tegmark, *How to make maps from CMB data without losing information*, *Astrophys. J. Lett.* **480** (1997) L87 [astro-ph/9611130].

[103] PLANCK collaboration, *Planck 2015 results. X. Diffuse component separation: Foreground maps*, *Astron. Astrophys.* **594** (2016) A10 [1502.01588].

[104] PLANCK collaboration, *Planck 2018 results. IV. Diffuse component separation*, *Astron. Astrophys.* **641** (2020) A4 [1807.06208].

[105] PLANCK collaboration, *Planck 2018 results. II. Low Frequency Instrument data processing*, *Astron. Astrophys.* **641** (2020) A2 [1807.06206].

[106] M. Gerbino, M. Lattanzi, M. Migliaccio, L. Pagano, L. Salvati, L. Colombo et al., *Likelihood methods for CMB experiments*, *Front. in Phys.* **8** (2020) 15 [1909.09375].

[107] M. Bass, C. DeCusatis, J. Enoch, V. Lakshminarayanan, G. Li, C. MacDonald et al., *Handbook of Optics, Third Edition Volume I: Geometrical and Physical Optics, Polarized Light, Components and Instruments(set)*, Handbook of Optics, McGraw-Hill Education (2009).

[108] A.J. Duivenvoorden, *Probing the early Universe with B-mode polarization: The Spider instrument, optical modelling and non-Gaussianity*, phd thesis, Stockholm University, Faculty of Science, Department of Physics., September, 2019.

[109] "qpoint documentation." https://github.com/arahlin/qpoint/blob/master/docs/qpoint.pdf.

[110] S. Pancharatnam, *Achromatic combinations of birefringent plates. Part I. An achromatic circular polarizer*, in *Proc. Indian Acad. Sci. 41*, p. 130–136, 1955, DOI.

[111] S. Pancharatnam, *Achromatic combinations of birefringent plates. Part II. An achromatic quarter-wave plate*, in *Proc. Indian Acad. Sci. 41*, p. 137–144, 1955, DOI.

[112] LITEBIRD collaboration, *Performance of a 200 mm Diameter Achromatic HWP with Laser-Ablated Sub-Wavelength Structures*, *J. Low Temp. Phys.* **211** (2023) 346.

[113] G. Pisano et al., *Development of large radii half-wave plates for CMB satellite missions*, *Proc. SPIE Int. Soc. Opt. Eng.* **9153** (2014) 915317 [1409.8516].

[114] G. Pisano, A. Ritacco, A. Monfardini, C. Tucker, P.A.R. Ade, A. Shitvov et al., *Development and application of metamaterial-based half-wave plates for the NIKA and NIKA2 polarimeters*, *Astron. Astrophys.* **658** (2022) A24 [2006.12081].

[115] A.J. Duivenvoorden, A.E. Adler, M. Billi, N. Dachlythra and J.E. Gudmundsson, *Probing frequency-dependent half-wave plate systematics for CMB experiments with full-sky beam convolution simulations*, *Mon. Not. Roy. Astron. Soc.* **502** (2021) 4526 [2012.10437].

[116] POLARBEAR collaboration, *A Measurement of the Degree Scale CMB B-mode Angular Power Spectrum with POLARBEAR*, *Astrophys. J.* **897** (2020) 55 [1910.02608].

[117] SPT-3G collaboration, *Measurements of the E-mode polarization and temperature-E-mode correlation of the CMB from SPT-3G 2018 data*, *Phys. Rev. D* **104** (2021) 022003 [2101.01684].

[118] S.A. Bryan et al., *Modeling and characterization of the SPIDER half-wave plate*, *Proc. SPIE Int. Soc. Opt. Eng.* **7741** (2010) 77412B [1006.3874].

[119] M.H. Abitbol, Z. Ahmed, D. Barron, R. Basu Thakur, A.N. Bender, B.A. Benson et al., *CMB-S4 Technology Book, First Edition*, *arXiv e-prints* (2017) arXiv:1706.02464 [1706.02464].

[120] S. Giardiello et al., *Detailed study of HWP non-idealities and their impact on future measurements of CMB polarization anisotropies from space*, *Astron. Astrophys.* **658** (2022) A15 [2106.08031].

[121] C. Bao et al., *The Impact of the Spectral Response of an Achromatic Half-Wave Plate on the Measurement of the Cosmic Microwave Background Polarization*, *Astrophys. J.* **747** (2012) 97 [1112.3057].

[122] T. Matsumura, *Mitigation of the spectral dependent polarization angle response for achromatic half-wave plate*, 1404.5795.

[123] C. Bao, C. Baccigalupi, B. Gold, S. Hanany, A. Jaffe and R. Stompor, *Maximum Likelihood Foreground Cleaning for Cosmic Microwave Background Polarimeters in the Presence of Systematic Effects*, *Astrophys. J.* **819** (2016) 12 [1510.08796].

[124] C. Vergès, J. Errard and R. Stompor, *Framework for analysis of next generation, polarized CMB data sets in the presence of Galactic foregrounds and systematic effects*, *Phys. Rev. D* **103** (2021) 063507 [2009.07814].

[125] S. di Serego Alighieri, *The conventions for the polarization angle*, *Exper. Astron.* **43** (2017) 19 [1612.03045].

[126] K.M. Górski, E. Hivon, A.J. Banday, B.D. Wandelt, F.K. Hansen, M. Reinecke et al., *HEALPix - A Framework for high resolution discretization, and fast analysis of data distributed on the sphere*, *Astrophys. J.* **622** (2005) 759 [astro-ph/0409513].

[127] A. Cooray, A. Melchiorri and J. Silk, *Is the cosmic microwave background circularly polarized?*, *Phys. Lett. B* **554** (2003) 1 [astro-ph/0205214].

[128] S. Alexander, J. Ochoa and A. Kosowsky, *Generation of Circular Polarization of the Cosmic Microwave Background*, *Phys. Rev. D* **79** (2009) 063524 [0810.2355].

[129] E. Bavarsad, M. Haghighat, Z. Rezaei, R. Mohammadi, I. Motie and M. Zarei, *Generation of circular polarization of the CMB*, *Phys. Rev. D* **81** (2010) 084035 [0912.2993].

[130] M. Sadegh, R. Mohammadi and I. Motie, *Generation of circular polarization in CMB radiation via nonlinear photon-photon interaction*, *Phys. Rev. D* **97** (2018) 023023 [1711.06997].

[131] K. Inomata and M. Kamionkowski, *Circular polarization of the cosmic microwave background from vector and tensor perturbations*, *Phys. Rev. D* **99** (2019) 043501 [1811.04957].

[132] A. Vahedi, J. Khodagholizadeh, R. Mohammadi and M. Sadegh, *Generation of Circular Polarization of CMB via Polarized Compton Scattering*, *JCAP* **01** (2019) 052 [1809.08137].

[133] S. Alexander, E. McDonough, A. Pullen and B. Shapiro, *Physics Beyond The Standard Model with Circular Polarization in the CMB and CMB-21cm Cross-Correlation*, *JCAP* **01** (2020) 032 [1911.01418].

[134] N. Bartolo, A. Hoseinpour, S. Matarrese, G. Orlando and M. Zarei, *CMB Circular and B-mode Polarization from New Interactions*, *Phys. Rev. D* **100** (2019) 043516 [1903.04578].

[135] M. Lembo, M. Lattanzi, L. Pagano, A. Gruppuso, P. Natoli and F. Forastieri, *Cosmic Microwave Background Polarization as a Tool to Constrain the Optical Properties of the Universe*, *Phys. Rev. Lett.* **127** (2021) 011301 [2007.08486].

[136] B. Feng, H. Li, M.-z. Li and X.-m. Zhang, *Gravitational leptogenesis and its signatures in CMB*, *Phys. Lett. B* **620** (2005) 27 [hep-ph/0406269].

[137] G. Patanchon, H. Imada, H. Ishino and T. Matsumura, *Effect of Instrumental Polarization with a Half-Wave Plate on the B-Mode Signal: Prediction and Correction*, arXiv e-prints (2023) [2308.00967].

[138] M. Tegmark, A. de Oliveira-Costa and A.J.S. Hamilton, *High resolution foreground cleaned cmb map from wmap*, *Phys. Rev. D* **68** (2003) 123523.

[139] B. Thorne, J. Dunkley, D. Alonso and S. Naess, *The Python Sky Model: software for simulating the Galactic microwave sky*, *Mon. Not. Roy. Astron. Soc.* **469** (2017) 2821 [1608.02841].

[140] T. Ghigna, T. Matsumura, G. Patanchon, H. Ishino and M. Hazumi, *Requirements for future CMB satellite missions: photometric and band-pass response calibration*, *JCAP* **11** (2020) 030 [2004.11601].

[141] "`FGBuster`." https://github.com/fgbuster/fgbuster.

[142] BeyondPlanck collaboration, *BEYONDPLANCK - VII. Bayesian estimation of gain and absolute calibration for cosmic microwave background experiments*, *Astron. Astrophys.* **675** (2023) A7 [2011.08082].

[143] C.H. Lineweaver, L. Tenorio, G.F. Smoot, P. Keegstra, A.J. Banday and P. Lubin, *The dipole observed in the COBE DMR four-year data*, *Astrophys. J.* **470** (1996) 38 [astro-ph/9601151].

[144] M. Piat, G. Lagache, J.P. Bernard, M. Giard and J.L. Puget, *Cosmic background dipole measurements with planck-high frequency instrument*, *Astron. Astrophys.* **393** (2002) 359 [astro-ph/0110650].

[145] C.L. Bennett, R.S. Hill, G. Hinshaw, M.R. Nolta, N. Odegard, L. Page et al., *First-Year Wilkinson Microwave Anisotropy Probe (WMAP) Observations: Foreground Emission*, *Astrophys. J.* **148** (2003) 97 [astro-ph/0302208].

[146] N. Katayama and E. Komatsu, *Simple foreground cleaning algorithm for detecting primordial B-mode polarization of the cosmic microwave background*, *Astrophys. J.* **737** (2011) 78 [1101.5210].

[147] H.K. Eriksen, J.B. Jewell, C. Dickinson, A.J. Banday, K.M. Gorski and C.R. Lawrence, *Joint Bayesian component separation and CMB power spectrum estimation*, *Astrophys. J.* **676** (2008) 10 [0709.1058].

[148] A. Lewis and A. Challinor, "CAMB: Code for Anisotropies in the Microwave Background." Astrophysics Source Code Library, record ascl:1102.026, Feb., 2011.

[149] A.A. Starobinsky, *A New Type of Isotropic Cosmological Models Without Singularity*, *Phys. Lett. B* **91** (1980) 99.

[150] PLANCK collaboration, *Planck 2018 results. XI. Polarized dust foregrounds*, *Astron. Astrophys.* **641** (2020) A11 [1801.04945].

[151] J.L. Weiland, G.E. Addison, C.L. Bennett, M. Halpern and G. Hinshaw, *An Examination of Galactic Polarization with Application to the Planck TB Correlation*, *Astrophys. J.* **893** (2020) 119 [1907.02486].

[152] K.M. Huffenberger, A. Rotti and D.C. Collins, *The Power Spectra of Polarized, Dusty Filaments*, *Astrophys. J.* **899** (2020) 31 [1906.10052].

[153] K. Komatsu, H. Ishino, N. Katayama, T. Matsumura and Y. Sakurai, *Design of a frequency-independent optic axis Pancharatnam-based achromatic half-wave plate*, *Journal of Astronomical Telescopes, Instruments, and Systems* **7** (2021) 034005.

[154] G. Pisano, C. Tucker, P.A.R. Ade, P. Moseley and M.W. Ng, *Metal mesh based metamaterials for millimetre wave and thz astronomy applications*, in *2015 8th UK, Europe, China Millimeter Waves and THz Technology Workshop (UCMMT)*, pp. 1–4, 2015, DOI.

[155] J. Delabrouille, J.F. Cardoso, M.L. Jeune, M. Betoule, G. Fay and F. Guilloux, *A full sky, low foreground, high resolution CMB map from WMAP*, *Astron. Astrophys.* **493** (2009) 835 [0807.0773].

[156] M. Remazeilles, A. Rotti and J. Chluba, *Peeling off foregrounds with the constrained moment ILC method to unveil primordial CMB B-modes*, *Mon. Not. Roy. Astron. Soc.* **503** (2021) 2478 [2006.08628].

[157] LITEBIRD collaboration, *Multiclustering needlet ILC for CMB B-mode component separation*, *Mon. Not. Roy. Astron. Soc.* **525** (2023) 3117 [2212.04456].

[158] A. Zonca, B. Thorne, N. Krachmalnicoff and J. Borrill, *The Python Sky Model 3 software*, *J. Open Source Softw.* **6** (2021) 3783 [2108.01444].

[159] Planck Collaboration, R. Adam, P.A.R. Ade, N. Aghanim, M. Arnaud, M. Ashdown et al., *Planck 2015 results. VII. High Frequency Instrument data processing: Time-ordered information and beams*, *Astron. Astrophys.* **594** (2016) A7 [1502.01586].

[160] L. Isserlis, *On a formula for the product-moment coefficient of any order of a normal frequency distribution in any number of variables*, *Biometrika* **12** (1918) 134.

# Acknowledgements

First of all, I would like to express my sincere gratitude to my supervisor, Eiichiro Komatsu, for his guidance and encouragement over the past four years. Thank you for giving me this amazing opportunity and allowing me to experience worlds I couldn't even imagine. It has been a privilege to work with you and to be your student.

I am also grateful to all the other brilliant researchers I was lucky enough to work with (especially Martin Reinecke, Paolo Campeti, Jon Gudmundsson, Adriaan Duivenvoorden, Nadia Dachlythra, Alexandre Adler, Tomotake Matsumura, and Tommaso Ghigna) for all their help and support in the projects we did together. I am especially grateful to Martin for going through my wonky code without ever flinching. Special thanks also go to Vyoma Muralidhara and Shaghaiegh Azyzy for proofreading parts of this thesis.

I am extremely grateful to Elisa Ferreira, Fabian Schmidt, José Luis Bernal, Julia Stadler and Luisa Lucie-Smith for being amazing examples of mentorship. Among other things, I am 100% sure that I couldn't have written this thesis without the 'jar of shame'.

I would also like to thank all the wonderful people I have met during my time in Munich. In particular, Laura for working with me on the 'most important project of the PhD', Vyoma for her endless kindness and patience in listening to my ramblings, the friends of rucolone for always helping me get out of my head, and Leonie for being Leonie.

I am deeply grateful to my family, the one I got and the one I chose. Thanks to my parents for their constant support, to my brother for teaching me the wonders of Timbits, and to my sister for being the best sorellam***a I could ever ask for. Thanks to Fabio, Martolina, Tiziana and Valerio, for still being there for me even though we have all changed so much. And finally, thanks to Luca for supporting me and loving me over all these years, I look forward to all our next adventures.

> *"now that you are free*
> *and the only obligation you are under*
> *is your own dreams*
> *what will you do*
> *with your time"*

Rupi Kaur - home body